\pdfoutput=1
\newcommand*{\ATLASLATEXPATH}{./}
\documentclass[cernpreprint,UKenglish,texlive=2016]{\ATLASLATEXPATH atlasdoc}
\usepackage[utf8]{inputenc}

\usepackage{\ATLASLATEXPATH atlaspackage}

\usepackage{\ATLASLATEXPATH atlasbiblatex}

\usepackage{\ATLASLATEXPATH atlasphysics}

\graphicspath{{./}{./}}

\usepackage{EXOT-2016-35-PAPER-defs}
\usepackage{norm}

\AtlasTitle{\LARGE Search for pair and single production of vectorlike quarks in final states with at least one $Z$ boson decaying into a pair of electrons or muons in $pp$ collision data collected with the ATLAS detector at $\sqrt{s}=13~\TeV$}

\author{The ATLAS Collaboration}

\PreprintIdNumber{CERN-EP-2018-145}
\AtlasJournalRef{Phys.\ Rev.\ D 98 (2018) 112010}
\AtlasDOI{10.1103/PhysRevD.98.112010}

\AtlasAbstract{
A search for vectorlike quarks is presented, which targets their decay into a $Z$ boson and a third-generation Standard Model quark. In the case of a vectorlike quark $T$ ($B$) with charge $+\sfrac{2}{3}e$ ($-\sfrac{1}{3}e$), the decay searched for is $T \rightarrow Zt$ ($B \rightarrow Zb$). Data for this analysis were taken during 2015 and 2016 with the ATLAS detector at the Large Hadron Collider and correspond to an integrated luminosity of \lumi\ of $pp$ collisions at $\sqrt{s}=13~\TeV$. The final state used is characterized by the presence of $b$-tagged jets, as well as a $Z$ boson with high transverse momentum, which is reconstructed from a pair of opposite-sign same-flavor leptons. Pair and single production of vectorlike quarks are both taken into account and are each searched for using optimized dileptonic exclusive and trileptonic inclusive event selections. In these selections, the high scalar sum of jet transverse momenta, the presence of high-transverse-momentum large-radius jets, as well as---in the case of the single-production selections---the presence of forward jets are used. No significant excess over the background-only hypothesis is found and exclusion limits at 95\% confidence level allow masses of vectorlike quarks of $m_T > 1030~\GeV$ ($m_T > 1210~\GeV$) and $m_B > 1010~\GeV$ ($m_B > 1140~\GeV$) in the singlet (doublet) model. In the case of 100\% branching ratio for $T\rightarrow Zt$ ($B\rightarrow Zb$), the limits are $m_T > 1340~\GeV$ ($m_B > 1220~\GeV$). Limits at 95\% confidence level are also set on the coupling to Standard Model quarks for given vectorlike quark masses.
}

\hypersetup{pdftitle={ATLAS document},pdfauthor={The ATLAS Collaboration}}

\usepackage{rotating}
\usepackage{xfrac}
\usepackage{array}
\newcolumntype{R}[1]{>{\raggedleft}p{#1}}
\newcolumntype{L}[1]{>{\raggedright}p{#1}}

\newcommand{\dilres}{PP~2$\ell$~0-1J}

\newcommand{\dilboost}{PP~2$\ell$~$\geq$2J}
\newcommand{\tripair}{PP~$\geq$3$\ell$}
\newcommand{\dilsing}{SP~2$\ell$}
\newcommand{\trising}{SP~$\geq$3$\ell$}

\newcommand{\capdilres}{(PP)~2$\ell$~0-1J}
\newcommand{\capdilboost}{(PP)~2$\ell$~$\geq$2J}
\newcommand{\captripair}{(PP)~$\geq$3$\ell$}
\newcommand{\capdilsing}{(SP)~2$\ell$}
\newcommand{\captrising}{(SP)~$\geq$3$\ell$}

\begin{document}

\maketitle

\tableofcontents

\clearpage

\section{Introduction}
\label{sec:intro}
In the Standard Model (SM), the electromagnetic and weak interactions arise from a $\mathrm{SU}(2)_L \times \mathrm{U}(1)_Y$ gauge symmetry that is spontaneously broken by the Englert--Brout--Higgs mechanism. Measurements at collider experiments are so far consistent with its predictions. However, the SM is believed to be only a low-energy approximation of a more fundamental theory due to several unanswered questions. For example, it cannot explain the matter--antimatter asymmetry in the universe and the origin of dark matter. When the SM is extrapolated to high energies, fine-tuning is required due to divergent corrections to the Higgs boson self-energy~\cite{Naturalness}. Solutions to this so-called ``hierarchy problem'' are proposed in several beyond-the-Standard Model (BSM) theories, which can be considered a first step toward a more fundamental theory of particle physics.

Since a large contribution to the fine-tuning originates from top-quark loop corrections, the hierarchy problem can be reduced in models predicting top-quark partners that mitigate the SM top quark's contribution: while a scalar top-quark partner appears in supersymmetry as the bosonic superpartner of the top quark, fermionic top-quark partners appear in theories with a new broken global symmetry, in which the Higgs boson is interpreted as a pseudo Nambu--Goldstone boson~\cite{StrongEWSB}, for example in Little Higgs~\cite{LittleHiggs,LittleHiggsRev} and Composite Higgs~\cite{CompHiggs1,CompHiggs2} models. In these models, the new symmetry corresponds to a new strong interaction, whose bound states include vectorlike quarks (VLQ). These are color-triplet spin-$1/2$ fermions, but in contrast to the chiral SM quarks their left- and right-handed components have the same properties under $\mathrm{SU}(2)_L \times \mathrm{U}(1)_Y$ transformations.

Only a limited set of possibilities exists for the quantum numbers of the VLQs if gauge invariance is required to be preserved~\cite{delAguila,delAguila:2000rc}. Their electric charge could be $+\sfrac{2}{3}e$ ($T$ quark), $-\sfrac{1}{3}e$ ($B$ quark), $+\sfrac{5}{3}e$ ($X$ quark) or $-\sfrac{4}{3}e$ ($Y$ quark), where $e$ is the elementary charge, and they could appear in electroweak singlets, ($T$) or ($B$), electroweak doublets, ($X$~$T$), ($T$~$B$), or ($B$~$Y$), or electroweak triplets, ($X$~$T$~$B$) or ($T$~$B$~$Y$). This paper focuses solely on the search for $T$ and $B$ quarks, which could couple to SM quarks by mixing~\cite{VLQmixing}. Although couplings of VLQs to first- and second-generation SM quarks are not excluded~\cite{Atre:2008iu,Atre:2011ae}, this paper searches for VLQs that couple exclusively to third-generation SM quarks. The couplings of $T$ and $B$ quarks can be described in terms of $\sin\theta_T$ and $\sin\theta_B$~\cite{Aguilar-Saavedra:2013qpa}, where $\theta_T$ and $\theta_B$ are the mixing angles with the top quark and the $b$-quark, respectively, or they can be described in terms of generalized couplings $\kappa_T$ and $\kappa_B$ of the $T$ or $B$ quark to third-generation SM quarks~\cite{Buchkremer:2013bha,Matsedonskyi:2014mna}.

Search strategies for VLQs have been proposed~\cite{ContinoServant,JA_TP,TP_guide,Aguilar-Saavedra:2013qpa,Backovic:2015bca} that focus either on the search for VLQ pair production via the strong interaction or on single production via the electroweak interaction. The decay of $T$ and $B$ quarks can either happen via the charged current, i.e.\ $T\rightarrow Wb$ and $B\rightarrow Wt$,\footnote{Throughout this document, decays that are written in a short form, for example $T\rightarrow Zt$ or $B\bar{B}\rightarrow ZbWt$, also refer to the corresponding antiparticle decays, i.e.\ $\bar{T}\rightarrow Z\bar{t}$, and are understood to include the proper $W$ boson charge and antifermion notation, i.e.\ $B\bar{B}\rightarrow ZbW^+\bar{t}$ and $B\bar{B}\rightarrow Z\bar{b}W^-t$.} or via flavor-changing neutral currents~\cite{delAguila2}, i.e.\ $T\rightarrow Zt$, $T\rightarrow Ht$, $B\rightarrow Zb$, and $B\rightarrow Hb$. Decays including non-SM particles are not excluded~\cite{Chala:2017xgc}, but are not considered in this paper, so that for $T$ and $B$ quarks the branching ratios (BR) to the three decay modes add up to unity. While the cross section for pair production is given by quantum chromodynamics, the single-production cross section explicitly depends on the coupling of the VLQ to SM quarks.

\begin{table}[t]
\centering
\caption{Overview of the requirements used in each channel to search for pair and single production of VLQs.}
\begin{tabular}{l||m{0.151\textwidth}|m{0.151\textwidth}|m{0.125\textwidth}||m{0.125\textwidth}|m{0.125\textwidth}}\toprule
 & \multicolumn{3}{c||}{Pair-production (PP) channels} & \multicolumn{2}{c}{Single-production (SP) channels} \\ \midrule
 & Dilepton with $\leq 1$ \ljet & Dilepton with $\geq 2$ \ljets & Trilepton & Dilepton & Trilepton \\
 & (\dilres) & (\dilboost) & (\tripair) & (\dilsing) & (\trising) \\ \midrule
 Leptons & \multicolumn{2}{c|}{ $= 2$} & \multicolumn{1}{c||}{$\geq 3$} & \multicolumn{1}{c|}{$= 2$} & \multicolumn{1}{c}{$\geq 3$} \\ \midrule
 \btagged\ jets & \multicolumn{2}{c|}{ $\geq 2$} & \multicolumn{1}{c||}{$\geq 1$} & \multicolumn{2}{c}{$\geq 1$} \\ \midrule
 \Ljets & \multicolumn{1}{c|}{$\leq 1$} & \multicolumn{1}{c|}{$\geq 2$} & \multicolumn{1}{c||}{--} & \multicolumn{1}{c|}{$\geq 1$ (top-tagged)} & \multicolumn{1}{c}{--} \\ \midrule
 Forward jets & \multicolumn{3}{c||}{--} & \multicolumn{2}{c}{$\geq 1$} \\ \midrule
 \ptll  & \multicolumn{2}{c|}{$> 250~\GeV$} & \multicolumn{1}{c||}{$> 200~\GeV$} & \multicolumn{1}{c|}{$> 200~\GeV$} & $> 150~\GeV$ \\ \midrule
 \multicolumn{6}{c}{Additional optimized kinematic requirements for each channel} \\
 \bottomrule
\end{tabular}
\label{tab:selection_overview}
\end{table}

The ATLAS and CMS Collaborations have searched for pair production of $T$ and $B$ quarks that decay into third-generation quarks in $pp$ collisions at $\sqrt{s} = 8~\TeV$~\cite{Aad:2015kqa,Aad:2015gdg,Aad:2014efa,Aad:2015mba,Khachatryan:2015gza,Khachatryan:2015oba} in all three possible decay modes of each of the VLQs. Current searches at $\sqrt{s} = 13~\TeV$ have used single-lepton final states to search for the $T\rightarrow Zt$ decay with the $Z$ boson decaying invisibly~\cite{Aaboud:2017qpr,Aaboud:2018xuw}, $T\rightarrow Wb$~\cite{Aaboud:2017zfn,Sirunyan:2017pks}, $T\rightarrow Ht$~\cite{Aaboud:2018xuw}, and $B\rightarrow Wt$~\cite{Aaboud:2017zfn,WtX}, general single-lepton final states with boosted $W$ and Higgs bosons~\cite{Sirunyan:2017usq}, final states with leptons with the same electric charge~\cite{Aaboud:2018xpj}, and all-hadronic final states~\cite{Aaboud:2018wxv}. The CMS Collaboration has also searched for pair production of $T$ and $B$ quarks in a combination of single-lepton final states, dilepton final states with the same electric charge and trilepton final states~\cite{Sirunyan:2018omb} at $\sqrt{s} = 13~\TeV$. These searches have set upper limits at 95\% confidence level (CL) on the VLQ pair-production cross section, also interpreted as lower limits on the VLQ mass, $\mVLQ$, depending on the VLQ BRs assumed. The most stringent limits in the case of the $T$ and $B$ singlets are 1.20~\TeV~\cite{Sirunyan:2018omb} and 1.17~\TeV~\cite{WtX,Sirunyan:2018omb}, respectively. In the case of 100\% BRs of $T$ to $Zt$ and $B$ to $Zb$, the most stringent limits are 1.30~\TeV~\cite{Sirunyan:2018omb} and 0.96~\TeV~\cite{Sirunyan:2018omb}, respectively. The searches at $\sqrt{s} = 13~\TeV$ are significantly more sensitive than the searches at $\sqrt{s} = 8~\TeV$ due to the larger expected pair-production cross sections at the higher center-of-mass energy. This paper includes searches for pair-produced VLQs at $\sqrt{s} = 13~\TeV$ in final states with more than one lepton which are particularly sensitive to the decays $T\rightarrow Zt$ and $B\rightarrow Zb$.

At large $\mVLQ$, the cross section for the single production of VLQs may be larger than the pair-production cross section because of the larger available phase space, even though single production is mediated by the weak interaction. However, the comparison of single- and pair-production cross sections depends on the assumed coupling to the SM quarks. Single production was searched for at $\sqrt{s} = 8~\TeV$~\cite{Aad:2015voa,Aad:2014efa,Aad:2016qpo} by the ATLAS and CMS Collaborations. At $\sqrt{s} = 13~\TeV$, the CMS Collaboration has searched for the decays $T\rightarrow Wb$~\cite{Sirunyan:2017tfc}, $T\rightarrow Ht$~\cite{Sirunyan:2016ipo,Khachatryan:2016vph}, $T\rightarrow Zt$~\cite{Sirunyan:2017ynj,Sirunyan:2017ezy}, $B\rightarrow Hb$~\cite{Sirunyan:2018fjh}, $B\rightarrow Zb$~\cite{Sirunyan:2017ezy}, and $B\rightarrow Wt$~\cite{Sirunyan:2018ncp}. In these searches, upper limits were set on the single-production cross section, which were also interpreted as upper limits on the coupling to SM quarks as a function of $\mVLQ$. Similarly to the case of pair production, the expected single-production cross sections are much larger at $\sqrt{s} = 13~\TeV$ than at $\sqrt{s} = 8~\TeV$, so that the searches at the higher center-of-mass energy are more sensitive. Searches for single-$T$-quark production at $\sqrt{s} = 13~\TeV$ were not performed before by the ATLAS Collaboration. As in the search for VLQ pair production, final states with more than one lepton are used, which are particularly sensitive to the decay $T\rightarrow Zt$.

\begin{figure}[p]
\centering
\subfloat[]{\includegraphics[width=.49\textwidth]{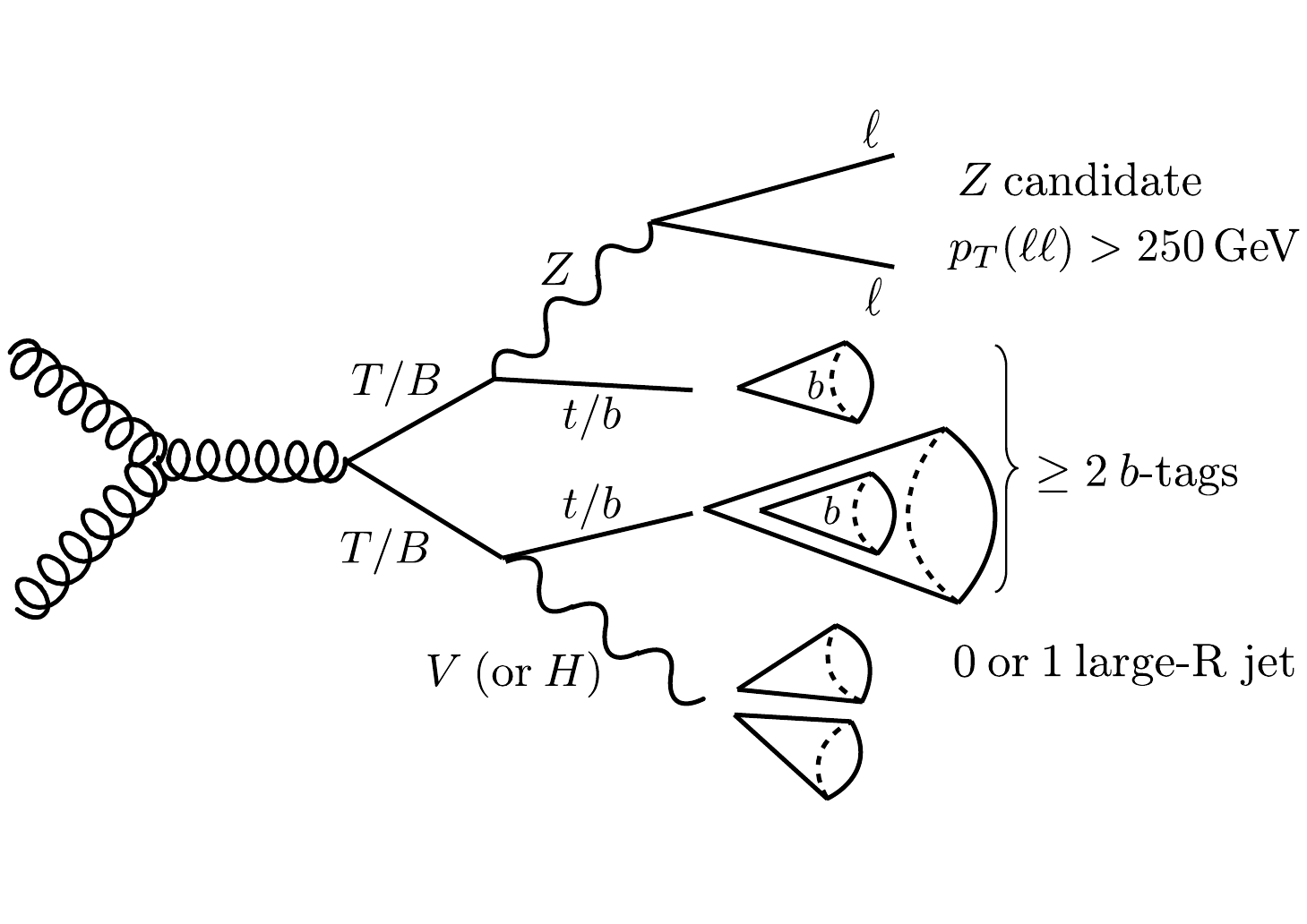}}
\subfloat[]{\includegraphics[width=.49\textwidth]{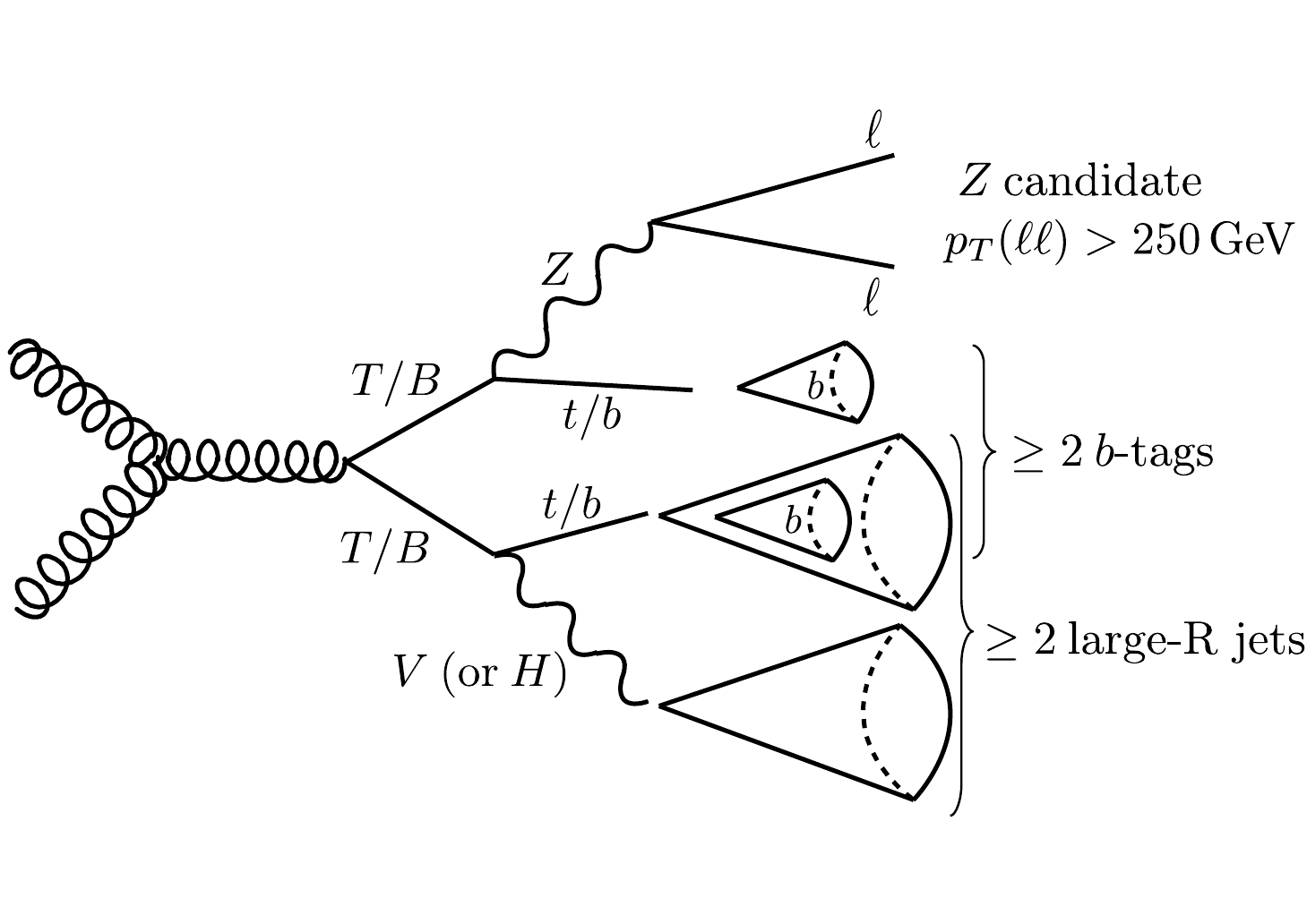}}\\
\subfloat[]{\includegraphics[width=.49\textwidth]{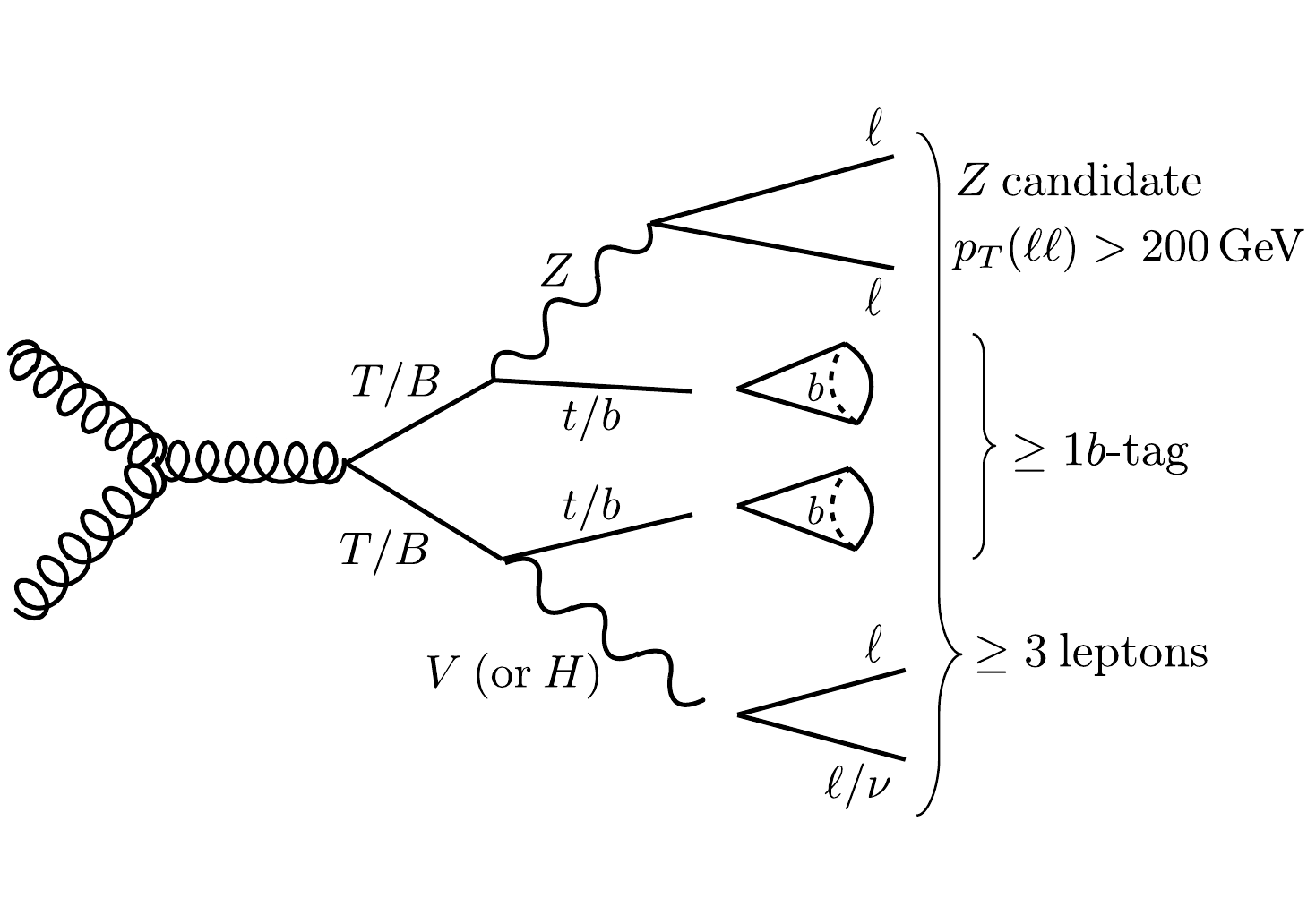}}\\
\subfloat[]{\includegraphics[width=.49\textwidth]{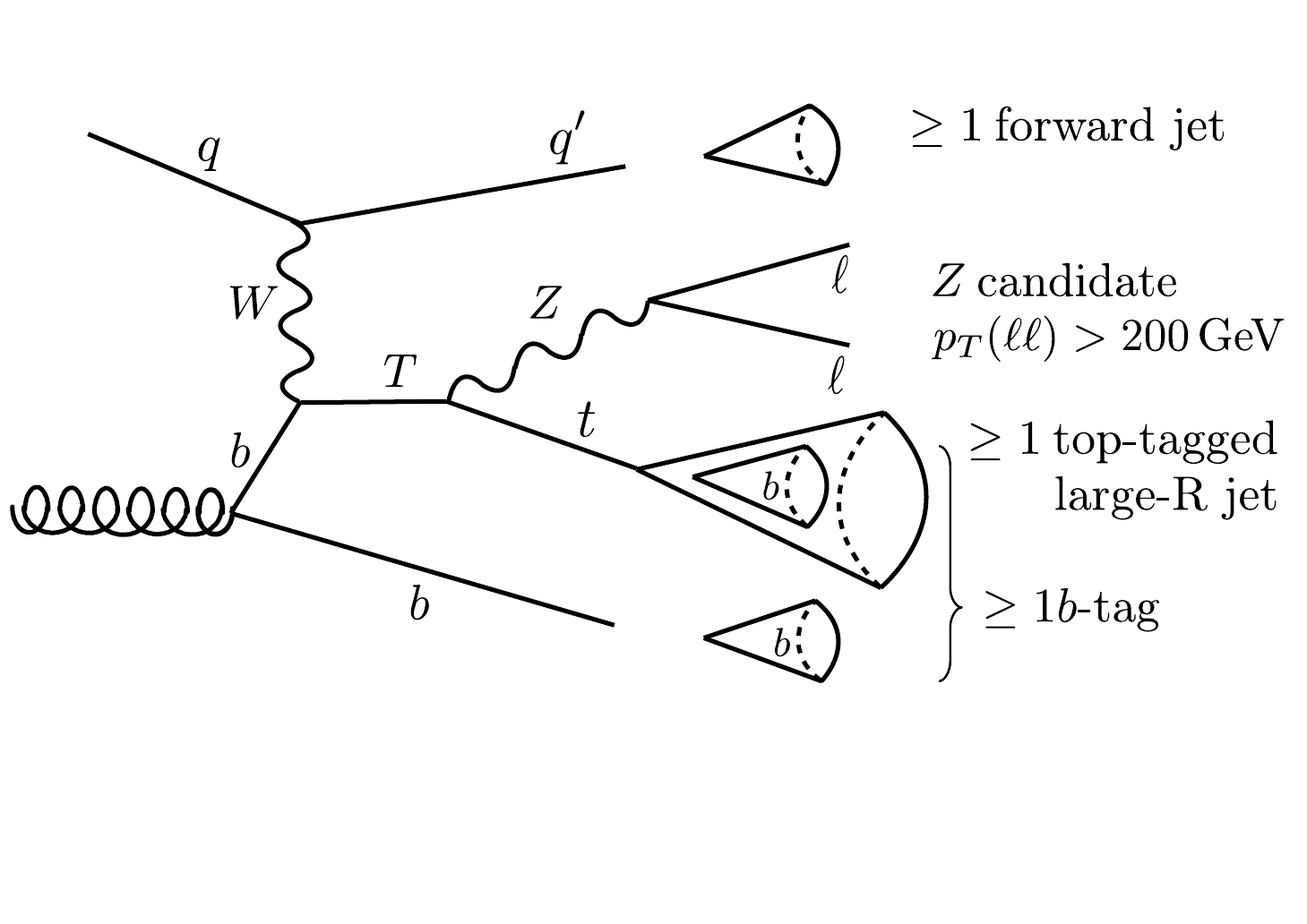}}
\subfloat[]{\includegraphics[width=.49\textwidth]{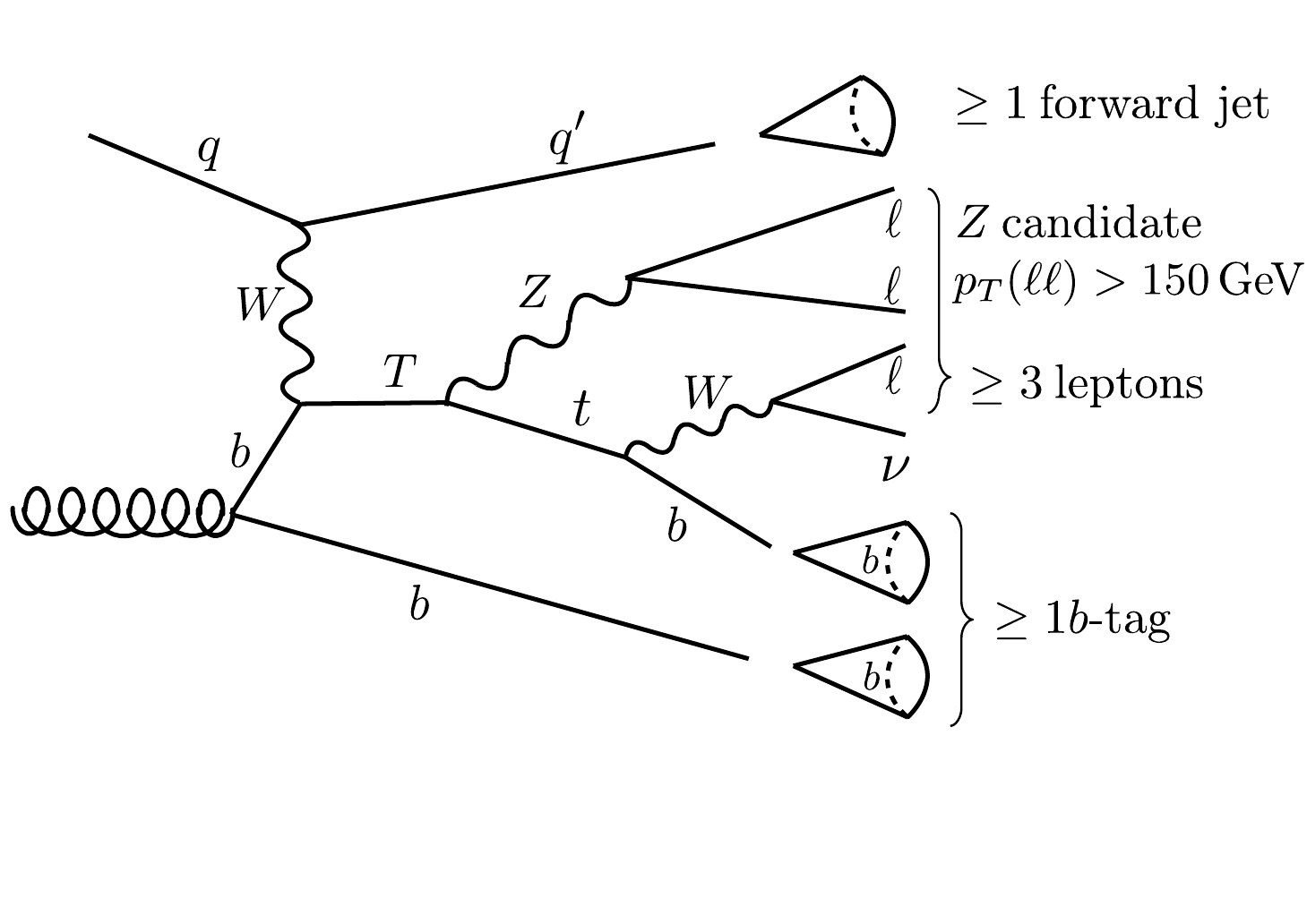}}
\caption{Sketches of the processes searched for in the pair-production channels in (a) dilepton final states with at most one \ljet\ (\dilres), (b) dilepton final states with at least two \ljets\ (\dilboost), and (c) final states with at least three leptons (\tripair), and sketches of the processes searched for in the single-production channels in (d) the dilepton final state (\dilsing), and (e) final states with at least three leptons (\trising). Only \sjets\ are \btagged. As the reconstruction of \sjets\ and \ljets\ is independent of each other, \btagged\ \sjets\ can overlap with \ljets.}
\label{fig:sketches}
\end{figure}

The analysis was performed with data collected in $pp$ collisions at $\sqrt{s} = 13~\TeV$, searching for the pair production of $T$ and $B$ quarks and for the single production of $T$ quarks in final states with at least one $Z$ boson. In the case of single production, the $T$ quark is hence expected to decay into $Zt$. In the case of pair production, the search targets only one VLQ decay into a $Z$ boson and a third-generation quark explicitly, so that it is particularly sensitive to all decays that include at least one $Z$ boson in the final state, i.e.\ not only $T\bar{T}\rightarrow ZtZt$ and $B\bar{B}\rightarrow ZbZb$, but also $T\bar{T}\rightarrow ZtWb$, $T\bar{T}\rightarrow ZtHt$, $B\bar{B}\rightarrow ZbWt$, and $B\bar{B}\rightarrow ZbHb$.

The overall analysis strategy is based on a search that was performed with $\sqrt{s} = 8~\TeV$ data~\cite{Aad:2014efa}, which exploited the leptonic $Z$ boson decays $Z\rightarrow e^+e^-$ and $Z\rightarrow\mu^+\mu^-$. Several improvements have been made, in particular new channels have been added and an event selection was used that was optimized for the higher $\sqrt{s}$ and a larger dataset. Five analysis channels are defined; three for the search for $T$ and $B$ pair production, and two for the search for single-$T$-quark production, as shown in \Tab{\ref{tab:selection_overview}}. An event preselection that is common to all channels is used, in which all events are required to include a $Z$ boson candidate, reconstructed from two same-flavor leptons ($e$, $\mu$) with opposite electric charge. The event selection in each channel was then optimized for a particular final state, as shown in \Fig{\ref{fig:sketches}}. First, the searches were split into pair- and single-production categories and then further into dilepton channels---requiring no lepton in addition to the leptons that are used to reconstruct the $Z$ boson candidate---and trilepton channels, in which at least three leptons are required. Since the VLQs are assumed to decay into third-generation SM quarks, the presence of $b$-tagged jets is exploited in order to discriminate the signal from SM background processes. Since the signal process includes high-energy hadronically-decaying massive resonances, \ljets\ (J) are used in the dilepton channels,  further enhancing the sensitivity of the search. In the dilepton single-production channel, top-tagging is used in order to identify \ljets\ originating from the hadronic decays of high-energy top quarks. Only \sjets\ are $b$-tagged. As the reconstruction of \sjets\ and \ljets\ is independent of each other, $b$-tagged \sjets\ can overlap with \ljets. In both single-production channels, the presence of a forward jet from the $t$-channel production is used to suppress the background. Due to the large expected values of $\mVLQ$, the transverse momentum\footnote{ATLAS uses a right-handed coordinate system with its origin at the nominal interaction point (IP) in the center of the detector and the $z$-axis along the beam pipe. The $x$-axis points from the IP to the center of the LHC ring, and the $y$-axis points upwards. Cylindrical coordinates $(r,\phi)$ are used in the transverse plane, $\phi$ being the azimuthal angle around the $z$-axis. The pseudorapidity is defined in terms of the polar angle $\theta$ as $\eta = -\ln \tan(\theta/2)$. Angular distance is measured in units of $\Delta R \equiv \sqrt{(\Delta\eta)^{2} + (\Delta\phi)^{2}}$. The transverse momentum is defined as $\pt = p\sin\theta = p/\cosh\eta$, and the transverse energy, \et, is defined analogously.} of the $Z$ boson, \ptll, is expected to be much larger in signal than in background events. More requirements, in particular on the event kinematics, were optimized in each channel, as discussed in \Sect{\ref{sec:selection}}. In the following, the three pair-production channels are referred to as the dilepton channel with at most one \ljet\ (\dilres), the dilepton channel with at least two \ljets\ (\dilboost), and the trilepton channel (\tripair). The two single-production channels are referred to as the dilepton channel (\dilsing), and the trilepton channel (\trising).

\section{The ATLAS detector}
The ATLAS detector~\cite{PERF-2007-01} at the LHC covers nearly the entire solid angle around the collision point. It consists of an inner tracking detector surrounded by a thin superconducting solenoid, electromagnetic and hadronic calorimeters, and a muon spectrometer incorporating three large superconducting toroidal magnets.

The inner-detector system (ID) is immersed in a \SI{2}{\tesla} axial magnetic field and provides charged-particle tracking in the range $|\eta| < 2.5$. The high-granularity silicon pixel detector covers the vertex region and typically provides four measurements per track, the first hit being normally in the innermost layer, the insertable B-layer~\cite{ATLAS-TDR-19}. It is followed by the silicon microstrip tracker which usually provides four two-dimensional measurement points per track. These silicon detectors are complemented by the transition radiation tracker, which enables radially extended track reconstruction up to $|\eta| = 2.0$. The transition radiation tracker also provides electron identification information based on the fraction of hits (typically 30 in total) above a higher energy-deposit threshold corresponding to transition radiation.

The calorimeter system covers the pseudorapidity range $|\eta| < 4.9$. Within the region $|\eta|< 3.2$, electromagnetic calorimetry is provided by barrel and endcap high-granularity lead/liquid-argon (LAr) electromagnetic calorimeters, with an additional thin LAr presampler covering $|\eta| < 1.8$, to correct for energy loss in material upstream of the calorimeters. Hadronic calorimetry is provided by the steel/scintillator-tile calorimeter, segmented into three barrel structures within $|\eta| < 1.7$, and two copper/LAr hadronic endcap calorimeters. The solid angle coverage is completed with forward copper/LAr and tungsten/LAr calorimeter modules optimized for electromagnetic and hadronic measurements respectively.

The muon spectrometer (MS) comprises separate trigger and high-precision tracking chambers measuring the deflection of muons in a magnetic field generated by superconducting air-core toroidal magnets. The field integral of the toroidal magnets ranges between \num{2.0} and \SI{6.0}{\tesla\metre} across most of the detector. A set of precision chambers covers the region $|\eta| < 2.7$ with three layers of monitored drift tubes, complemented by cathode strip chambers in the forward region, where the background is highest. The muon trigger system covers the range $|\eta| < 2.4$ with resistive plate chambers in the barrel, and thin gap chambers in the endcap regions.

A two-level trigger system is used in order to select interesting events~\cite{Aaboud:2016leb}. The first-level trigger is implemented in hardware and uses a subset of detector information to reduce the event rate to a design value of at most \SI{100}{\kHz}. This is followed by a software-based trigger which reduces the event rate to about \SI{1}{\kHz}.

\section{Data and Monte Carlo samples}
\label{sec:samples}
For this search, $pp$ collision data collected with the ATLAS detector during 2015 and 2016 at $\sqrt{s} = 13~\TeV$ were used, corresponding to an integrated luminosity of \lumi. Only events taken during stable beam conditions, and for which all relevant components of the detector were operational, are considered. Events are required to have a primary vertex with at least two tracks with a minimum \pt\ of 400~\MeV\ each. If several such vertices exist, the vertex with the highest $\sum_\mathrm{tracks} \pt^2$ is chosen as the hard-scatter vertex~\cite{ATL-PHYS-PUB-2015-026}. Events are rejected if they satisfy the criteria~\cite{ATLAS-CONF-2015-029} designed to reject beam-induced background and backgrounds from cosmic-ray showers and calorimeter noise. Several single-lepton triggers with different \pt\ thresholds were used for electrons and muons depending on the data-taking period. For data collected in 2015, the thresholds are 24, 60 and 120~\GeV\ for electrons and 20 and 50~\GeV\ for muons, where lepton isolation requirements are applied to the lowest-\pt\ triggers to reduce their rate. For the highest-\pt\ electron trigger, the identification criteria are relaxed. For data-taking in 2016, the thresholds were raised slightly to 26, 60 and 140~\GeV\ for electrons and 26 and 50~\GeV\ for muons.

The main sources of background in this search are \zjets\ and \ttbar\ production in the case of the dilepton channels and diboson ($WZ$, $WW$, $ZZ$) and $\ttbar+X$ production in the case of the trilepton channels, where $\ttbar+X$ is dominated by \ttbar\ production with associated vector bosons ($\ttbar+V$, $V=W$ or $Z$) but also includes $t\bar{t}t\bar{t}$ and $\ttbar WW$ production. Smaller sources of background are also considered,\footnote{In the figures in this paper the smaller backgrounds are grouped together and are denoted by ``Other.''} which include single-top and triboson production ($WWW$, $WWZ$, $WZZ$, $ZZZ$). The background contribution from $t\bar{t}H$ production was found to be negligible and is not considered in this search. For all background and signal processes, Monte Carlo (MC) samples were generated and the detector response was simulated in \texttt{GEANT4}~\cite{Agostinelli:2002hh} with a full model of the ATLAS detector~\cite{SOFT-2010-01}, unless stated otherwise. The simulations include the contributions from additional $pp$ collisions in the same or an adjacent bunch crossing (pileup). Corrections for trigger and object-identification efficiencies, and for $b$-tagging misidentification efficiencies, as well as for energy and momentum scales and resolutions of the objects were applied to the simulated samples, based on the differences observed between data and MC samples in reference processes. A summary of the background samples used in this paper is shown in \Tab{\ref{tab:samples}}.

The \zjets\ process was simulated with \SHERPAV{2.2.1}~\cite{Gleisberg:2008ta, Hoeche:2009rj, Gleisberg:2008fv, Schumann:2007mg} using the NNPDF3.0~\cite{Ball:2014uwa} next-to-next-to-leading-order (NNLO) set of parton distribution functions (PDFs), and normalized to the NNLO cross section in QCD\footnote{The order in perturbation theory refers to QCD throughout this paper.} calculated with FEWZ~\cite{Anastasiou:2003ds} and the MSTW 2008~\cite{Martin:2009iq,Martin:2009bu,Martin:2010db} NNLO PDF set. The \ttbar\ process was simulated with the POWHEG method~\cite{Nason:2004rx, Frixione:2007vw} implemented in \POWHEGBOXV{v2}~\cite{Alioli:2010xd, Campbell:2014kua} using the NNPDF3.0 NNLO PDF set. \POWHEGBOX was interfaced with \PYTHIAV{8}~\cite{Sjostrand:2007gs} with the A14 set of tuned\footnote{MC programs that model non-perturbative effects, such as the parton shower, hadronization, and multiple parton interaction need to be fit (``tuned'') to collider data, as the values of these parameters cannot be derived from first principles.} parameters~\cite{ATL-PHYS-PUB-2014-021} and the NNPDF2.3 leading order (LO) PDF set~\cite{Ball:2012cx} for parton showering and hadronization. The $h_{\mathrm{damp}}$ parameter\footnote{The $h_{\mathrm{damp}}$ parameter controls the transverse momentum of the first additional gluon emission beyond the Born configuration. The main effect of choosing $h_{\mathrm{damp}} = 1.5 m_t$ is to regulate the high-\pT\ emission against which the \ttbar\ system recoils.} in \POWHEGBOX was set to $1.5m_t$~\cite{ATL-PHYS-PUB-2017-007}, where $m_t = 172.5~\GeV$. The sample was normalized to the NNLO cross section including resummation of next-to-next-to-leading logarithmic (NNLL) soft gluon terms with \textsc{TOP++}~\cite{Czakon:2011xx,Czakon:2013goa, Beneke:2011mq, Cacciari:2011hy, Czakon:2012pz, Czakon:2012zr, Baernreuther:2012ws}. The PDF and $\alphas$ uncertainties were calculated using the PDF4LHC prescription~\cite{Botje:2011sn} with the MSTW 2008 NNLO, CT10 NNLO~\cite{Lai:2010vv,Gao:2013xoa} and NNPDF2.3 5f FFN PDF sets, added in quadrature to the scale uncertainty. The diboson processes were simulated with \SHERPAV{2.2.1} for up to one additional parton at next-to-leading order (NLO) and up to three additional partons at LO using \textsc{Comix}~\cite{Gleisberg:2008fv} and \textsc{OpenLoops}~\cite{Cascioli:2011va}, and merged with the \SHERPA\ parton shower~\cite{Schumann:2007mg} according to the ME+PS@NLO prescripton~\cite{Hoeche:2012yf}. The NNPDF3.0 NNLO PDF set was used and the samples were normalized to the NLO cross sections calculated with \SHERPA. The $\ttbar+V$ processes were simulated with \MGMCatNLO~\cite{Alwall:2014hca} using the NNPDF3.0 NLO PDF set. \MGMCatNLO was interfaced with \PYTHIAV{8} with the A14 set of tuned parameters and the NNPDF2.3 LO PDF set for parton showering and hadronization. The $\ttbar+V$ samples were normalized to the NLO cross section calculated with \MGMCatNLO. The single-top processes were simulated with \POWHEGBOXV{v1}~\cite{Alioli:2009je, Re:2010bp} using the CT10 PDF set. \POWHEGBOX was interfaced to \PYTHIAV{6}~\cite{Sjostrand:2006za} with the Perugia 2012~\cite{Skands:2010ak} set of tuned parameters and the CTEQ6L1 PDF set~\cite{Pumplin:2002vw}. The single-top samples were normalized to NLO cross sections with additional NNLL soft gluon terms~\cite{Kidonakis:2011wy,Kidonakis:2010tc,Kidonakis:2010ux}.
The triboson processes were simulated using \SHERPAV{2.1} using the CT10 PDF set, and normalized to the NLO cross sections calculated with \SHERPA. The $\ttbar\ttbar$ and $\ttbar+WW$ processes were simulated with \MADGRAPHV{5} and \PYTHIAV{8} using the NNPDF2.3 LO PDF set and the A14 set of tuned parameters, and were normalized to the NLO cross section calculated with \MGMCatNLO. Additional MC samples were generated for the evaluation of systematic uncertainties due to the choice of factorization and renormalization scales, generator, and parton shower program for the \zjets, \ttbar, $\ttbar+V$ and diboson background processes. These samples are described in \Sect{\ref{sec:systematics}}.

The pair production of VLQs was simulated at LO with \PROTOS~\cite{PROTOS} using the NNPDF2.3LO PDF set. \PROTOS\ was interfaced to \PYTHIAV{8} with the A14 set of tuned parameters. Samples were produced for $\mVLQ$ in the range of 500 to 1400~\GeV. Steps of 50~\GeV\ were used in the range from 700 to 1200~\GeV, and steps of 100~\GeV\ otherwise. The samples were generated in the singlet models for $T$ and $B$ quarks, but samples at $\mVLQ$ of 700, 900 and 1200~\GeV\ were also generated in the ($T$~$B$) doublet model in order to test kinematic differences between singlet and doublet models. In the singlet models, the BRs are independent of the mixing angles between VLQ and SM quarks for small values of the mixing angles and hence only a function of $\mVLQ$. With this assumption, for large $\mVLQ$, the BRs approach the relative proportions of 50:25:25 for the $W$:$Z$:$H$ decay modes in the singlet model for the $T$ quark as well as for the $B$ quark. In the ($X$~$T$) doublet and ($B$~$Y$) doublet models, the BRs approach the relative proportions of 50:50 for the $Z$:$H$ decays of the $T$ quark and $B$ quark, respectively. The same holds for the ($T$~$B$) doublet model if the top quark mixes much more strongly with its VLQ partner than the bottom quark, a natural scenario for the SM Yukawa couplings~\cite{JA_TP}. However, kinematic differences may exist between the singlet and doublet models. The samples generated for the ($T$~$B$) doublet were used to verify that such kinematic differences have a small impact on the analysis, and therefore the difference between the two cases is only a change in the BRs. Thus, the singlet model samples were also used for the doublet case, reweighting the yields for each decay mode to obtain the expected observables for any given BR. The pair-production cross sections were calculated with \textsc{TOP++} at NNLO+NNLL using the MSTW 2008 NNLO PDF set.

The single production of $T$ quarks was simulated using \MADGRAPHV{5} with the ``VLQ'' UFO model~\cite{UFO}, which implements the Lagrangian described in Ref~\cite{Buchkremer:2013bha}, using the NNPDF2.3 LO PDF set. \MADGRAPH\ was interfaced to \PYTHIAV{8} with the A14 set of tuned parameters. Only the decay $T\rightarrow Zt$ was considered. Samples were generated with a $T$ quark produced via $Wb$ and also via $Zt$ interactions. Since production via the $Zt$ interaction is suppressed due to the required top quark in the initial state, single-VLQ production refers to  production via the $Wb$ interaction in the remainder of this paper, unless stated otherwise. Samples were generated for $\mVLQ$ in the range from 700 to 2000~\GeV, with steps of 100~\GeV\ (200~\GeV) in the range 700--1600~\GeV\ (1600--2000~\GeV), with a benchmark coupling of $\kappa_T = 0.5$ for the $Wb$ and $Zt$ interactions. Additional samples were generated with alternative values of $\kappa_T = 0.1$ and $1.0$ in order to study the effect of a varying $T$-quark width on kinematic distributions.

The single-production cross sections were calculated~\cite{Matsedonskyi:2014mna} at NLO and in narrow-width approximation for $c_W = 1$, with the coupling $c_W$ defined in Ref.~\cite{Matsedonskyi:2014mna} and corresponding to $\kappa_T$ up to numerical constants. In order to predict the cross section for different values of $c_W$, they are multiplied by $c_W^2$. It was shown in the context of this analysis that the chirality of the coupling has a negligible impact on the sensitivity of the analysis and hence $c_W$ is taken as the sum in quadrature of the left- and right-handed couplings $c_{W,L}$ and $c_{W,R}$, i.e.\ $c_W = \sqrt{c^2_{W,L} + c^2_{W,R}}$. The cross section is additionally corrected for width effects calculated with \MADGRAPHV{5}, assuming that the ratio of NLO and LO cross sections remains approximately the same for a non-vanishing $T$-quark width. The cross section is then multiplied by the BR for the decay into $Zt$ in the singlet model, which is $\approx 25$\% in the range of VLQ masses investigated in this analysis. The benchmark coupling of $\kappa_T = 0.5$ corresponds to a coupling of the $T$ quark to the $W$ boson $c_W = 0.45$.

\begin{table}[h!]
\centering
\caption{List of background Monte Carlo samples used, giving information about the matrix-element generator, the parton shower program to which it is interfaced and its set of tuned parameters (``tune'', if applicable), the PDF sets used in the matrix element (ME), and the order in QCD of the cross-section calculation.}
\begin{tabular}{l|c|c|c|c}
\toprule
& Generator & Shower program & PDF set (ME) & Cross section \\
&           & and tune & \\ \midrule
 \zjets & \SHERPAV{2.2.1} & \SHERPAV{2.2.1} & NNPDF3.0 NNLO & NNLO \\ \midrule
 \ttbar & \POWHEGBOXV{v2} & \PYTHIAV{8}, A14 & NNPDF3.0 NNLO & NNLO+NNLL \\ \midrule
 Diboson & \SHERPAV{2.2.1} & \SHERPAV{2.2.1} & NNPDF3.0 NNLO & NLO \\ \midrule
 $\ttbar+V$ ($W$/$Z$) & \MGMCatNLO & \PYTHIAV{8}, A14 & NNPDF3.0 NLO & NLO \\ \midrule
 $\ttbar+WW$ & \MADGRAPHV{5} & \PYTHIAV{8}, A14 & NNPDF2.3 LO & NLO \\ \midrule
 $\ttbar\ttbar$ & \MADGRAPHV{5} & \PYTHIAV{8}, A14 & NNPDF2.3 LO & NLO \\ \midrule
 Single top & \POWHEGBOXV{v1} & \PYTHIA{8},  & CT10 & NLO+NNLL \\
            &                & Perugia 2012 &      &  \\ \midrule
 Triboson & \SHERPAV{2.1} & \SHERPAV{2.1} & CT10 & NLO \\ 
\bottomrule
\end{tabular}
\label{tab:samples}
\end{table}

\section{Object reconstruction}
Reconstructed electrons, muons and jets are used. Jets are reconstructed with the anti-$k_t$ algorithm~\cite{Cacciari:2008gp} with a radius parameter of 0.4 (\sjets) and with a parameter of 1.0 (\ljets). A $b$-tagging algorithm is applied to \sjets, and a top-tagging algorithm is applied to \ljets. Moreover, missing transverse momentum (\met) is used for the definition of one signal-enriched region and one background-enriched region. For electrons, muons and jets, an overlap-removal procedure based on their proximity in $\eta$--$\phi$ space is used, as described at the end of this section.

Electrons are reconstructed~\cite{ATLAS-CONF-2016-024} from energy clusters in the electromagnetic calorimeter with ID tracks matched to them. Their energy is calibrated~\cite{PERF-2013-05,ATL-PHYS-PUB-2016-015}, and they are required to fulfill the ``tight likelihood'' identification criteria~\cite{ATLAS-CONF-2016-024}. Electrons are required to have a minimum transverse energy, \et, of at least 28~\GeV\ and to be within the fiducial region $|\eta_{\mathrm{cluster}}|<2.47$, excluding the barrel--endcap transition region, $1.37 < |\eta_{\mathrm{cluster}}| < 1.52$. Electron tracks must point to the primary vertex, which is ensured by requiring that the track's impact parameter significance is smaller than 5, and that $|z_0 \cdot \sin\theta|$ is smaller than 0.5~mm, where $z_0$ is the distance along the $z$-axis between the primary vertex and the track's point of closest approach. In order to suppress background from electrons originating from hadron decays and from hadrons that are misidentified as electrons, an isolation criterion is applied that requires the scalar sum of the \pt\ of the tracks which point to the primary vertex within a cone around the electron (but excluding its track) be less than 6\% of its \et. A variable cone size~\cite{Rehermann:2010vq} of $\Delta R = \min\left(10~\GeV/\et, 0.2\right)$ is used.

Muons are reconstructed~\cite{PERF-2015-10} from combined tracks in the MS and the ID. Their transverse momentum, \pt, is calibrated~\cite{PERF-2015-10}, and they are required to fulfill the ``medium'' identification criteria~\cite{PERF-2015-10}. Muons must have a minimum \pt\ of 28~\GeV\ and they must be within the fiducial region $|\eta|<2.5$. Muon tracks must point to the primary vertex, which is ensured by requiring that the track's impact parameter significance is smaller than 3, and that $|z_0 \cdot \sin\theta|$ is smaller than 0.5~mm. In order to suppress background from muons originating from hadron decays, an isolation criterion similar to that for electrons is applied: the scalar sum of the \pt\ of the tracks around the muon which point to the primary vertex, excluding the muon track, must be less than 6\% of its \pt, using a variable cone size of $\Delta R = \min\left(10~\GeV/\pt, 0.3\right)$.

\Sjets\ are reconstructed from topological clusters of calorimeter cells~\cite{PERF-2014-07,ATL-PHYS-PUB-2015-036} with the anti-$k_t$ algorithm using \textsc{FastJet}~\cite{Cacciari:2011ma} with a radius parameter of 0.4. \Sjets\ are calibrated to the jet energy scale (JES) at particle level~\cite{PERF-2016-04} and are required to be within the fiducial volume $|\eta|<4.5$. \Sjets\ with $|\eta|<2.5$ must have a minimum \pt\ of 25~\GeV\ and forward jets, $2.5 < |\eta| < 4.5$, must have a minimum \pt\ of 35~\GeV\ to reduce contributions from pileup. For \sjets\ with $|\eta|<2.4$ and $\pt<60~\GeV$, pileup contributions are suppressed by the use of the jet vertex tagger~\cite{PERF-2014-03}. \Sjets\ within $|\eta|<2.5$ are \btagged\ using the MV2c10 algorithm~\cite{ATL-PHYS-PUB-2016-012}, for which several basic $b$-tagging-algorithms~\cite{PERF-2012-04} are combined in a boosted decision tree. The MV2c10 algorithm is used such that it provides a $b$-tagging efficiency of $\sim 77\%$ for $b$-jets,\footnote{Jets originating from the hadronization of gluons and light quarks ($u$-, $d$-, $s$- and $c$-quarks) are called light jets in this document. Jets originating from the hadronization of $b$-quarks are called $b$-jets.} and a rejection factor\footnote{The rejection factor is defined as the inverse of the $b$-tagging efficiency for non-$b$-jets.} of $\sim 6$ for $c$-jets and $\sim 130$ for other light jets, based on simulated \ttbar\ events.

\Ljets\ are also reconstructed from topological clusters with the anti-$k_t$ algorithm, but with a radius parameter of 1.0. In contrast to the \sjet\ calibration, the topological clusters that are used as inputs to the \ljet\ reconstruction take into account corrections for the calorimeter's response to hadrons and other effects~\cite{Barillari:2009zza}. Contributions to \ljets\ from pileup and the underlying event are removed by applying trimming~\cite{Krohn:2009th} with parameters that were optimized for separating \ljets\ that originate from hadronic decays of high-energy massive resonances~\cite{PERF-2015-03,ATL-PHYS-PUB-2015-033,ATL-PHYS-PUB-2015-053} from those that originate from $b$-quarks, light quarks or gluons. \Ljets\ are calibrated to the JES at particle level~\cite{PERF-2012-02}. They are required to have a minimum \pt\ of 200~\GeV\ and to be within the fiducial region $|\eta|<2.0$. The mass of \ljets\ is calculated from a combination of calorimeter and tracking information~\cite{ATLAS-CONF-2016-035}. It is calibrated~\cite{ATLAS-CONF-2016-035} and required to be at least 50~\GeV, which suppresses contributions from $b$-jets and light jets in favor of \ljets\ that originate from hadronic decays of high-energy $W$ bosons, $Z$ bosons, Higgs bosons and top quarks. In the \dilsing\ channel (\Sect{\ref{sec:dilepsing}}), top-tagging is used to identify hadronic decays of high-energy top quarks. It is based on a combination~\cite{ATL-PHYS-PUB-2015-053} of the \ljet\ mass and the $N$-subjettiness~\cite{Thaler:2010tr,Thaler:2011gf} ratio $\tau_{32} = \tau_3 / \tau_2$, calculated in the ``winner-take-all'' mode~\cite{Larkoski:2014uqa}. This top-tagger provides an efficiency of $\sim 80\%$ for hadronically decaying top quarks with a \pt\ of at least 200~\GeV\ with a varying background rejection of $\sim 20$ at $\pt = 200~\GeV$ that decreases to $\sim 4$ at $\pt = 1~\TeV$, as estimated with simulated dijet events.

In order to avoid double-counting of tracks or energy deposits and in order to improve the identification of the different reconstructed objects, a sequential overlap-removal procedure is used. In the first step, electrons that share a track with a muon are removed. In the second step, any \sjet\ is removed that has a $\Delta R$ to an electron that is smaller than 0.2, and in the third step, electrons are removed if they are closer than 0.4 to any remaining \sjet. Finally, \sjets\ that have a $\Delta R < 0.04 + 10~\GeV/\pt(\mu)$ to a muon are removed if they have at most two associated tracks with $\pt(\mathrm{track}) > 0.5~\GeV$, otherwise the muon is removed. \Sjets\ and \ljets\ are not subject to an overlap-removal procedure, because the analysis strategies in all channels are designed such that the energy deposits in \ljets\ and \sjets\ are not counted twice, as explained in the following lines: in the trilepton channels, \ljets\ are not used (\Sect{\ref{sec:trilepair}} and \ref{sec:trilepsing}); in the dilepton pair-production channels, \ljets\ are only used for the classification of events (\Sect{\ref{sec:dilres}} and \ref{sec:dilboost}); in the dilepton single-production channel, \sjets\ are only used for the classification of events, but not for the calculation of the discriminating variable (\Sect{\ref{sec:dilepsing}}).

Missing transverse momentum is only used for the reduction of the contribution from \TTbar\ pair production in one search region for single-$T$-quark production (\Sect{\ref{sec:dilepsing}}) and for the definition of one background-enriched region (\Sect{\ref{sec:dilboost}}), and it is calculated from the vectorial sum of the transverse momenta of reconstructed and calibrated leptons and \sjets~\cite{Aaboud:2018tkc}, with the overlap between these objects removed. The calculation also includes the contributions from tracks in the ID that are matched to the primary vertex but are not associated with any of the reconstructed objects.

\section{Event selection and background control regions}
\label{sec:selection}
Five different channels are analyzed, each searching for either pair production or single production of VLQs, as introduced in \Sect{\ref{sec:intro}} and visualized in \Fig{\ref{fig:sketches}}. In each channel, event-selection criteria were optimized for maximum sensitivity to benchmark processes by studying expected 95\% CL exclusion limits. In the pair-production channels, the mass reach for $T$ and $B$ quarks in the singlet and doublet models was maximized. While the search focuses on the decay of one VLQ to a $Z$ boson and a third-generation SM quark, a high sensitivity to all three $T$- and $B$-quark decay modes is ensured by choosing these benchmark models, because the second VLQ is not only allowed to decay into a $Z$ boson, but also into a $W$ boson or a Higgs boson in association with a third-generation SM quark. In the single-production channels, the sensitivity to single-$T$-quark production via the exchange of a $W$ boson with $\kappa_T = 0.5$ was optimized.

\begin{figure}[p]
\centering
\subfloat[]{\includegraphics[width=.49\textwidth]{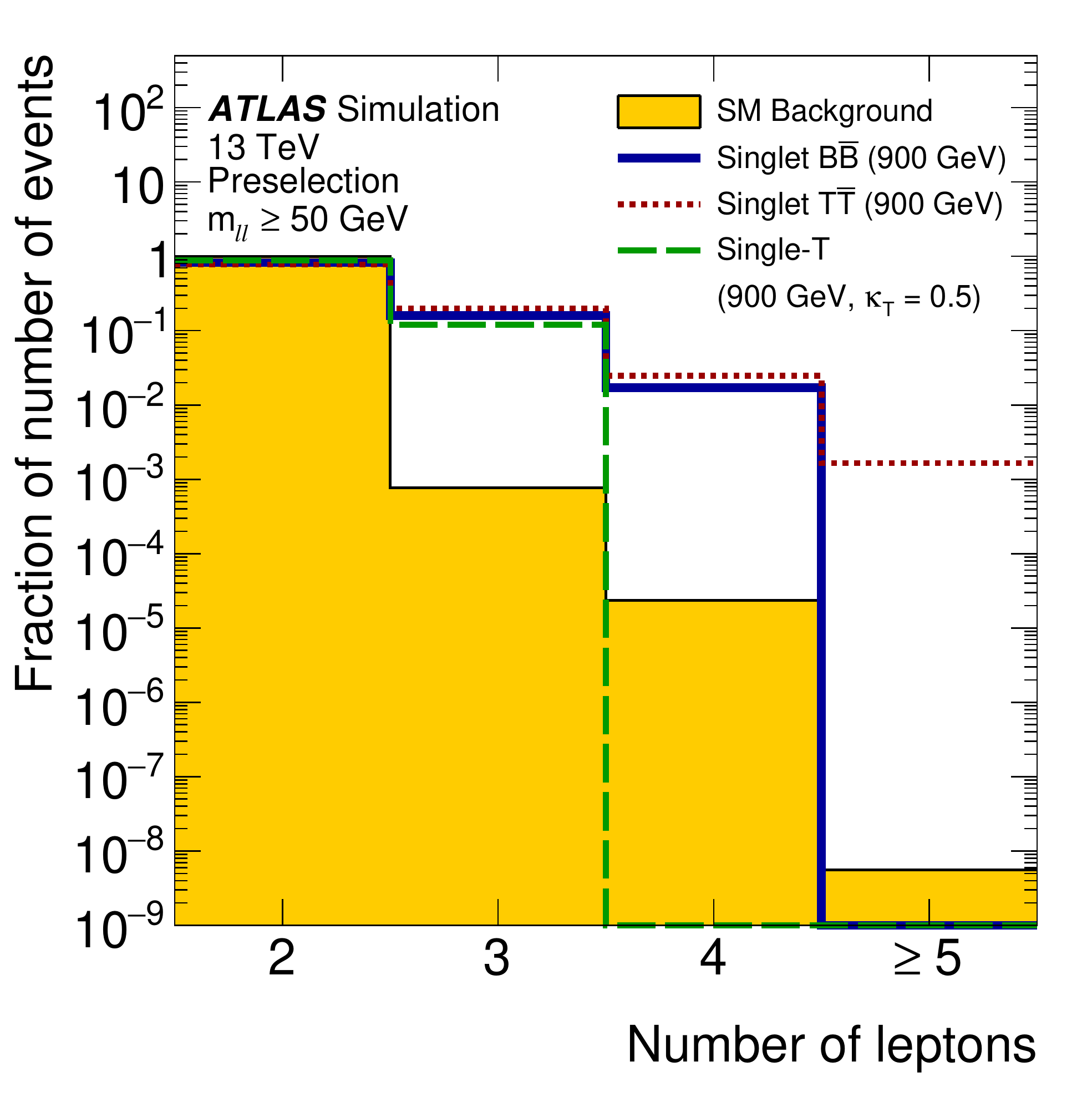}\label{fig:preselection_nlep}}
\subfloat[]{\includegraphics[width=.49\textwidth]{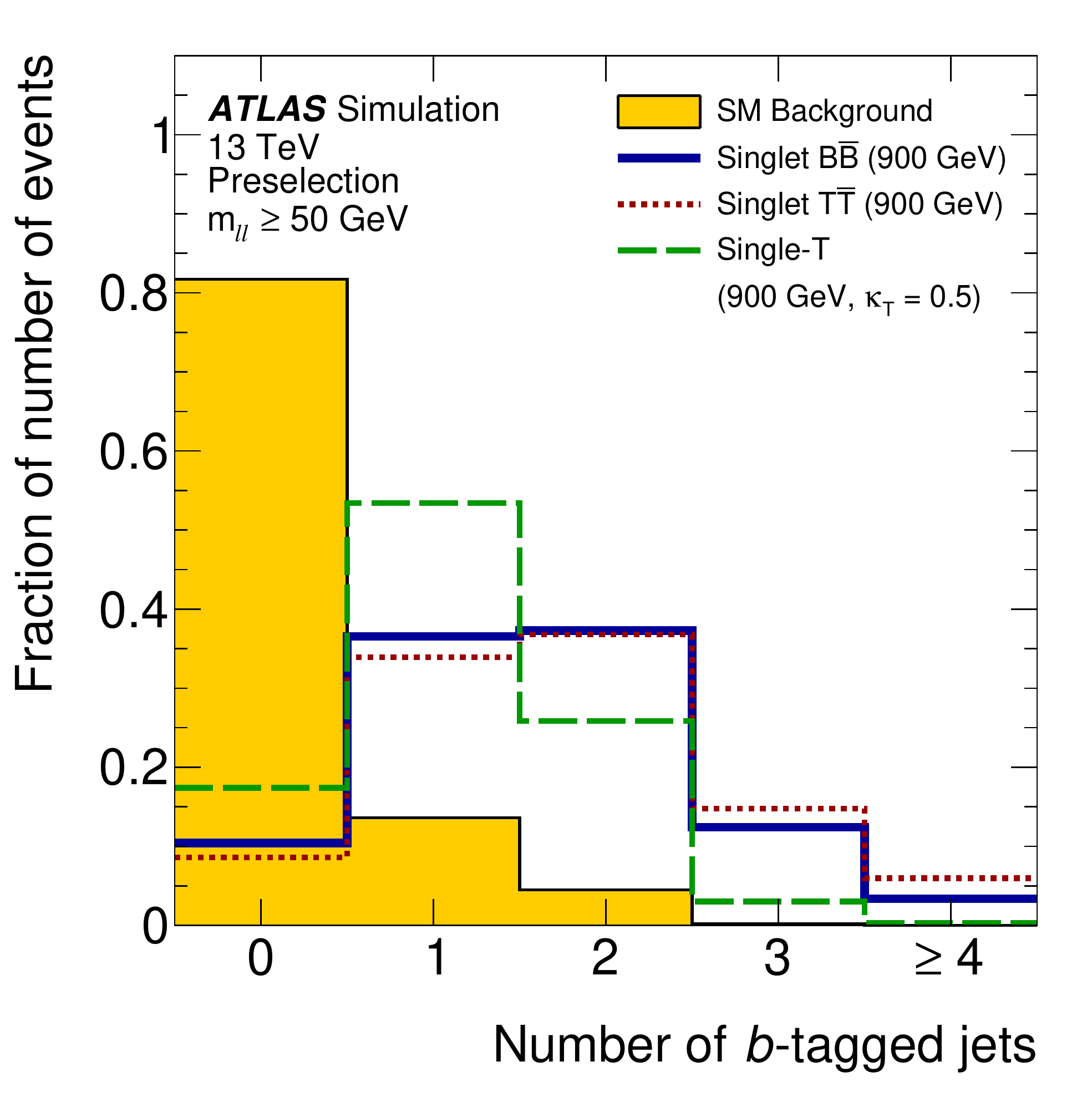}\label{fig:preselection_ntag}}\\
\subfloat[]{\includegraphics[width=.49\textwidth]{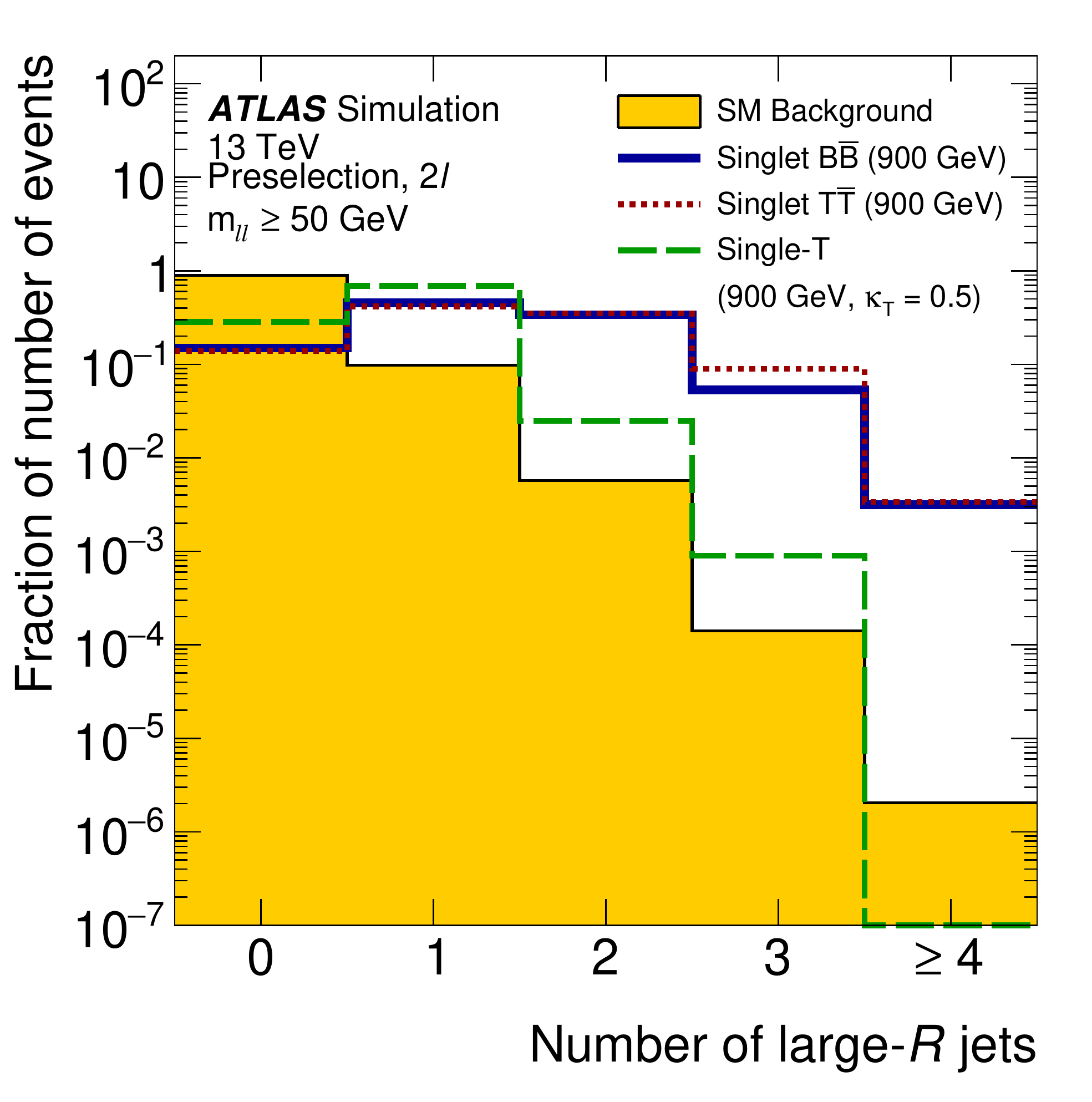}\label{fig:preselection_nLRJ}}
\subfloat[]{\includegraphics[width=.49\textwidth]{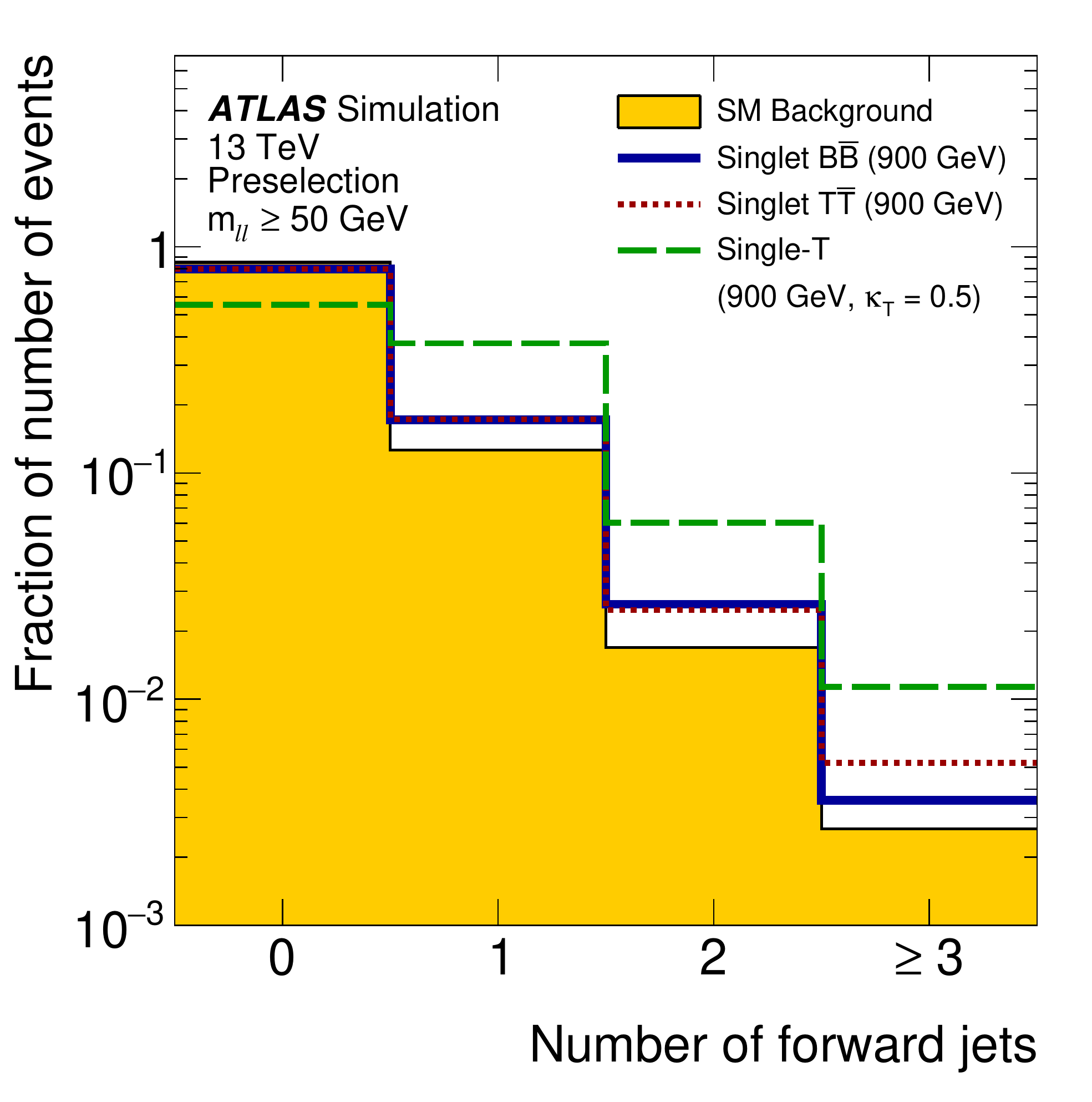}\label{fig:preselection_nFWJ}}
\caption{Distributions of the sum of all background processes (solid area) and of benchmark signal processes (lines), based on MC simulations after preselection and requiring $\mll > 50~\GeV$: (a) the number of leptons, (b) the number of \btagged\ jets, (c) the number of \ljets\ in events with exactly two charged leptons, and (d) the number of forward jets. The signal processes shown are $B$- and $T$-quark pair production in the singlet model and single-$T$-quark production with a coupling of $\kappa_T = 0.5$, each with a mass of $\mVLQ = 900~\GeV$. All distributions are normalized to unit area. The last bin contains the overflow.}
\label{fig:preselection}
\end{figure}

A preselection common to the channels was used as the basis for these optimizations. This preselection requires the presence of a $Z$ boson candidate that is constructed from two leptons with opposite-sign electric charge. In all events, at least two leptons of the same flavor with $\pt > 28~\GeV$ and with opposite-sign electric charge are required. Out of all such lepton pairs in an event, a $Z$ boson candidate is defined by the pair with invariant mass closest to the mass of the $Z$ boson. Events in which this invariant mass is larger than 400~\GeV\ are removed because they are very unlikely to occur in any of the considered signal processes. In addition, at least two \sjets\ with $\pt>25~\GeV$ must be present. In the \dilsing\ channel, this last criterion is replaced by a requirement on the presence of at least one \ljet\ with $\pt>200~\GeV$ and $m>50~\GeV$.

In \Fig{\ref{fig:preselection}}, normalized distributions after preselection are shown for the sum of all background processes, which are estimated from MC simulations, as well as for benchmark signal models for pair and single VLQ production. In \Fig{\ref{fig:preselection_nlep}}, the distribution of the number of leptons is shown. By selecting events with exactly two leptons, a high signal efficiency is achieved. In events with at least three leptons, however, the signal-to-background ratio is significantly improved. The searches for pair and single production are hence split into complementary dilepton and trilepton channels. The distribution of the number of \btagged\ jets is shown in \Fig{\ref{fig:preselection_ntag}}. A higher number of \btagged\ jets is characteristic of the signal processes, and at least one or two \btagged\ jets are required in the event selection, depending on the channel. The distribution of the number of \ljets\ is shown in \Fig{\ref{fig:preselection_nLRJ}} for events that contain exactly two charged leptons. Signal events show a higher number of \ljets\ than background events, which in signal mostly originate from the hadronic decays of boosted top quarks, $W$ bosons, $Z$ bosons or Higgs bosons. The presence of \ljets\ is used in the dilepton channels to suppress backgrounds and hence improve the sensitivity to the signal. In order to achieve a high signal efficiency, in the pair-production case, two complementary dilepton channels are defined, one for events with at most one \ljet\ and one for events with at least two \ljets. In the trilepton channels, \ljet\ requirements are not used because the presence of at least three leptons suppresses the backgrounds efficiently. In \Fig{\ref{fig:preselection_nFWJ}}, the forward-jet multiplicity is shown. The single-production process often features a forward jet from $t$-channel production. The presence of a forward jet is hence used in the single-production searches to separate the signal from the background.

The event selection criteria in the different channels are defined in Sections~\ref{sec:dilres}--\ref{sec:trilepsing}. In each channel, these signal regions (SR) are complemented by a set of control regions (CR), which are rich in the main background processes. The CRs are used to check the modeling of the background and to improve the background prediction in the SRs by a combined fit of CRs and SRs (\Sect{\ref{sec:result}}). In the design of the CRs, not only a high purity of the respective background processes was aimed for, but also a large number of background events, as well as kinematic properties of the background events that resemble those of the events in the SRs. Each CR was checked to ensure that it was not sensitive to any signal process.

All SRs and CRs defined in the three pair-production channels (Sections~\ref{sec:dilres}--\ref{sec:trilepair}) are orthogonal (i.e. they have no common events), so that the results in these channels can be combined (\Sect{\ref{sec:result}}). The same holds for all SRs and CRs in the single-production channels (Sections~\ref{sec:dilepsing}--\ref{sec:trilepsing}), which are also combined (\Sect{\ref{sec:result}}). Orthogonality is not ensured between pair- and single-production regions. However, single-production channels include requirements designed to suppress the pair-production signal in their SRs.

\subsection{Search strategy: \dilres}
\label{sec:dilres}
Two orthogonal channels are defined for the pair-production search in dilepton final states, one with at least two \ljets, described in \Sect{\ref{sec:dilboost}} (\dilboost), and one with at most one \ljet\ (\dilres), described in this section. All such \ljets\ are required to have a \pt\ of at least 200~\GeV\ and a mass of at least 50~\GeV\ after trimming. Due to the mass requirement, hadronic decays of boosted top quarks, and of $W$, $Z$, and Higgs bosons are efficiently selected and jets that originate from the hadronization of high-\pt\ light quarks, $b$-quarks or gluons are suppressed. While in the \dilboost\ channel background processes are strongly suppressed,  the signal efficiency is also reduced so that a complementary channel optimized for events with at most one \ljet\ provides additional sensitivity to the signal.

The definitions of the SRs in the \dilres\ channel are summarized in \Tab{\ref{tab:dilres_sel}}. Two SRs are defined, for which the preselection and the presence of exactly two leptons are required. The mass of the $Z$ boson candidate, built from the two leptons, \mll, must be within a 10~\GeV\ window around the $Z$ boson mass, $\mz$. At least two \btagged\ jets must be present\footnote{\Sjets\ that are \btagged\ and \ljets\ may overlap in $\eta$--$\phi$ space, but no requirement is made on the proximity of \btagged\ \sjets\ and \ljets.}, which strongly reduces the background contribution from the production of a $Z$ boson in association with light jets. The sensitivity of the channel is improved by defining two SRs, one for events without any \ljet\ and one for events with exactly one \ljet. Since in the signal process the $Z$ boson is produced in the decay of a massive VLQ, the \pt\ of the $Z$ boson candidate, \ptll, is on average much larger than in the background processes, so \ptll\ is  required to be larger than 250~\GeV\ for both SRs. Moreover, the scalar sum of the transverse momenta of all \sjets\ in the event, \htj, is on average much larger for signal events than for background events, because the quarks from the decay chain of the massive VLQ result in high-\pt\ jets. Therefore, the \htj\ distribution is  used in the statistical analysis (\Sect{\ref{sec:result}}) to search for an excess of data over the background prediction, with a signal expected to result in an excess for large values of \htj. In addition, a minimum \htj\ value of 800~\GeV\ is required for both SRs.

\begin{table}[h!]
\centering
\caption{Definition of the control and signal regions for the \dilres\ channel.}
\begin{tabular}{c|c|c|c}
\toprule
\textbf{$\boldmath\ttbar\unboldmath$ CR} & \textbf{$\boldmath Z\unboldmath$+jets CR} & \textbf{0-\ljetbf\ SR} & \textbf{1-\ljetbf\ SR} \\
\midrule
\multicolumn{4}{c}{Preselection} \\\midrule
\multicolumn{4}{c}{$= 2$ leptons} \\\midrule
$|\mll-\mz| > 10~\GeV$ & \multicolumn{3}{c}{$|\mll-\mz| < 10~\GeV$} \\
and $\mll > 50~\GeV$ & \multicolumn{3}{c}{} \\\midrule
\multicolumn{4}{c}{$\geq 2$ \btagged\ jets} \\\midrule
\multicolumn{2}{c|}{$\leq 1$ \ljet} & $= 0$ \ljets & $= 1$ \ljet \\\midrule
$\ptll < 600~\GeV$ & \multicolumn{3}{c}{$\ptll > 250~\GeV$} \\\midrule
$\htj>200~\GeV$& $200~\GeV < \htj < 800~\GeV$ & \multicolumn{2}{c}{$\htj > 800~\GeV$} \\
\bottomrule
\end{tabular}
\label{tab:dilres_sel}
\end{table}

The main background processes are from $Z$+jets production containing two jets which originate from the hadronization of $b$-quarks and \ttbar\ production with a dileptonic final state. The background from \ttbar\ production is strongly suppressed by requiring \mll\ to be close to \mz. In both main background processes, no hadronically decaying massive resonances are present, so that the SR with exactly one \ljet\ has a higher signal-to-background ratio than the SR without a \ljet. The contributions from all background processes are strongly reduced by the requirements on \ptll\ and \htj.

In order to validate the modeling of the main background processes, CRs are defined for the $Z$+jets and \ttbar\ processes. A summary of the CR definitions is given in \Tab{\ref{tab:dilres_sel}}. The $Z$+jets CR is defined by the same criteria as the SRs, except for the \ljets\ and \htj\ criteria. Events with no \ljets\ and events with exactly one \ljet\ are considered together and \htj\ is required to be in the range 200--800~\GeV, ensuring that the CR is almost free of a potential signal. The resulting CR sample is expected to be $88\%$ $Z$+jets events. The \ttbar\ CR is defined by requiring the same preselection, lepton multiplicity, and \btagged-jet multiplicity criteria as in the SRs. However, the mass of the $Z$ boson candidate, \mll, must be outside of a 10~\GeV\ window around the $Z$ boson mass, \mz. In addition, \mll\ is required to be larger than 50~\GeV, because events with lower \mll\ do not stem mainly from \ttbar\ production, but from Drell--Yan production in association with jets. Also in the \ttbar\ CR, events without \ljets\ and events with exactly one \ljet\ are considered together. In contrast to the definition of the SRs, the \pt\ of the $Z$ boson candidate is required to be less than 600~\GeV\ in order to ensure that the CR does not contain signal contributions from potential VLQ pair production with two leptons that do not stem from the decay of a $Z$ boson, such as $\TTbar\rightarrow HtWb$. Moreover, the lower bound on \htj\ is lowered to 200~\GeV\ in order to increase the number of events in the CR and to test the modeling of the full \htj\ distribution. The resulting CR sample is expected to be $93\%$ \ttbar\ events. The ratio of expected signal and background events in the $Z$+jets (\ttbar) CR is as low as 0.0005 (0.003) for \BBbar\ and \TTbar\ production in the singlet model for $\mVLQ = 900~\GeV$.

\subsection{Search strategy: \dilboost}
\label{sec:dilboost}
In addition to the \dilres\ channel, a second dilepton channel was optimized for events with at least two \ljets\ (\dilboost) in order to exploit the presence of highly boosted, hadronically decaying massive resonances in the signal processes.

The definition of the SR in the \dilboost\ channel is summarized in \Tab{\ref{tab:dilboost_sel}}. The same requirements as in the \dilres\ channel are imposed: the preselection, the presence of exactly two leptons with \mll\ within a 10~\GeV\ window around \mz, and the presence of at least two \btagged\ jets. In addition, at least two \ljets\ are required in each event. Also in this channel, the large expected values for \ptll\ and \htj\ are exploited to discriminate the signal from the background processes. The optimized requirements are $\ptll > 250~\GeV$ and $\htj > 1150~\GeV$. In order to search for an excess of data over the background prediction, the invariant mass of the $Z$ boson candidate and the highest-\pt\ \btagged\ jet, $m_{Zb}$, is used as a discriminating variable. In the search for \BBbar\ production, $m_{Zb}$ would show a resonant structure around $\mVLQ$ if VLQs were present, because it often corresponds to the reconstructed mass of the VLQ. Also, in the search for \TTbar\ production, this variable shows very good discrimination between signal and background, with the signal resulting in larger values of $m_{Zb}$ than the background.

\begin{table}[h!]
\centering
\caption{Definition of the control regions and the signal region for the \dilboost\ channel.}
\begin{tabular}{c|c|c}
\toprule
\textbf{$\boldmath\ttbar\unboldmath$ CR} & \textbf{$\boldmath Z\unboldmath$+jets CR} & \textbf{SR} \\
\midrule
\multicolumn{3}{c}{Preselection} \\\midrule
\multicolumn{3}{c}{$=2$ leptons} \\\midrule
$|\mll-\mz| > 10~\GeV$ and $\mll > 50~\GeV$ & \multicolumn{2}{c}{$|\mll-\mz| < 10~\GeV$} \\\midrule
\multicolumn{3}{c}{$\geq 2$ \btagged\ jets} \\\midrule
\multicolumn{3}{c}{$\geq 2$ \ljets} \\\midrule
$\ptll < 600~\GeV$ & -- & $\ptll > 250~\GeV$ \\\midrule
-- & $\htj < 1150~\GeV$ &  $\htj > 1150~\GeV$ \\\midrule
$\met < 200~\GeV$ & \multicolumn{2}{c}{--} \\\midrule
$\Delta R (\ell\ell, \mbox{highest-\pt\ \ljet}) < 2.0$ or $>2.8$ & \multicolumn{2}{c}{--} \\
\bottomrule
\end{tabular}
\label{tab:dilboost_sel}
\end{table}

The main background processes are $Z$+jets production with two jets originating from the hadronization of $b$-quarks, and \ttbar\ production in the dileptonic decay mode. As in the \dilres\ channel, \ttbar\ production is strongly suppressed by requiring \mll\ to be close to the mass of the $Z$ boson, and the contributions from all background processes are significantly reduced by the requirements on \ptll\ and \htj. The contributions from $Z$+jets production and dileptonic \ttbar\ decays are efficiently reduced by the presence of two \ljets, because no massive hadronically decaying resonance is present in these processes.

For the two main background processes, $Z$+jets and \ttbar\ production, CRs are defined. A summary of the CR definitions is given in \Tab{\ref{tab:dilboost_sel}}. Similarly to the \ttbar\ CR in the \dilres\ channel, the definition of the \ttbar\ CR is based on the requirement that \mll\ must be outside a 10~\GeV\ window around \mz\ but must still fulfill $\mll > 50~\GeV$. In order to suppress potential signal contributions in the CR, \ptll\ is required to be smaller than 600~\GeV. The requirement on \htj\ is removed, which increases the number of events in the CR. In addition, \met\ is required to be smaller than 200~\GeV, which reduces potential signal contributions from VLQ pair production with two leptons that do not stem from the decay of a $Z$ boson, but for example from the decay of $W$ bosons from the VLQ decay chain. Moreover, the $\Delta R$ between the $Z$ boson candidate and the highest-\pt\ \ljet\ is required to be smaller than 2.0 or larger than 2.8, which further reduces the contributions from a potential signal because in signal events the highest-\pt\ \ljet\ and the $Z$ boson candidate are typically not back-to-back due to the presence of additional final-state particles. The resulting CR sample is expected to be $82\%$ \ttbar\ events. The CR for the $Z$+jets process is defined by the same criteria as in the SR, but the requirement on \htj\ is inverted in order to remove potential signal contributions, and the requirement on \ptll\ is removed in order to increase the number of events in the CR. The resulting CR sample is expected to be only $64\%$ $Z$+jets events, but also $17\%$ \ttbar\ events. The ratio of expected signal and background events in the $Z$+jets (\ttbar) CR is as low as 0.03 (0.06) for \BBbar\ and 0.04 (0.04) for \TTbar\ production in the singlet model for $\mVLQ = 900~\GeV$.

\subsection{Search strategy: \tripair}
\label{sec:trilepair}
The trilepton pair-production channel (\tripair) is sensitive to signal events in which at least one lepton appears in addition to the leptons from the $Z$ boson decay that originates from $T\rightarrow Zt$ or $B\rightarrow Zb$. Additional leptons can originate from the decay of the other VLQ, such as in $B\rightarrow Wt\rightarrow\ell\nu_\ell bqq^\prime$ or $T\rightarrow Ht\rightarrow bb\ell\nu_\ell b$. In \TTbar\ production, an additional lepton can also originate from the $T\rightarrow Zt$ decay itself, if the top quark decays into $\ell\nu_\ell b$.

The definition of the SR is summarized in \Tab{\ref{tab:trilepair_sel}}. Events must pass the preselection, and they must have at least three leptons including a $Z$ boson candidate with \mll\ within a 10~\GeV\ window around \mz. Only one \btagged\ jet is required, because background contributions are already strongly reduced by the requirement of at least one additional lepton. Relaxing the $b$-tagging requirement compared to the dilepton channels improves the sensitivity to the signal processes because of the larger signal efficiency. As in the dilepton channels, a large transverse momentum of the $Z$ boson candidate is required, $\ptll > 200~\GeV$. In order to search for an excess of data over the background prediction, the scalar sum of the \sjet\ and lepton transverse momenta, \htjl, is used. In contrast to the use of \htj\ in the \dilres\ channel (\Sect{\ref{sec:dilres}}), the lepton transverse momenta are added to the discriminating variable \htjl, which exploits the \pt\ of all leptons in order to discriminate the signal from the background in addition to the use of \ptll, which is constructed from only two leptons.

\begin{table}[h!]
\centering
\caption{Definition of the control regions and the signal region for the \tripair\ channel.}
\begin{tabular}{c|c|c}
\toprule
\textbf{Diboson CR} & \textbf{$\boldmath\ttbar+X\unboldmath$ CR} & \textbf{SR} \\\midrule
\multicolumn{3}{c}{Preselection}\\\midrule
\multicolumn{3}{c}{$\geq 3$ leptons} \\\midrule
\multicolumn{3}{c}{$|\mll - \mz| < 10~\GeV$} \\\midrule
= 0 \btagged\ jets & \multicolumn{2}{c}{$\geq 1$ \btagged\ jets} \\\midrule
-- & $\ptll \leq 200~\GeV$ & $\ptll \geq 200~\GeV$\\
\bottomrule
\end{tabular}
\label{tab:trilepair_sel}
\end{table}

The main background processes are diboson, in particular $WZ$ and $ZZ$, production, and $\ttbar+X$ production (dominated by $\ttbar+Z$ production), which can both result in events with three leptons. The diboson background is strongly reduced by the $b$-tagging requirement, so that only diboson events with additional $b$-jets or mis-tagged light jets pass the event selection. Both main backgrounds are suppressed by the requirement on \ptll, because in background events $Z$ boson candidates rarely have a large transverse momentum.

For the two main background processes, diboson and $\ttbar+X$ production, CRs are defined and summarized in \Tab{\ref{tab:trilepair_sel}}. The diboson CR is defined by the same criteria as the SR, except for the $b$-tagging and \ptll\ requirements. No \btagged\ jets are allowed in the diboson CR, which reduces contributions from $\ttbar+X$ production and from a potential VLQ signal. The \ptll\ requirement is removed in order to further increase the number of diboson events in the CR. The resulting CR is expected to consist of $92\%$ diboson events, mainly from $WZ$ production. The $\ttbar+X$ CR is defined by inverting only the \ptll\ requirement, which removes contributions from a potential VLQ signal. The resulting CR sample is expected to consist mainly of $\ttbar+X$ and diboson events in similar proportions ($39\%$ and $43\%$, respectively). The ratio of expected signal and background events in the diboson ($\ttbar+X$) CR is as low as 0.001 (0.004) for \BBbar\ and 0.001 (0.006) for \TTbar\ production in the singlet model for $\mVLQ = 900~\GeV$.

\subsection{Search strategy: \dilsing}
\label{sec:dilepsing}
The production of a single $T$ quark results in a signature with fewer high-\pt\ objects than in \TTbar\ production. As a result it is more difficult to separate it from the background. However, a forward-jet from the $t$-channel production is often present, which can be exploited to strongly reduce the contributions from background processes. The final state from the decay of a single $T\rightarrow Zt$ with a leptonic $Z$ boson decay consists of the two leptons from the $Z$ boson, a forward jet and the decay products of the top quark. While the leptonic top-quark decay, $t\rightarrow \ell\nu_\ell b$, is used in the trilepton single-production channel (\trising), described in \Sect{\ref{sec:trilepsing}}, the hadronic decay, $t\rightarrow qq^\prime b$, is used in the dilepton channel (\dilsing), described in this section.

The definition of the SR is summarized in \Tab{\ref{tab:dilepsing_sel}}. Events are required to pass the preselection with the minimum requirement of two \sjets\ replaced by the presence of at least one \ljet. Events must have exactly two leptons that form a $Z$ boson candidate with an invariant mass within a 10~\GeV\ window around \mz. In this channel, a minimum \pt\ of the $Z$ boson candidate is also required, $\ptll > 200~\GeV$. At least one \btagged\ jet is required in the event. Although a second $b$-quark from  gluon splitting (\Fig{\ref{fig:sketches}}) is present in the signal, only in a fraction of signal events is a second \btagged\ jet  found within the $|\eta|$ acceptance of the ID. The hadronically decaying top quark originating from the $T$-quark decay often has such a large \pt\ that the top-quark decay products are contained within one \ljet. Top-tagging is used to discriminate \ljets\ from hadronic top-quark decays in single-$T$-quark production from the main background process, $Z$+jets production, which can only fulfill this requirement if a quark or gluon jet is falsely top-tagged (mis-tags). At least one forward jet is required in each event, which is a characteristic property of single-$T$-quark production. In order to search for an excess of data over the background prediction, the invariant mass of the $Z$ boson candidate and the highest-\pt\ top-tagged \ljet, $m_{Zt}$, is used, which, if VLQs were present, would show a resonant structure around $\mVLQ$. In order to facilitate the interpretation of the search for single-$T$-quark production, the potential signal contribution from \TTbar\ production is reduced by requiring $\htj+\met<m_{Zt}$. This requirement has an efficiency of $\approx 20\%$ for \TTbar\ pair production in the singlet model in the mass range 800--1400~\GeV, while maintaining an efficiency of 90--95\% for single-$T$-quark production with $\kappa_T = 0.5$ across the whole mass range studied.

\begin{table}[ht!]
\centering
\caption{Definition of the control regions and the signal region for the \dilsing\ channel.}
\begin{tabular}{c|c|c}
\toprule
\textbf{0-$\boldmath b\unboldmath$-tagged-jet CR} & \textbf{$\boldmath\geq 1\unboldmath$-$\boldmath b\unboldmath$-tagged-jet CR} & \textbf{SR} \\\midrule
\multicolumn{3}{c}{Preselection with $\geq 1$ \ljet} \\\midrule
\multicolumn{3}{c}{$=2$ leptons}\\\midrule
\multicolumn{3}{c}{$|\mll - \mz| < 10~\GeV$} \\\midrule
\multicolumn{3}{c}{$\ptll > 200~\GeV$} \\\midrule
$=0$ \btagged\ jets & \multicolumn{2}{c}{$\geq 1$ \btagged\ jets} \\\midrule
\multicolumn{2}{c|}{$\geq 1$ loose-not-tight top-tagged \ljet} & $\geq 1$ top-tagged \ljet \\\midrule
\multicolumn{2}{c|}{--} & $\geq 1$ forward jet \\\midrule
\multicolumn{3}{c}{$\htj + \met < m_{Zt}$} \\
\bottomrule
\end{tabular}
\label{tab:dilepsing_sel}
\end{table}

The main background process is $Z$+jets production, which mainly passes the event selection in the SR if it contains jets that originate from the hadronization of $b$-quarks. The $Z$ boson is mostly produced with low values of \pt, so that the \ptll\ requirement strongly reduces this background. In addition, the requirement of at least one top-tagged \ljet\ efficiently suppresses the contribution from $Z$+jets production, because it does not contain top quarks and can only fulfill the top-tagging requirement through mis-tags. Similarly, the requirement of at least one forward jet reduces the $Z$+jets background, because forward jets are not characteristic for the main production mode of this process.

For the $Z$+jets production background, two CRs are defined. One CR, called 0-\btagged-jet CR, requires that no \btagged\ jets be present, allowing to correct the modeling of $Z$+jets production in a region that is kinematically close to the SR. In a second CR, called $\geq 1$-\btagged-jet CR, the modeling of $Z$+jets production in association with \btagged\ jets is controlled. If good data-MC agreement is observed in both CRs consistently, this provides confidence in the overall modeling of $Z$+jets production. A summary of the CR definitions is given in \Tab{\ref{tab:dilepsing_sel}}. Both CRs are based on the SR with changes to the top-tagging, $b$-tagging and forward-jet requirements. For both CRs, the top-tagging requirement is changed, so that there must be at least one \ljet\ that fails the top-tagging requirements on $\tau_{32}$ but fulfills the top-tagging requirements on the \ljet\ mass. Out of these \ljets, called ``loose-not-tight top-tagged'', the \ljet\ with the largest \pt\ is used in the calculation of $m_{Zt}$ in the CRs. The change in the top-tagging requirement enriches the CRs in $Z$+jets production in comparison with a potential signal contribution. In both CRs, the forward-jet requirement is removed, which increases the number of events in the CRs. Finally, in the 0-\btagged-jet CR, no \btagged\ jet is allowed, while in the $\geq 1$-\btagged-jet CR the same $b$-tagging requirement as in the SR is used. The resulting samples in the CRs are expected to be $96\%$ and $91\%$ $Z$+jets events, respectively, and to contain a negligible amount of a potential single-$T$-quark signal. The ratio of expected signal and background events in the 0-\btagged-jet ($\geq 1$-\btagged-jet) CR is as low as 0.001 (0.02) for single-$T$-quark production with $\mVLQ = 900~\GeV$ with $\kappa_T = 0.5$. As the CRs do not contain requirements on the number of forward jets and make use of a modified top-tagging requirement (loose-not-tight), the modeling of the $Z$+jets background was cross-checked in another region with no \btagged\ jets, but requiring the presence of at least one forward jet and using the nominal top-tagging algorithm. The modeling of the distributions of kinematic properties was found to be consistent between the CRs and the cross-check region and a small difference observed between the overall numbers of events was assigned as a systematic uncertainty (\Sect{\ref{sec:systematics}}).

\subsection{Search strategy: \trising}
\label{sec:trilepsing}
The search for single-$T$-quark production in the trilepton channel (\trising) is sensitive to the decay $T\rightarrow Zt\rightarrow \ell\ell\ell\nu_\ell b$, featuring an additional lepton from the top-quark decay. It is hence complementary to the \dilsing\ channel (\Sect{\ref{sec:dilepsing}}).

The definition of the SR is summarized in \Tab{\ref{tab:trilepsing_sel}}. Events must pass the preselection, and they must have at least three leptons including a $Z$ boson candidate with \mll\ within a 10~\GeV\ window around \mz. In this channel, a minimum \pt\ of the $Z$ boson candidate is also required, $\ptll > 150~\GeV$. As in the \dilsing\ channel (\Sect{\ref{sec:dilepsing}}), at least one \btagged-jet and at least one forward jet are required. In order to suppress background contributions in which leptons have lower \pt\ on average than in the signal, the transverse momentum of the highest-\pt\ lepton in each event, $\max\pt^\ell$, must be larger than 200~\GeV. As in the \dilsing\ channel, the potential signal contribution from \TTbar\ production is reduced in the search for single-$T$-quark production. In the \trising\ channel, this is achieved by requiring that \htj\ multiplied by the number of \sjets\ in the event is smaller than 6~\TeV. This requirement has an efficiency of 50--30\% for \TTbar\ pair production in the singlet model in the mass range 800--1400~\GeV, while maintaining an efficiency of $\approx 95\%$ for single-$T$-quark production with $\kappa_T = 0.5$ across the whole mass range studied. In order to search for an excess of data over the background prediction, \htjl\ is used, as in the \tripair\ channel (\Sect{\ref{sec:trilepair}}).

\begin{table}[h!]
\centering
\caption{Definition of the control regions and the signal region for the \trising\ channel.}
\begin{tabular}{c|c|c}
\toprule
\textbf{Diboson CR} & \textbf{$\boldmath t\bar t+X\unboldmath$ CR} & \textbf{SR} \\\midrule
\multicolumn{3}{c}{Preselection}\\\midrule
\multicolumn{3}{c}{$\geq$ 3 leptons} \\\midrule
\multicolumn{3}{c}{$|\mll - \mz| < 10~\GeV$} \\\midrule
\multicolumn{2}{c|}{--} & $\ptll > 150~\GeV$\\\midrule
= 0 \btagged\ jets & \multicolumn{2}{c}{$\geq 1$ \btagged\ jets} \\\midrule
-- & $=0$ forward jets & $\geq 1$ forward jets\\\midrule
-- & $28~\GeV < \max\pt^\ell < 200~\GeV$ & $\max\pt^\ell > 200~\GeV$ \\\midrule
\multicolumn{3}{c}{$\htj\,\times\,\left(\mbox{number of \sjets}\right)$ < 6~\TeV} \\
  \bottomrule
  \end{tabular}
  \label{tab:trilepsing_sel}
\end{table}

The main background processes are diboson production with additional $b$-quarks and $\ttbar+X$ production (dominated by $\ttbar+Z$ production). The contributions of these backgrounds are strongly reduced by the requirements on \ptll\ and $\max\pt^\ell$, as well as by requiring at least one forward jet, because forward jets are not characteristic for these processes.

For the two main background processes, diboson and $\ttbar+X$ production, two CRs are defined and summarized in \Tab{\ref{tab:trilepsing_sel}}. The diboson CR is defined following the criteria in the SR, but the requirements on \ptll, $\max\pt^\ell$ and the presence of at least one forward jet are removed in order to increase the number of events in the CR. In addition, no \btagged\ jet is allowed in the diboson CR. The resulting CR sample is expected to be $92\%$ diboson events and to contain a negligible number of potential signal events. The $\ttbar+X$ CR is based on the SR by inverting the requirement on $\max\pt^\ell$ and by requiring that no forward jet is present. These changes remove potential signal contributions. In addition, the requirement on \ptll\ is removed in order to increase the number of events in the CR. The resulting CR sample is expected to consist mainly of $\ttbar+X$ and diboson events in similar proportions ($40\%$ and $44\%$, respectively). The ratio of expected signal and background events in the diboson ($\ttbar+X$) CR is as low as 0.002 (0.007) for single-$T$-quark production with $\mVLQ = 900~\GeV$ with $\kappa_T = 0.5$.

\FloatBarrier

\section{Systematic uncertainties}
\label{sec:systematics}
Systematic uncertainties are divided into experimental uncertainties, mostly related to the uncertainty in the modeling of the detector response in the simulation, and theoretical uncertainties, related to the theoretical modeling of the background processes in the MC simulation. Experimental uncertainties on the signal efficiencies and the signal shape of the discriminating variables are also taken into account.

Systematic uncertainties are evaluated by varying each source by $\pm 1 \sigma$ of its uncertainty. As a result, the predicted background and signal event yields in the different CRs and SRs can vary as well as the predicted shapes of the discriminating variables in these regions. For some sources only one systematic variation is defined. In such cases, the effect on the yields and shapes are symmetrized in order to construct the corresponding variation in the other direction.

The uncertainty in the integrated luminosity of the analyzed dataset is 2.1\%. It is derived following a methodology similar to that in Ref.~\cite{DAPR-2013-01} from a calibration of the luminosity scale using $x$--$y$ beam-separation scans in August 2015 and May 2016.

Uncertainties in electron and muon trigger, reconstruction and identification efficiencies are derived from data using $Z\rightarrow e^+e^-$ decays~\cite{ATLAS-CONF-2016-024} and $Z\rightarrow \mu^+\mu^-$ decays~\cite{PERF-2015-10}. Uncertainties in the electron (muon) energy (momentum) calibration and resolution are also derived using $Z\rightarrow \ell^+\ell^-$ events~\cite{ATL-PHYS-PUB-2016-015,PERF-2015-10}.

Uncertainties in the \sjet\ energy scale are evaluated from MC simulations and from data using multijet, $Z$+jets, and $\gamma$+jets events~\cite{PERF-2016-04}. Additional \sjet\ uncertainties arise from the jet energy resolution~\cite{ATL-PHYS-PUB-2015-015}, which are also derived from multijet, $Z$+jets and $\gamma$+jets events and from the jet vertex tagger.

Uncertainties in the $b$-tagging efficiency of \sjets\ are derived from data~\cite{PERF-2012-04} for $b$-jets, $c$-jets, and other light jets. For the derivation of the $b$-tagging efficiency and its uncertainty for $b$-jets, dileptonic \ttbar\ events are used~\cite{Aaboud:2018xwy}. Additional uncertainties are derived using MC simulations for the extrapolation of this efficiency beyond the kinematic reach of the calibration.

Uncertainties in the \ljet\ energy scale, mass and $N$-subjettiness ratio $\tau_{32}$ are derived from a comparison of the calorimeter-to-track-jet ratio in data and MC simulations~\cite{PERF-2012-02,ATLAS-CONF-2017-063}. While the uncertainty in the mass is taken to be correlated with the uncertainty in the energy scale, the $\tau_{32}$ uncertainty is taken to be uncorrelated with these two. The uncertainty in the resolutions of the \ljet\ energy, mass and $\tau_{32}$ is estimated by comparing the prediction from the nominal MC simulations with simulations where the resolution is 20\% poorer.

The electron, muon and \sjet\ uncertainties are propagated to the calculation of the \met. Additional uncertainties are assigned to contributions to the \met\ calculation that arise from tracks which are matched to the primary vertex and not associated with any object~\cite{Aaboud:2018tkc}.

All MC distributions are reweighted so that the distribution of the average number of interactions per bunch crossing corresponds to the distribution in data. In order to assess the associated systematic uncertainty, the reweighting is varied within its uncertainty.

A 5\% uncertainty is assigned to the cross section for \zjets\ production~\cite{ATL-PHYS-PUB-2016-002}. Additional uncertainties in the selection efficiency and in the shape of the final discriminant due to the theoretical modeling of the \zjets\ process are evaluated by comparing the nominal \SHERPA\ sample with alternative samples, normalized to the same cross section. An uncertainty due to the choice of generator and parton shower is assigned by comparing the nominal sample with a sample generated with \MGMCatNLO\ and the NNPDF3.0 NLO PDF set, and showered with \PYTHIAV{8} and using the A14 set of tuned parameters with the NNPDF2.3 LO PDF set. An uncertainty due to the scale choice is evaluated by varying the renormalization and factorization scales in the nominal sample independently by factors of 2 and 0.5. The assigned uncertainty is based on the largest deviations from the nominal sample observed in each bin of the final discriminant. An uncertainty due to the choice of PDF set is evaluated by comparing the nominal \SHERPA\ sample using the NNPDF3.0 NLO PDF set with samples using the MMHT2014 NNLO~\cite{Harland-Lang:2014zoa} and CT14 NNLO PDF sets~\cite{Dulat:2015mca}. The largest observed deviations from the nominal sample in each bin of the final discriminant are used to assign the uncertainty.

The uncertainty in the cross section for \ttbar\ production is assigned as $+5.6\%$/$-6.1\%$~\cite{ATL-PHYS-PUB-2016-004}. Also for \ttbar\ production, additional uncertainties in the selection efficiency and in the shape of the final discriminant are assigned by comparing the nominal sample with alternative MC samples. An uncertainty due to the choice of generator is evaluated from a comparison of the nominal \POWHEGBOX\ sample with a sample generated with \MGMCatNLO\ with the NNPDF3.0 NLO PDF set, and showered with \PYTHIAV{8} using the A14 set of tuned parameters and the NNPDF2.3 LO PDF set. An uncertainty due to the choice of shower model is assigned by comparing the nominal sample, showered by \PYTHIAV{8}, with an alternative sample showered by \HERWIGV{7}~\cite{Bahr:2008pv,Bellm:2015jjp} with the H7-UE-MMHT set of tuned parameters and the MMHT PDF set. The uncertainties due to the choice of renormalization and factorization scales are evaluated by independently varying the scales by factors of 2 and 0.5. The largest differences observed in each bin of the final discriminant are assigned as the systematic uncertainty for these two scales. An uncertainty due to the choice of PDF set is evaluated by comparing the nominal sample with samples generated with the MMHT2014 NLO and CT14 NLO PDF sets. The largest observed deviations from the nominal sample in each bin of the final discriminant are used to assign the uncertainty.

An uncertainty of 6\% is assigned to the cross section for diboson production~\cite{ATL-PHYS-PUB-2016-002}. As with the \zjets\ and \ttbar\ processes, alternative MC samples are used to assess additional uncertainties in the selection efficiency and in the shape of the final discriminant of the diboson processes. In order to assess the uncertainty due to the choice of renormalization and factorization scales, the nominal \SHERPA\ samples are compared with alternative samples with the scales varied independently by factors of 2 and 0.5 and the largest observed differences in each bin of the final discriminant are assigned as the uncertainty. An uncertainty due to the choice of PDF set is assessed by comparing the nominal samples, generated with the NNPDF3.0 NNLO PDF set, with samples generated with the MMHT2014 NNLO and CT14 NNLO PDF sets. The largest deviations in each bin of the final discriminant are used to assign the uncertainty.

For the $\ttbar+V$ processes, uncertainties of $+13\%$/$-12\%$ are assigned for the $\ttbar+W$ production cross section and of $+10$/$-12\%$ for the $\ttbar+Z$ production cross section~\cite{ATL-PHYS-PUB-2016-005}. For the assessment of additional uncertainties in the selection efficiency and in the shape of the final discriminant of the $\ttbar+V$ processes, the nominal samples are compared with alternative MC samples. An uncertainty due to the choice of generator is assigned by comparing the nominal sample with a sample generated with \SHERPAV{2.2} and the NNPDF3.0 NLO PDF set. For these samples, a fast simulation of the ATLAS detector~\cite{SOFT-2010-01} was used, which relies on a parametrization of the calorimeter response~\cite{ATL-PHYS-PUB-2010-013}. The nominal sample was additionally produced with the fast simulation configuration and the relative differences observed in the comparison with the samples with varied scales are assigned as the systematic uncertainty. An uncertainty due to the parton shower is assigned by comparing the nominal sample with samples with a varied amount of initial-state radiation. These alternative samples were produced with fast detector simulation and the procedure to assign a systematic uncertainty is again based on the relative difference observed in comparison with the nominal sample obtained with fast detector simulation in each bin of the final discriminant.

Backgrounds due to misidentified electrons and muons play a minor role in this analysis, because such leptons typically have low transverse momentum and are hence strongly suppressed by the SR requirements, in particular by the lower thresholds for \ptll\ in the different channels. However, in the \ttbar\ CRs in the \dilres\ and \dilboost\ channels and in the \zjets\ CR in the \dilboost\ channel, low-\ptll\ events are included. Similarly, $Z$+jets and \ttbar\ events could contribute to the CRs and SRs in the \tripair\ and \trising\ channels due to misidentified leptons. The maximum observed difference between data and MC simulations in the lepton \pt\ spectra in the CRs is 25\%. This is assigned as an uncertainty to $Z$+jets and \ttbar\ events in the trilepton channels and to \ttbar\ events with $\ptll < 200~\GeV$ in the \dilres\ and \dilboost\ channels.

No $b$-tagged jet are allowed in the diboson CRs for the \tripair\ and \trising\ channels (\Sect{\ref{sec:trilepair}} and \Sect{\ref{sec:trilepsing}}).  While this requirement ensures a high purity in diboson processes, it differs from the requirements in the SRs. An uncertainty of 50\% is assigned to the production of diboson events in association with $b$-quarks, motivated by the precision of measurements of $W$- and $Z$-boson production in association with $b$-quarks~\cite{STDM-2012-11,STDM-2012-15}.

In order to ensure a large number of events in the CRs for the dilepton single-production search, the SR forward-jet requirement is removed (\Sect{\ref{sec:dilepsing}}). A cross-check was performed in a region that only differs from the SR by a veto on $b$-tagged jets. While the modeling of the shapes of kinematic variables in this region is satisfactory, the 11\% difference in the overall number of events between data and background expectation is assigned as an additional uncertainty in the SR due to the forward-jet requirement.

The uncertainties on the reconstructed objects and the luminosity also affect the predictions for VLQ pair and single production. No further uncertainties on the signal processes were considered. As discussed in \Sect{\ref{sec:samples}}, the MC samples for VLQ pair production were generated in the singlet model and alternative BR hypotheses for $T$ and $B$ quarks are obtained by reweighting the singlet BRs to the alternative BRs. This procedure is validated by comparing kinematic distributions of the nominal VLQ pair production samples with alternative samples that were generated in the ($T$~$B$) doublet model. After reweighting both to the same BRs, no large differences were observed between these samples. Hence, the reweighting procedure is considered validated and no systematic uncertainty is assigned.

\section{Results}
\label{sec:result}
In each channel, a binned likelihood fit is performed to the discriminating variable. Control and signal regions are fit simultaneously and systematic uncertainties are included in the fit as a set of nuisance parameters (NP), $\theta$. The likelihood function $L(\mu,\theta)$ consists of Poisson probabilities for each bin in the discriminating variable in each region, and a Gaussian or log-normal distribution for each NP. In the likelihood fit, the signal cross section $\sigma$ is parametrized by multiplying the predicted cross section with a correction factor $\mu$, called the signal-strength factor, which is a free parameter of the fit. In a background-only fit $\mu$, and hence $\sigma$, is set to zero. For the  combined control and signal region fit the modeling of the main background processes was adjusted during the fit via changes in the NPs, so that the background prediction in the signal regions is improved. The binning of the discriminating variable in the different channels was chosen in order to retain as much shape information about the distribution as possible given the number of background MC events in each bin.

The effect of each single source of systematic uncertainty is treated as correlated across all regions and processes with the following two exceptions. In order to avoid that CRs with high statistical power are able to constrain NPs in very different regions of phase space, for the uncertainties associated with misidentified leptons, separate NPs are defined for the different CRs and SRs in each channel, and for the uncertainties related to the choice of MC generator and hadronization model, separate NPs are defined for each channel. Different sources of systematic uncertainty are treated as uncorrelated with each other, except for the case of the \ljet\ scale uncertainties affecting the \pt\ and mass, which are treated as 100\% correlated. In addition to the systematic uncertainties discussed in \Sect{\ref{sec:systematics}}, an additional NP is added for each bin in the discriminating variable in each region due to the statistical uncertainty of the MC samples.

\subsection{Results: \dilres}
The observed number of events in the SRs and CRs and the expected number of events for the different background contributions are shown in \Tab{\ref{tab:dilres_pre}} for the \dilres\ channel. Also shown is the expected number of events for \BBbar\ and \TTbar\ production in the singlet model for $\mVLQ = 900~\GeV$. The signal efficiencies for these benchmarks are 0.060\% (0.013\%) for \BBbar\ (\TTbar) in the 0-\ljet\ SR and 0.33\% (0.16\%) in the 1-\ljet\ SR, and include the branching ratios of the VLQ as well as of its decay products, including the decay $Z\rightarrow\ell^+\ell^-$.

\begin{table}[p]
\centering
        \caption{Observed number of events in data and \emph{pre-fit} expected number of signal and background events in the control and signal regions for the \dilres\ channel, i.e.\ before the fit to data. For the signal, the expected number of events for the \BBbar\ and \TTbar\ benchmark processes with $\mVLQ = 900~\GeV$ is shown for the singlet model. Statistical uncertainties from the limited size of MC samples and systematic uncertainties are added in quadrature. The uncertainty in the ratio of the observed and expected numbers of events contains the systematic uncertainties and the statistical uncertainty of the prediction from Poisson fluctuations.}
        \begin{tabular}{l|R{1cm}@{\hspace{1.5pt}}c@{\hspace{1.5pt}}p{1cm}|R{1cm}@{\hspace{1.5pt}}c@{\hspace{1.5pt}}p{1cm}|R{1cm}@{\hspace{1.5pt}}c@{\hspace{1.5pt}}p{1cm}|R{1cm}@{\hspace{1.5pt}}c@{\hspace{1.5pt}}p{1cm}}
                         \toprule

                &\multicolumn{3}{c|}{$t\bar{t}$ CR} & \multicolumn{3}{c|}{$Z$+jets CR} &  \multicolumn{3}{c|}{0-\ljet\ SR} &    \multicolumn{3}{c}{1-\ljet\ SR}   \\\midrule
Singlet \BBbar (900~GeV)&    \num{16.2} & $\pm$ & \num{1.0}   &   \num{2.29} & $\pm$ & \num{0.31}  &   \num{1.94} & $\pm$ & \num{0.27}  &   \num{10.6} & $\pm$ & \num{0.8}   \\
Singlet \TTbar (900~GeV)&    \num{14.9} & $\pm$ & \num{0.9}   &   \num{1.81} & $\pm$ & \num{0.21}  &   \num{0.43} & $\pm$ & \num{0.09}  &    \num{5.1} & $\pm$ & \num{0.4}   \\
\midrule
    $Z$+jets    &    \num{1090} & $\pm$ & \num{310}   &    \num{630} & $\pm$ & \num{190}   &     \num{21} & $\pm$ & \num{9}     &     \num{43} & $\pm$ & \num{21}    \\
   $t\bar{t}$   &   \num{30000} & $\pm$ & \num{8000}  &      \num{8} & $\pm$ & \num{4}     &      \num{2} & $\pm$ & \num{5}     &      \num{2} & $\pm$ & \num{5}     \\
   Single top   &     \num{640} & $\pm$ & \num{60}    &    \num{5.3} & $\pm$ & \num{0.6}   &   \num{0.40} & $\pm$ & \num{0.23}  &   \num{0.71} & $\pm$ & \num{0.24}  \\
  $t\bar{t}+X$  &     \num{199} & $\pm$ & \num{26}    &     \num{37} & $\pm$ & \num{7}     &   \num{0.55} & $\pm$ & \num{0.23}  &    \num{4.6} & $\pm$ & \num{1.4}   \\
    Diboson     &     \num{44} & $\pm$ & \num{16}     &     \num{37} & $\pm$ & \num{12}    &    \num{0.9} & $\pm$ & \num{0.4}   &    \num{3.1} & $\pm$ & \num{1.9}   \\
\midrule
   Total Bkg.   &   \num{32000} & $\pm$ & \num{8000}  &    \num{710} & $\pm$ & \num{190}   &     \num{24} & $\pm$ & \num{9}     &     \num{54} & $\pm$ & \num{21}    \\
\midrule
      Data      &  \multicolumn{3}{c|}{\num{32216}}   &   \multicolumn{3}{c|}{\num{699}}   &   \multicolumn{3}{c|}{\num{35}}    &    \multicolumn{3}{c}{\num{51}}    \\
\midrule
   Data/Bkg.    &         1.00 & $\pm$ & 0.26         &        0.98 & $\pm$ & 0.26         &         1.4 & $\pm$ & 0.6          &         1.0 & $\pm$ & 0.4          \\
\bottomrule
        \end{tabular}
            \label{tab:dilres_pre}

\vspace{1cm}
        \caption{Observed number of events in data and \emph{post-fit} expected number of background events in the control and signal regions for the \dilres\ channel, i.e.\ after the fit to the data \htj\ distributions under the background-only hypothesis. The uncertainty in the expected number of events is the full uncertainty from the fit, from which the uncertainty in the ratio of the observed and expected numbers of events is calculated.}
        \begin{tabular}{l|R{1cm}@{\hspace{1.5pt}}c@{\hspace{1.5pt}}p{1cm}|R{1cm}@{\hspace{1.5pt}}c@{\hspace{1.5pt}}p{1cm}|R{1cm}@{\hspace{1.5pt}}c@{\hspace{1.5pt}}p{1cm}|R{1cm}@{\hspace{1.5pt}}c@{\hspace{1.5pt}}p{1cm}}
                         \toprule

              &\multicolumn{3}{c|}{$t\bar{t}$ CR} & \multicolumn{3}{c|}{$Z$+jets CR} & \multicolumn{3}{c|}{0-\ljet\ SR} &    \multicolumn{3}{c}{1-\ljet\ SR}   \\\midrule
   $Z$+jets   &   \num{1100} & $\pm$ & \num{100}   &   \num{622} & $\pm$ & \num{34}   &   \num{21.6} & $\pm$ & \num{2.6}   &     \num{43} & $\pm$ & \num{4}     \\
  $t\bar{t}$  &   \num{30200} & $\pm$ & \num{600}  &   \num{8.7} & $\pm$ & \num{3.0}  &    \num{3.1} & $\pm$ & \num{2.3}   &    \num{2.4} & $\pm$ & \num{2.0}   \\
  Single top  &    \num{630} & $\pm$ & \num{50}    &   \num{5.2} & $\pm$ & \num{0.6}  &   \num{0.40} & $\pm$ & \num{0.23}  &   \num{0.72} & $\pm$ & \num{0.20}  \\
 $t\bar{t}+X$ &    \num{197} & $\pm$ & \num{22}    &    \num{36} & $\pm$ & \num{6}    &   \num{0.60} & $\pm$ & \num{0.26}  &    \num{4.5} & $\pm$ & \num{1.2}   \\
   Diboson    &     \num{44} & $\pm$ & \num{6}     &    \num{37} & $\pm$ & \num{4}    &   \num{0.87} & $\pm$ & \num{0.24}  &    \num{3.1} & $\pm$ & \num{0.7}   \\
\midrule
  Total Bkg.  &   \num{32100} & $\pm$ & \num{700}  &   \num{709} & $\pm$ & \num{33}   &   \num{26.5} & $\pm$ & \num{3.2}   &     \num{54} & $\pm$ & \num{4}     \\
\midrule
     Data     &  \multicolumn{3}{c|}{\num{32216}}  &  \multicolumn{3}{c|}{\num{699}}  &   \multicolumn{3}{c|}{\num{35}}    &    \multicolumn{3}{c}{\num{51}}    \\
\midrule
  Data/Bkg.   &       1.003 & $\pm$ & 0.020        &       0.99 & $\pm$ & 0.05        &        1.32 & $\pm$ & 0.16         &        0.95 & $\pm$ & 0.08         \\
\bottomrule
        \end{tabular}
            \label{tab:dilres_post}

\end{table}

A fit of the background prediction to the \htj\ distributions in data was performed. The post-fit yields are shown in \Tab{\ref{tab:dilres_post}}. The uncertainty in the background prediction is significantly reduced in all regions compared to the pre-fit value (\Tab{\ref{tab:dilres_pre}}). The overall $Z$+jets (\ttbar) normalization is adjusted by a factor of \normfzjetszjcrtwolr\ (\normfttbarttcrtwolr) in the $Z$+jets (\ttbar) CR. The ratios of the post-fit and pre-fit background yields are consistent with unity in all regions.

\begin{figure}[p]
\centering
\subfloat[]{\includegraphics[width=.49\textwidth]{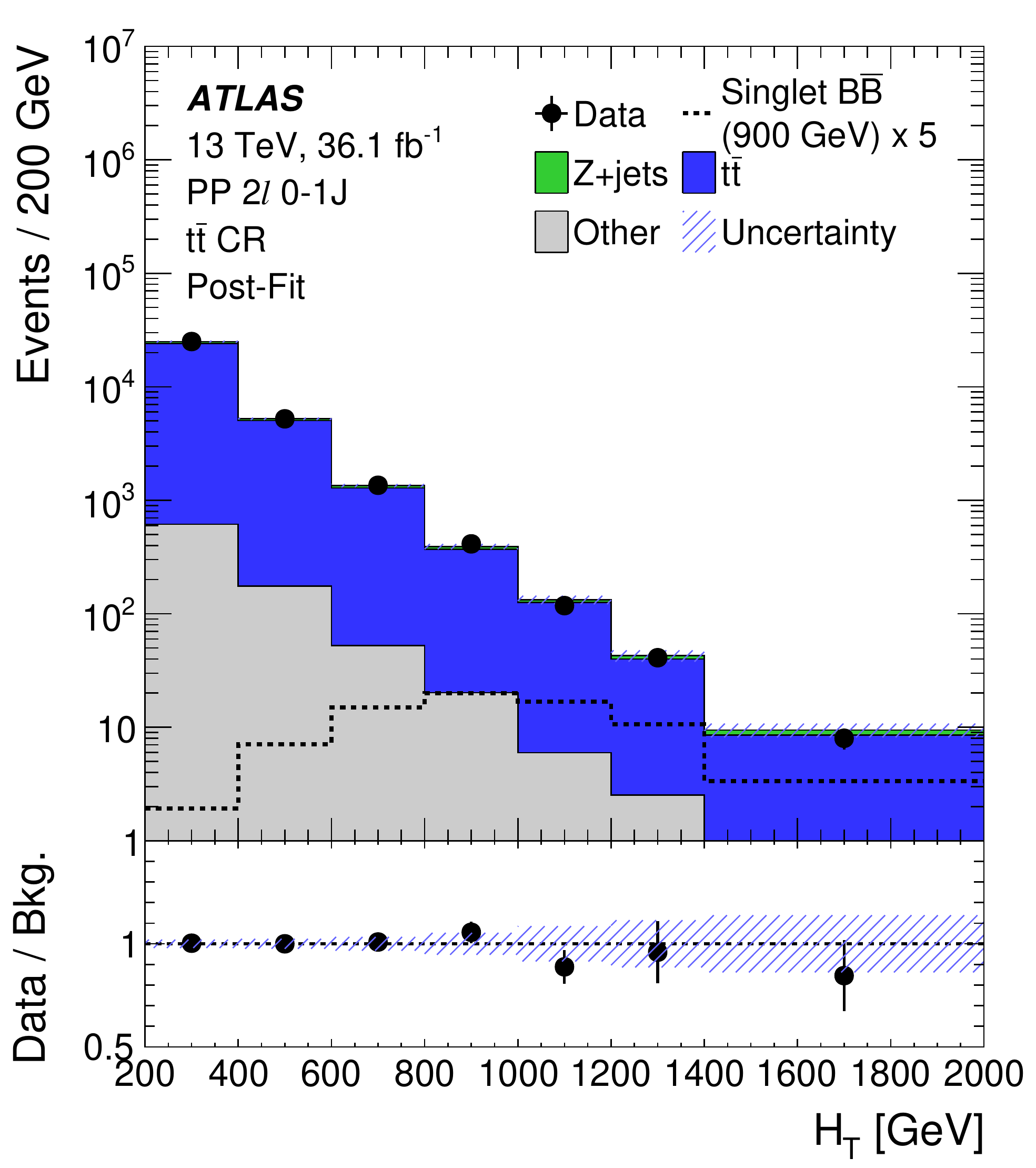}}
\subfloat[]{\includegraphics[width=.49\textwidth]{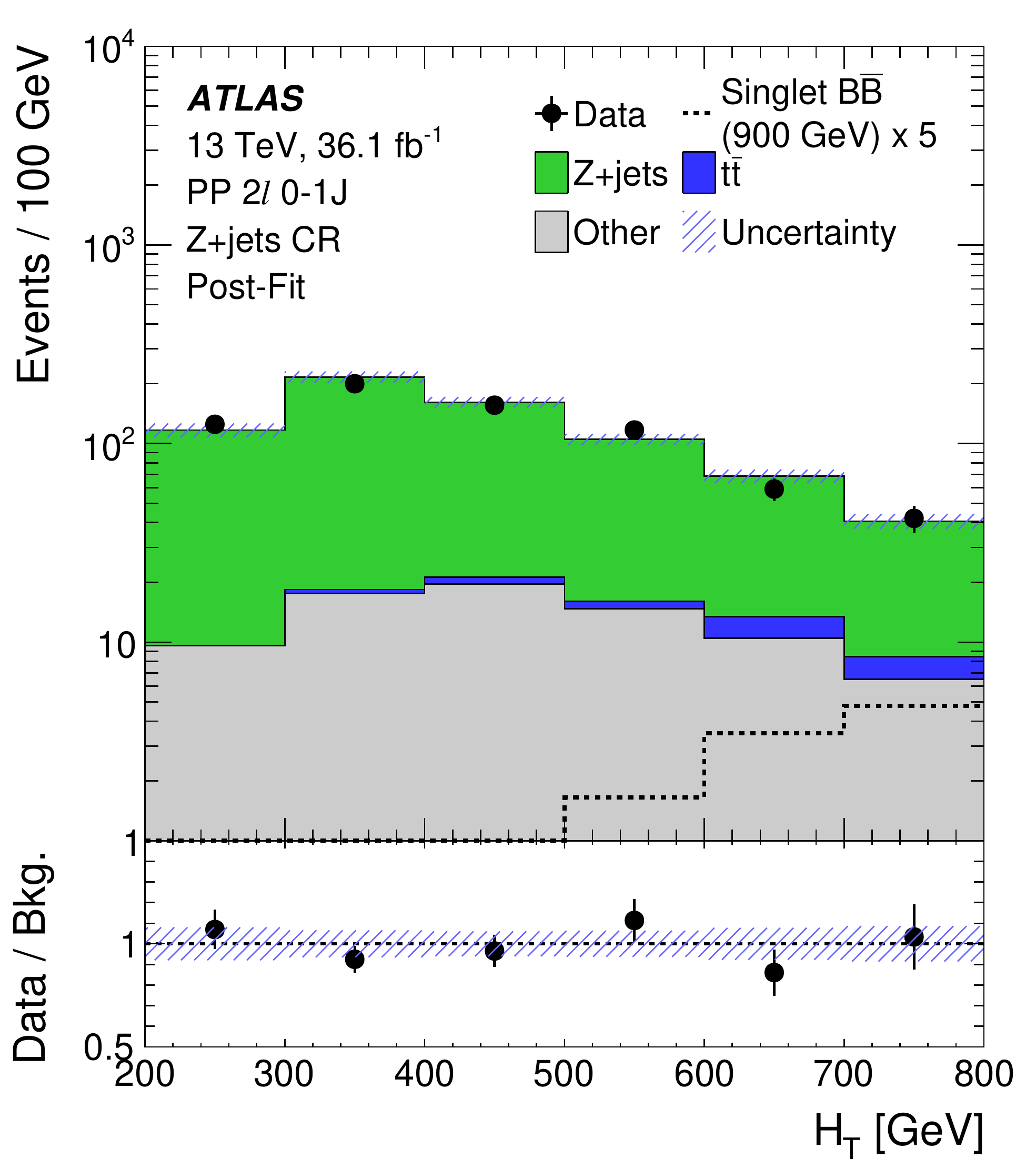}}\\
\subfloat[]{\includegraphics[width=.49\textwidth]{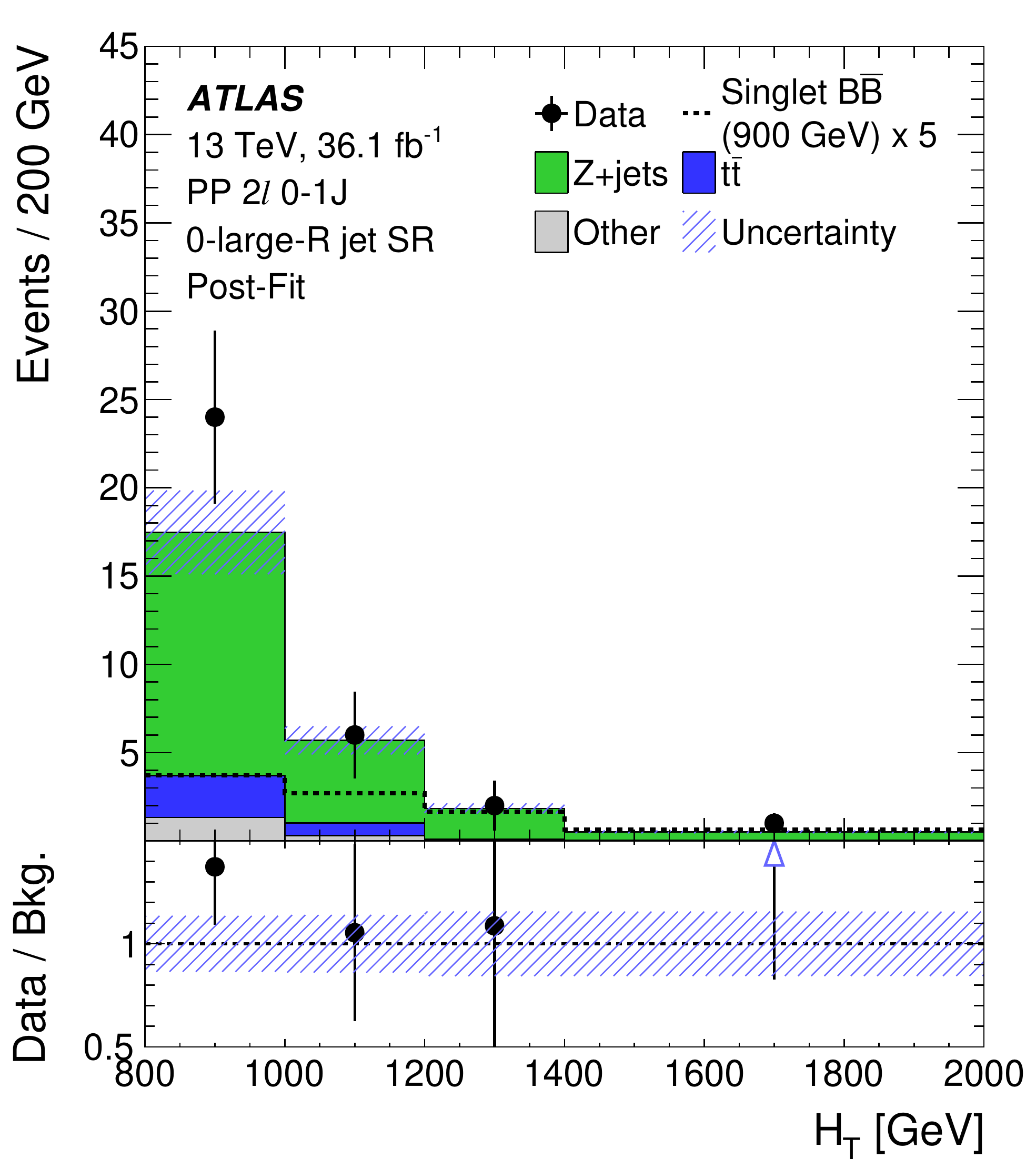}\label{fig:dilres_discr_SR0J}}
\subfloat[]{\includegraphics[width=.49\textwidth]{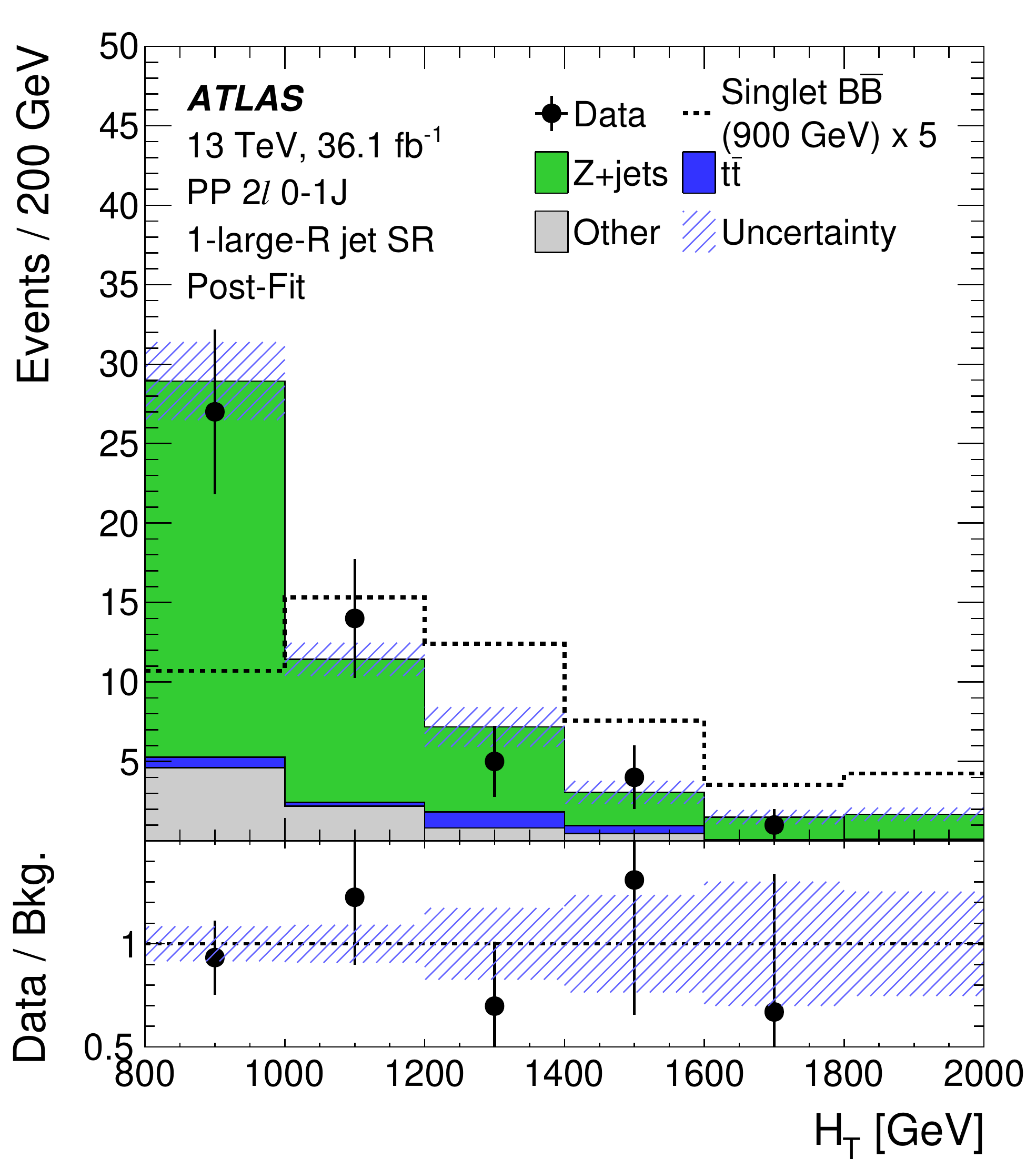}\label{fig:dilres_discr_SR1J}}
\caption{Comparison of the distribution of the scalar sum of \sjet\ transverse momenta, \htj, between data and the background prediction in (a) the \ttbar\ control region, (b) the $Z$+jets control region, (c) the 0-\ljet-signal region, and (d) the 1-\ljet-signal region of the pair-production \capdilres\ channel. The background prediction is shown \emph{post-fit}, i.e.\ after the fit to the data \htj\ distributions under the background-only hypothesis. The last bin contains the overflow. An upward pointing triangle in the ratio plot indicates that the value of the ratio is beyond scale. An example distribution for a \BBbar\ signal in the singlet model with $\mVLQ = 900~\GeV$ is overlaid. For better visibility, it is multiplied by a factor of five. The data are compatible with the background-only hypothesis.}
\label{fig:dilres_discr}
\end{figure}

\begin{figure}[p]
\centering
\subfloat[]{\includegraphics[width=.49\textwidth]{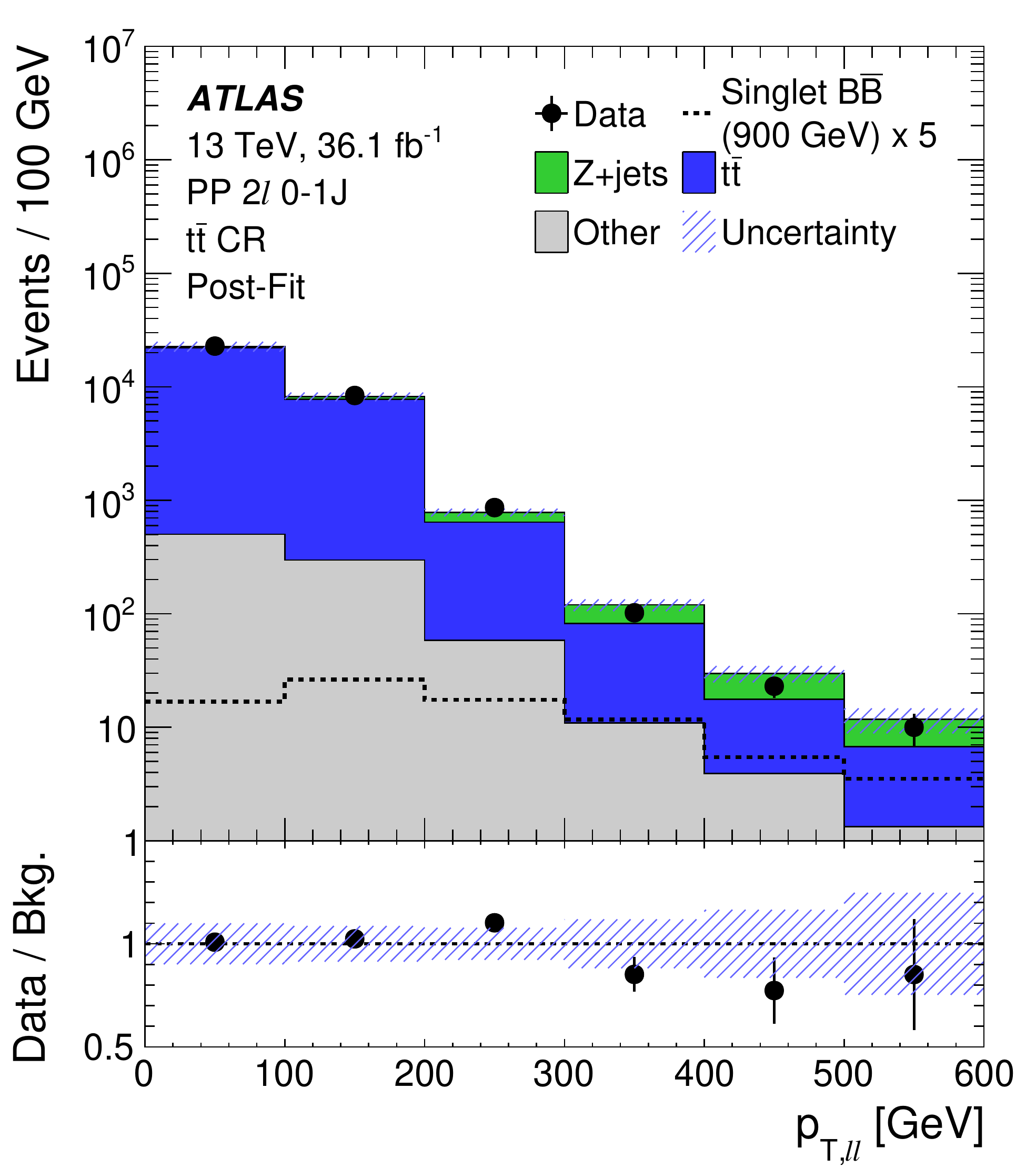}}
\subfloat[]{\includegraphics[width=.49\textwidth]{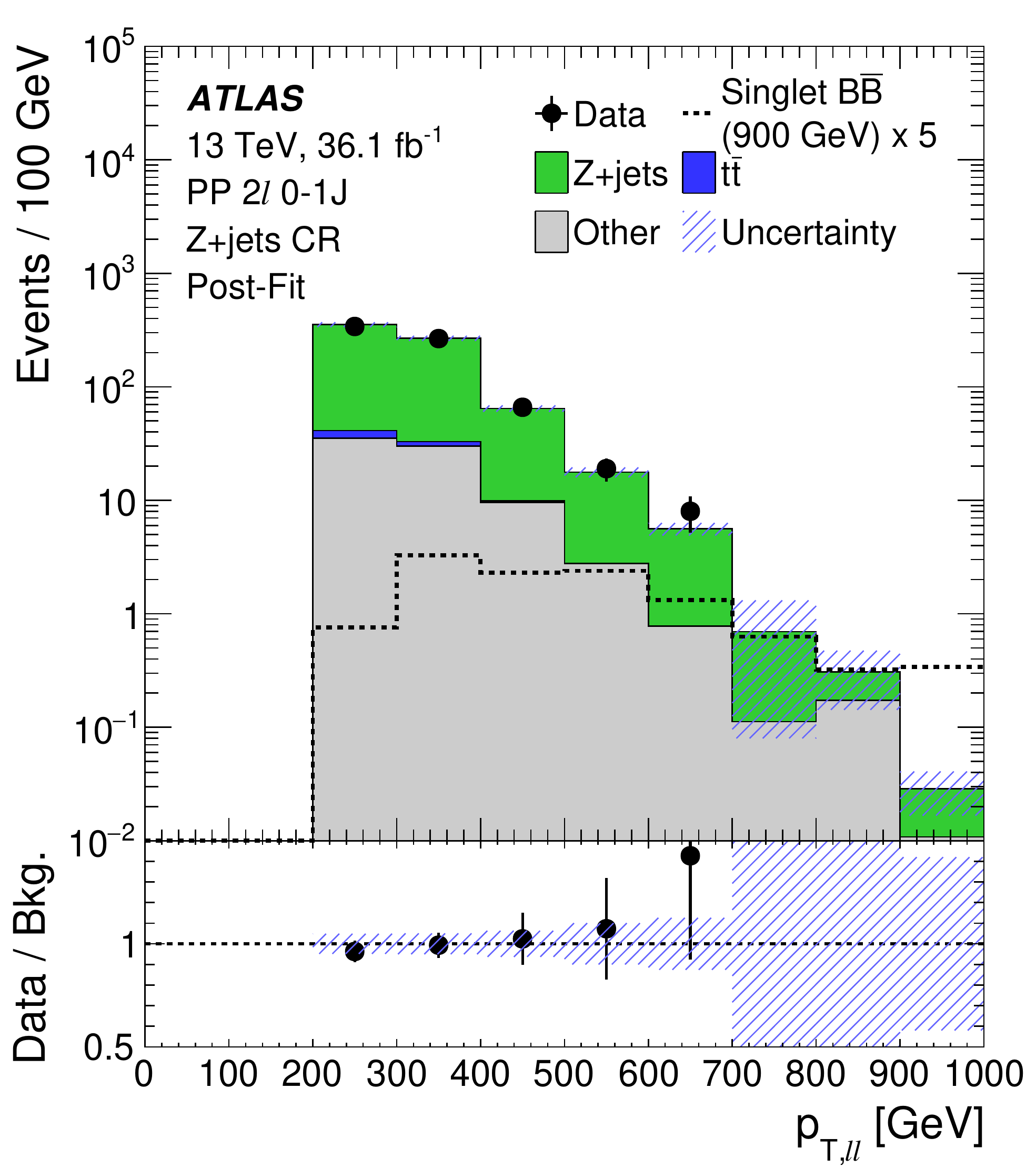}}\\
\subfloat[]{\includegraphics[width=.49\textwidth]{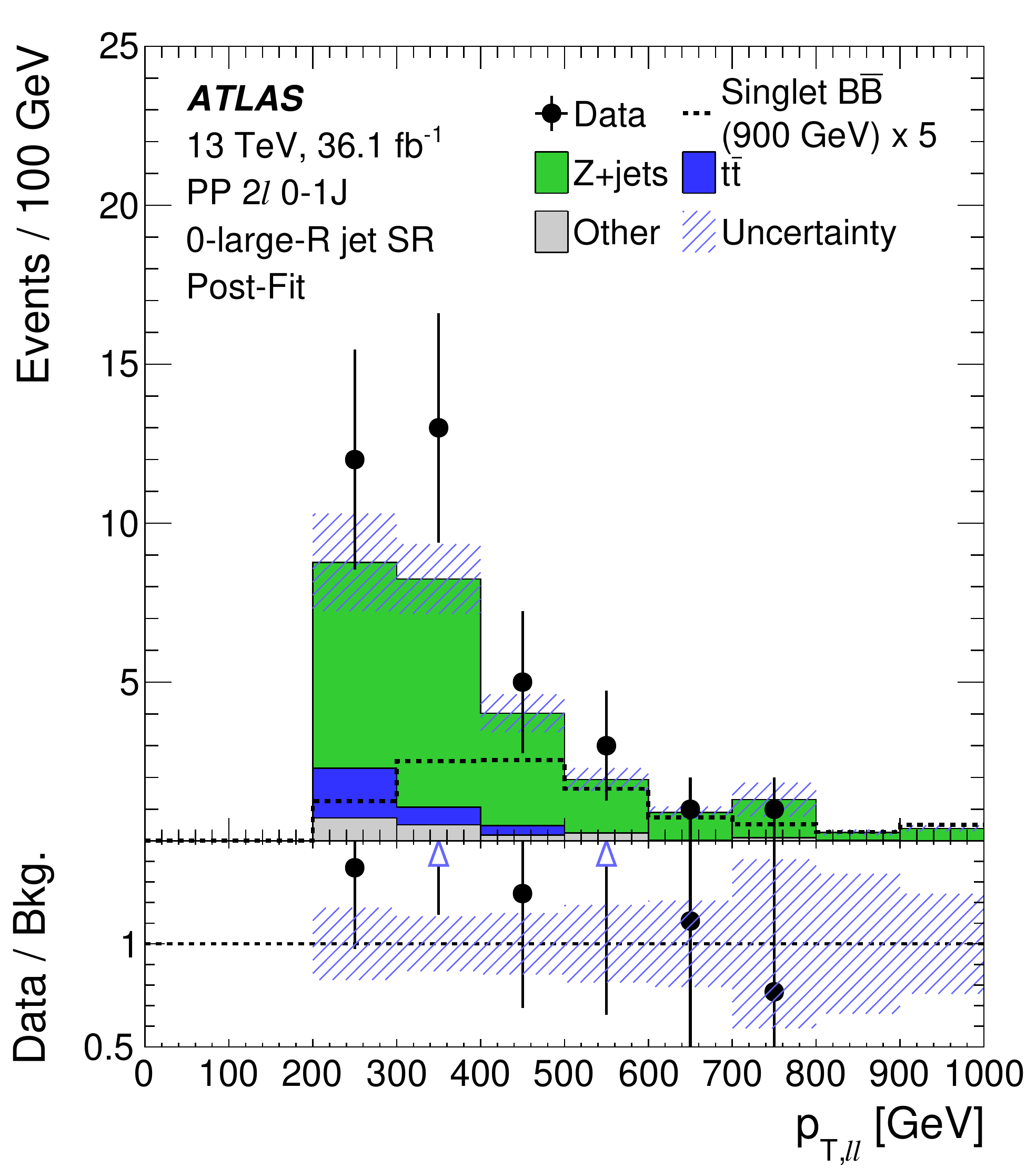}}
\subfloat[]{\includegraphics[width=.49\textwidth]{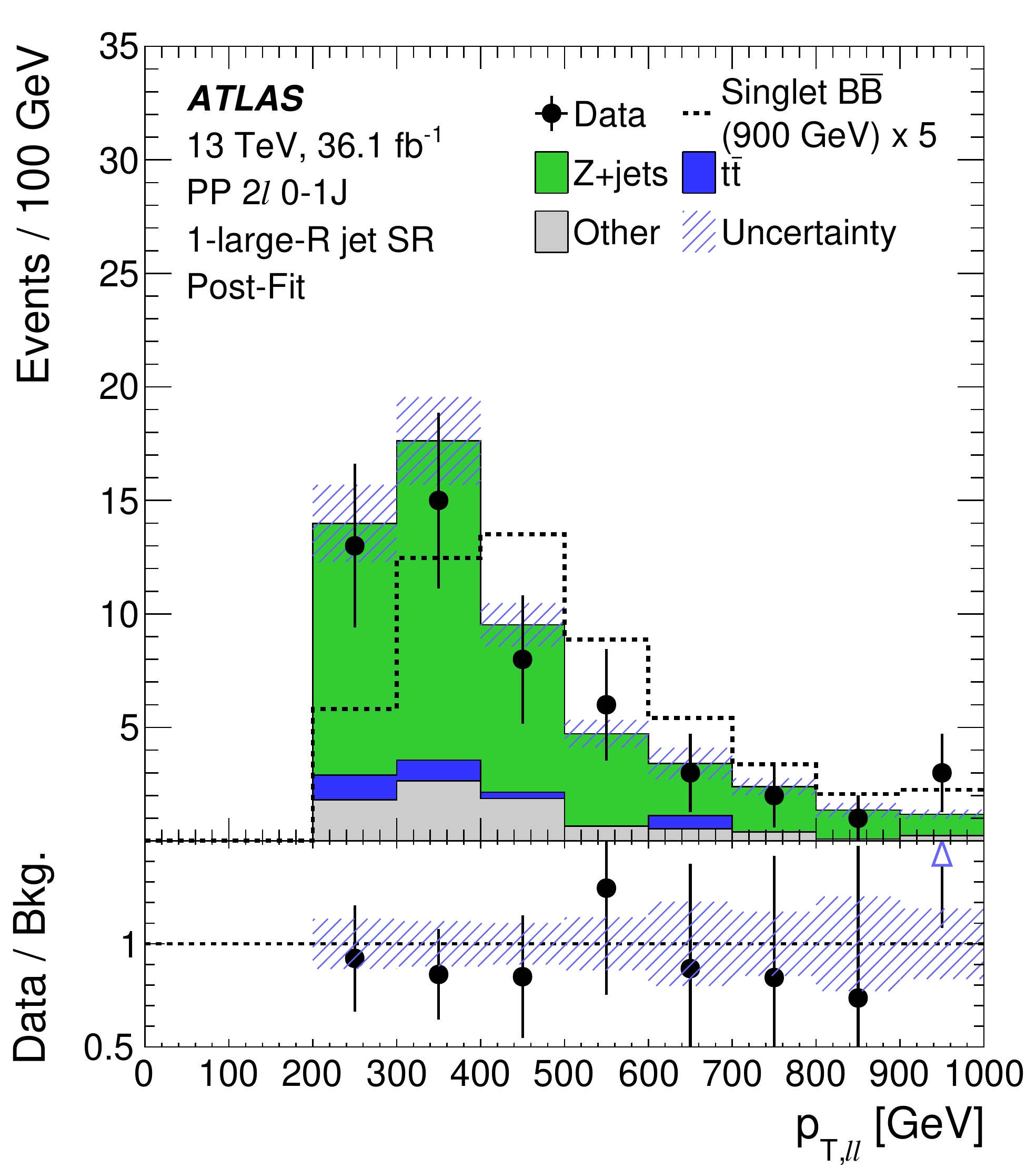}}
\caption{Comparison of the distribution of the transverse momentum of the $Z$ boson candidate, \ptll, between data and the background prediction in (a) the \ttbar\ control region, (b) the $Z$+jets control region, (c) the 0-\ljet-signal region, and (d) the 1-\ljet-signal region of the pair-production \capdilres\ channel. The background prediction is shown \emph{post-fit}, i.e.\ after the fit to the data \htj\ distributions under the background-only hypothesis. The last bin contains the overflow. An upward pointing triangle in the ratio plot indicates that the value of the ratio is beyond scale. An example distribution for a \BBbar\ signal in the singlet model with $\mVLQ = 900~\GeV$ is overlaid. For better visibility, it is multiplied by a factor of five. The data are compatible with the background-only hypothesis.}
\label{fig:dilres_Zpt}
\end{figure}

In \Fig{\ref{fig:dilres_discr}}, the \htj\ distribution is shown in the CRs and SRs for data and the background prediction after the fit. The VLQ pair-production signal would be expected to result in an excess of data over the background prediction at large values of \htj, as shown in \Fig{\ref{fig:dilres_discr_SR0J}} and \Fig{\ref{fig:dilres_discr_SR1J}}. The modeling of the main backgrounds was validated by comparing the distributions of kinematic variables and object multiplicities between data and background prediction in each respective CR. As an example, the \ptll\ distribution is shown in \Fig{\ref{fig:dilres_Zpt}} in the two CRs and the two SRs. The background prediction is shown after the fit to the \htj\ distribution. Good agreement between data and the background prediction is observed in both kinematic variables in the CRs, validating the background prediction.

\subsection{Results: \dilboost}
The observed and expected yields in the SR and the CRs and the expected number of events for the different background contributions are shown in \Tab{\ref{tab:dilboost_pre}} for the \dilboost\ channel. Also shown is the expected number of events for \BBbar\ and \TTbar\ production in the singlet model for $\mVLQ = 900~\GeV$. The signal efficiency in the SR for these benchmarks is 0.28\% for both \BBbar\ and \TTbar production, and includes the branching ratios of the VLQ as well as of its decay products, including the decay $Z\rightarrow\ell^+\ell^-$.

\begin{table}[p]
\centering
        \caption{Observed number of events in data and \emph{pre-fit} expected number of signal and background events in the control regions and the signal region for the \dilboost\ channel, i.e.\ before the fit to data. For the signal, the expected number of events for the \BBbar\ and \TTbar\ benchmark processes with $\mVLQ = 900~\GeV$ is shown for the singlet model. Statistical uncertainties from the limited size of MC samples and systematic uncertainties are added in quadrature. The uncertainty in the ratio of the observed and expected numbers of events contains the systematic uncertainties and the statistical uncertainty of the prediction from Poisson fluctuations.}
        \begin{tabular}{l|R{1cm}@{\hspace{1.5pt}}c@{\hspace{1.5pt}}p{1cm}|R{1cm}@{\hspace{1.5pt}}c@{\hspace{1.5pt}}p{1cm}|R{1cm}@{\hspace{1.5pt}}c@{\hspace{1.5pt}}p{1cm}}
                         \toprule

                &\multicolumn{3}{c|}{$t\bar{t}$ CR} & \multicolumn{3}{c|}{$Z$+jets CR} &          \multicolumn{3}{c}{SR}        \\\midrule
Singlet \BBbar (900~GeV)&   \num{3.00} & $\pm$ & \num{0.34}  &   \num{2.65} & $\pm$ & \num{0.28}  &     \num{9.2} & $\pm$ & \num{0.6}    \\
Singlet \TTbar (900~GeV)&   \num{2.25} & $\pm$ & \num{0.17}  &    \num{3.9} & $\pm$ & \num{0.4}   &     \num{9.2} & $\pm$ & \num{0.6}    \\
\midrule
    $Z$+jets    &     \num{11} & $\pm$ & \num{5}     &     \num{66} & $\pm$ & \num{22}    &       \num{8} & $\pm$ & \num{4}      \\
   $t\bar{t}$   &     \num{80} & $\pm$ & \num{70}    &     \num{18} & $\pm$ & \num{14}    &     \num{2.0} & $\pm$ & \num{3.4}    \\
   Single top   &    \num{1.5} & $\pm$ & \num{0.8}   &   \num{0.61} & $\pm$ & \num{0.34}  &   \num{0.010} & $\pm$ & \num{0.010}  \\
  $t\bar{t}+X$  &    \num{4.3} & $\pm$ & \num{0.9}   &   \num{14.4} & $\pm$ & \num{2.9}   &     \num{1.3} & $\pm$ & \num{0.4}    \\
    Diboson     &   \num{0.74} & $\pm$ & \num{0.20}  &    \num{4.1} & $\pm$ & \num{1.0}   &     \num{0.9} & $\pm$ & \num{0.4}    \\
\midrule
   Total Bkg.   &    \num{100} & $\pm$ & \num{70}    &    \num{103} & $\pm$ & \num{26}    &      \num{12} & $\pm$ & \num{5}      \\
\midrule
      Data      &   \multicolumn{3}{c|}{\num{112}}   &   \multicolumn{3}{c|}{\num{100}}   &     \multicolumn{3}{c}{\num{9}}      \\
\midrule
   Data/Bkg.    &         1.2 & $\pm$ & 0.8          &        0.98 & $\pm$ & 0.26         &          0.7 & $\pm$ & 0.4           \\
\bottomrule
        \end{tabular}
            \label{tab:dilboost_pre}

\vspace{1cm}
        \caption{Observed number of events in data and \emph{post-fit} expected number of background events in the control regions and the signal region for the \dilboost\ channel, i.e.\ after the fit to the data $m_{Zb}$ distributions under the background-only hypothesis. The uncertainty in the expected number of events is the full uncertainty from the fit, from which the uncertainty in the ratio of the observed and expected numbers of events is calculated.}
        \begin{tabular}{l|R{1cm}@{\hspace{1.5pt}}c@{\hspace{1.5pt}}p{1cm}|R{1cm}@{\hspace{1.5pt}}c@{\hspace{1.5pt}}p{1cm}|R{1cm}@{\hspace{1.5pt}}c@{\hspace{1.5pt}}p{1cm}}
                         \toprule

                      &\multicolumn{3}{c|}{$t\bar{t}$ CR} & \multicolumn{3}{c|}{$Z$+jets CR} &          \multicolumn{3}{c}{SR}        \\\midrule
       $Z$+jets       &    \num{9.0} & $\pm$ & \num{2.3}   &     \num{60} & $\pm$ & \num{10}    &     \num{6.5} & $\pm$ & \num{2.2}    \\
      $t\bar{t}$      &     \num{95} & $\pm$ & \num{12}    &     \num{20} & $\pm$ & \num{6}     &     \num{2.2} & $\pm$ & \num{1.5}    \\
      Single top      &    \num{1.5} & $\pm$ & \num{0.6}   &   \num{0.63} & $\pm$ & \num{0.28}  &   \num{0.016} & $\pm$ & \num{0.011}  \\
     $t\bar{t}+X$     &    \num{4.5} & $\pm$ & \num{0.8}   &   \num{14.7} & $\pm$ & \num{2.7}   &     \num{1.3} & $\pm$ & \num{0.4}    \\
       Diboson        &   \num{0.74} & $\pm$ & \num{0.20}  &    \num{4.2} & $\pm$ & \num{0.8}   &     \num{0.9} & $\pm$ & \num{0.4}    \\
\midrule
      Total Bkg.      &    \num{111} & $\pm$ & \num{12}    &    \num{100} & $\pm$ & \num{10}    &    \num{10.9} & $\pm$ & \num{2.7}    \\
\midrule
         Data         &   \multicolumn{3}{c|}{\num{112}}   &   \multicolumn{3}{c|}{\num{100}}   &     \multicolumn{3}{c}{\num{9}}      \\
\midrule
      Data/Bkg.       &        1.01 & $\pm$ & 0.11         &        1.00 & $\pm$ & 0.10         &         0.83 & $\pm$ & 0.21          \\
\bottomrule
        \end{tabular}
            \label{tab:dilboost_post}

\end{table}

A fit of the background prediction to the $m_{Zb}$ distributions in data was performed. The post-fit yields are shown in \Tab{\ref{tab:dilboost_post}}. The uncertainty in the background prediction was significantly reduced in all regions compared to the pre-fit value (\Tab{\ref{tab:dilboost_pre}}). The overall $Z$+jets (\ttbar) normalization was adjusted by a factor of \normfzjetszjcrtwolb\ (\normfttbarttcrtwolb) in the $Z$+jets (\ttbar) CR. The ratios of the post-fit and pre-fit background yields are consistent with unity in all regions.

\begin{figure}[p]
\centering
\subfloat[]{\includegraphics[width=.49\textwidth]{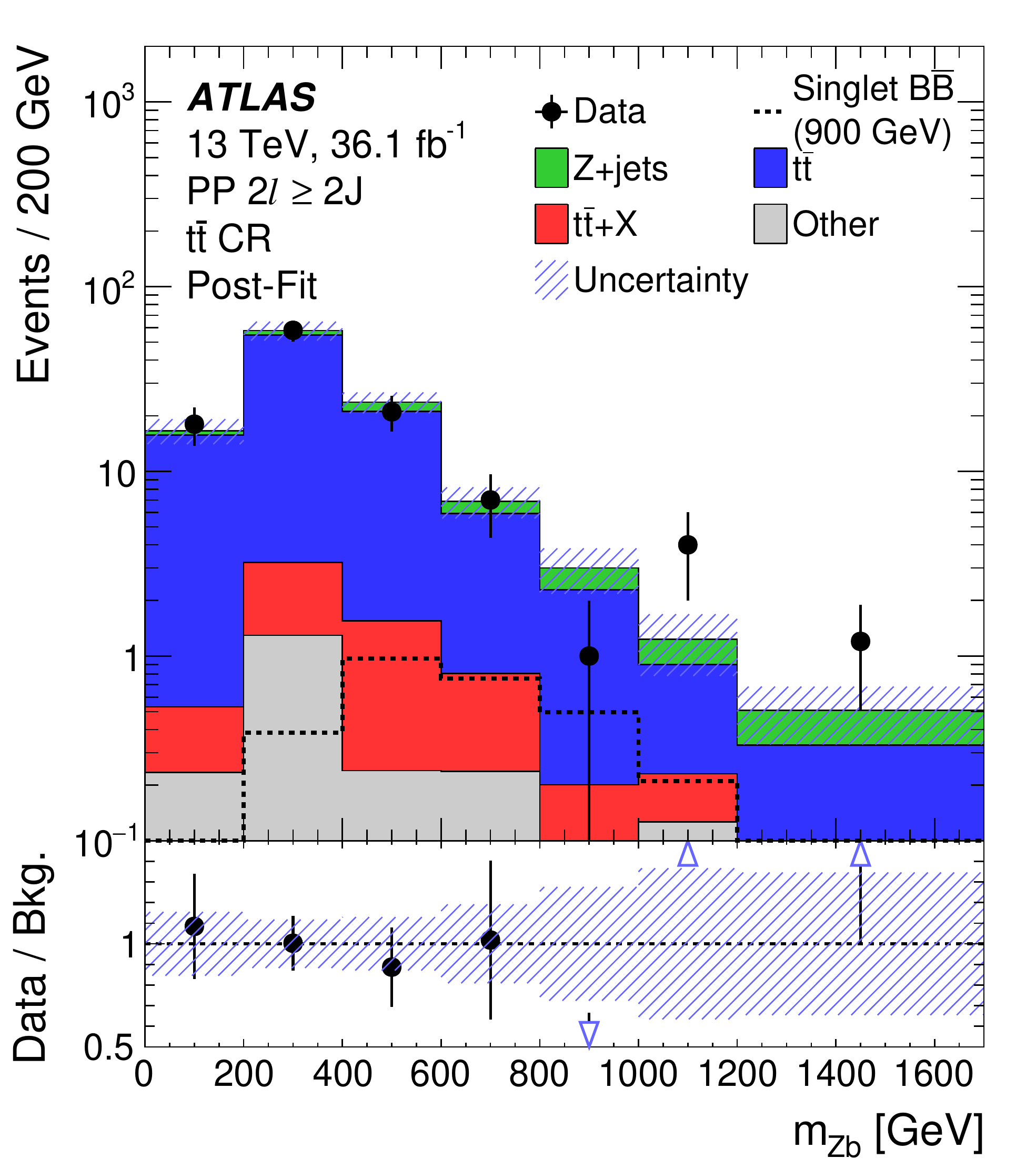}}
\subfloat[]{\includegraphics[width=.49\textwidth]{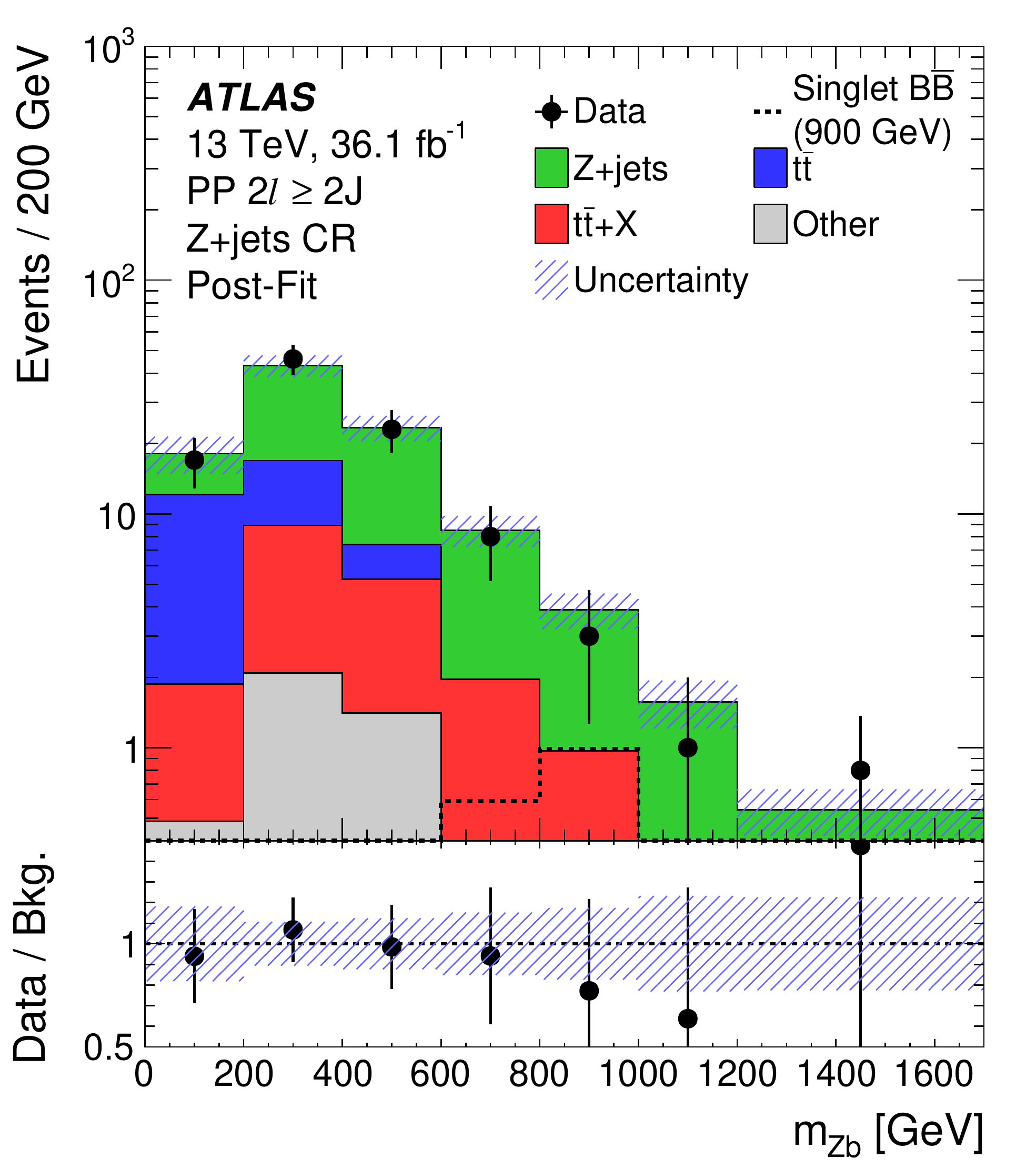}}\\
\subfloat[]{\includegraphics[width=.49\textwidth]{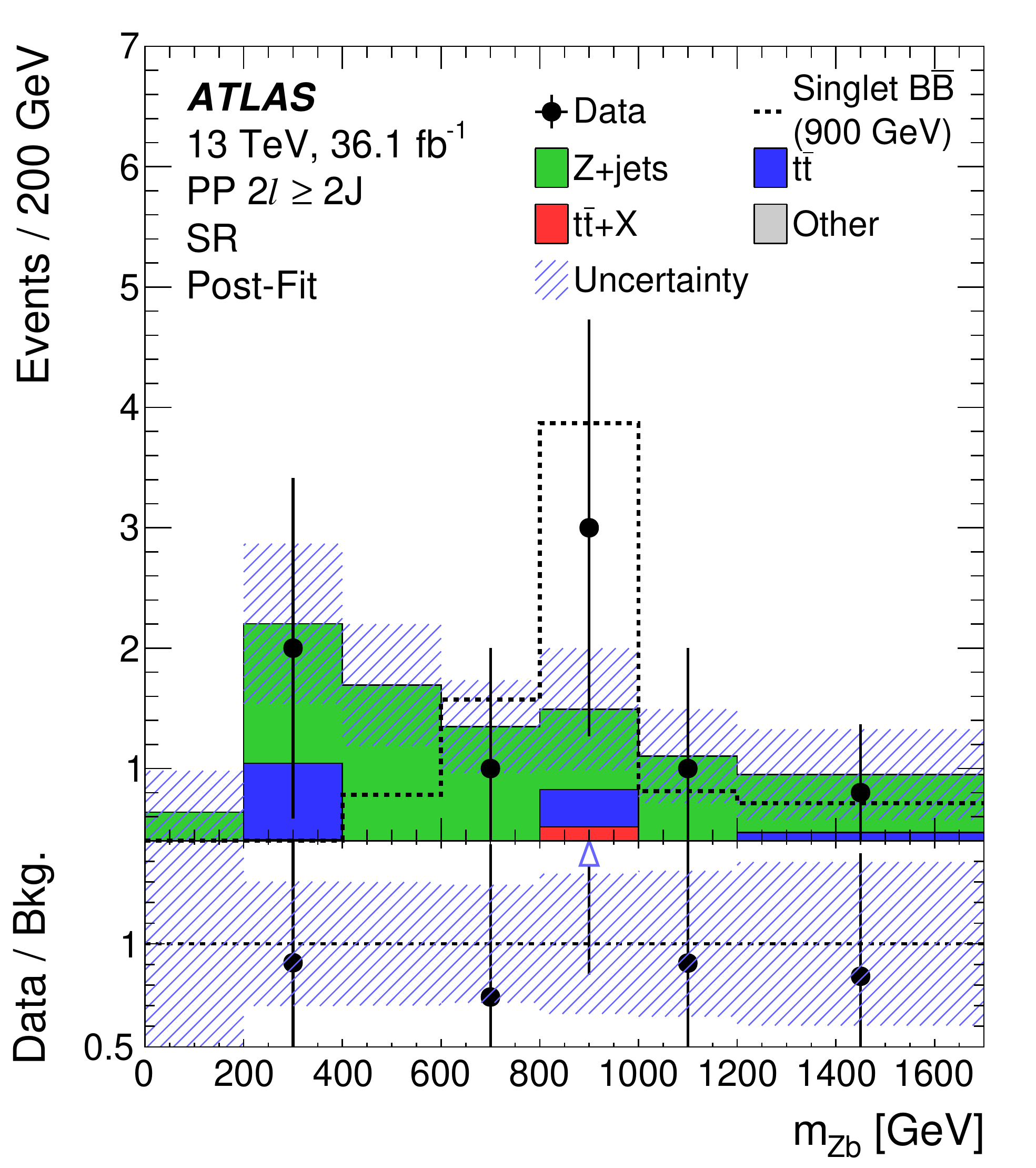}\label{fig:dilboost_discr_SR}}
\caption{Comparison of the distribution of the invariant mass of the $Z$ boson candidate and the highest-\pt\ \btagged\ jet, $m_{Zb}$, between data and the background prediction in (a) the \ttbar\ control region, (b) the $Z$+jets control region, and (c) the signal region of the pair-production \capdilboost\ channel. The background prediction is shown \emph{post-fit}, i.e.\ after the fit to the data $m_{Zb}$ distributions under the background-only hypothesis. The last bin contains the overflow. An upward or downward pointing triangle in the ratio plot indicates that the value of the ratio is beyond scale. An example distribution for a \BBbar\ signal in the singlet model with $\mVLQ = 900~\GeV$ is overlaid. The data are compatible with the background-only hypothesis.}
\label{fig:dilboost_discr}
\end{figure}

\begin{figure}[p]
\centering
\subfloat[]{\includegraphics[width=.49\textwidth]{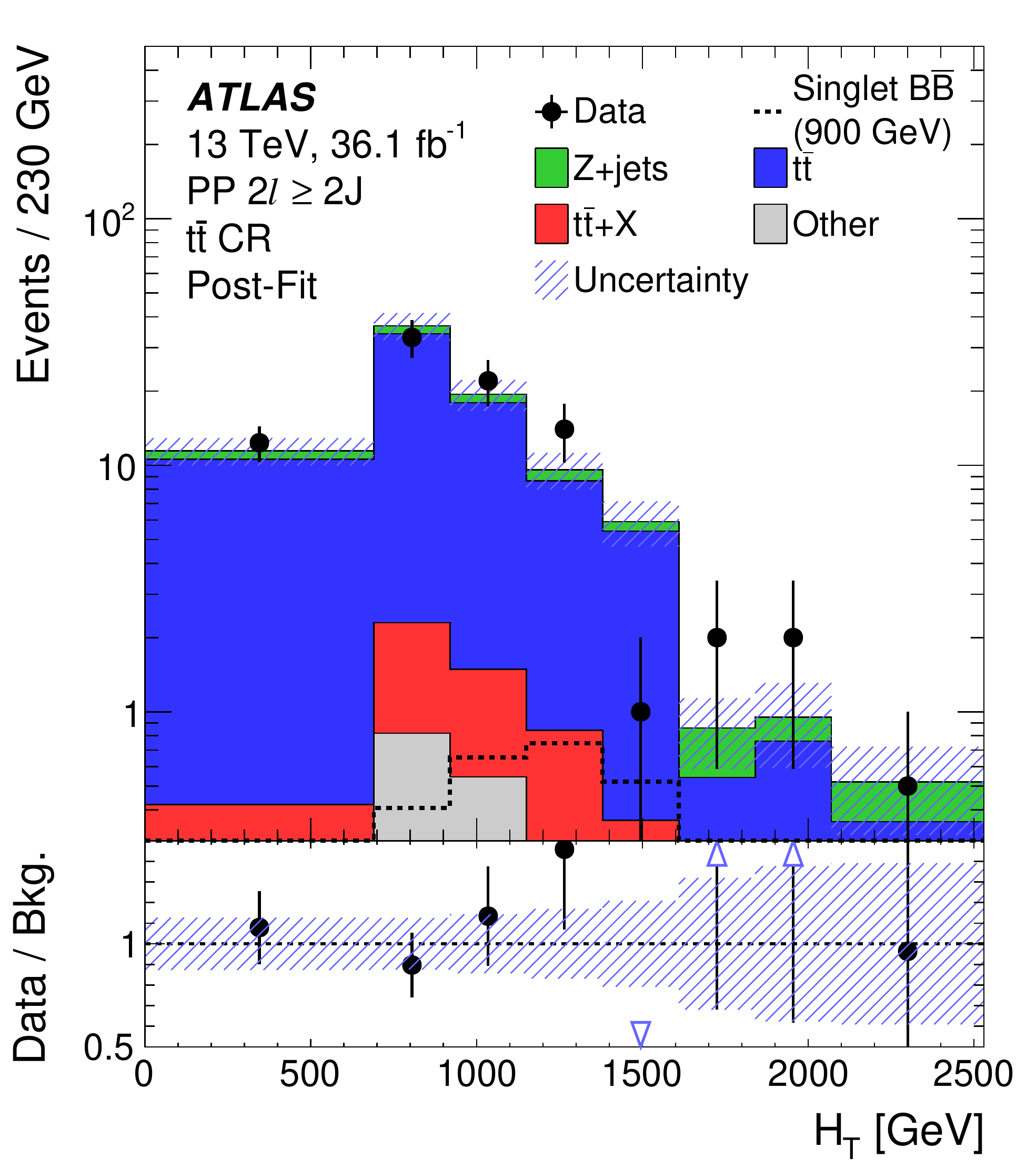}}
\subfloat[]{\includegraphics[width=.49\textwidth]{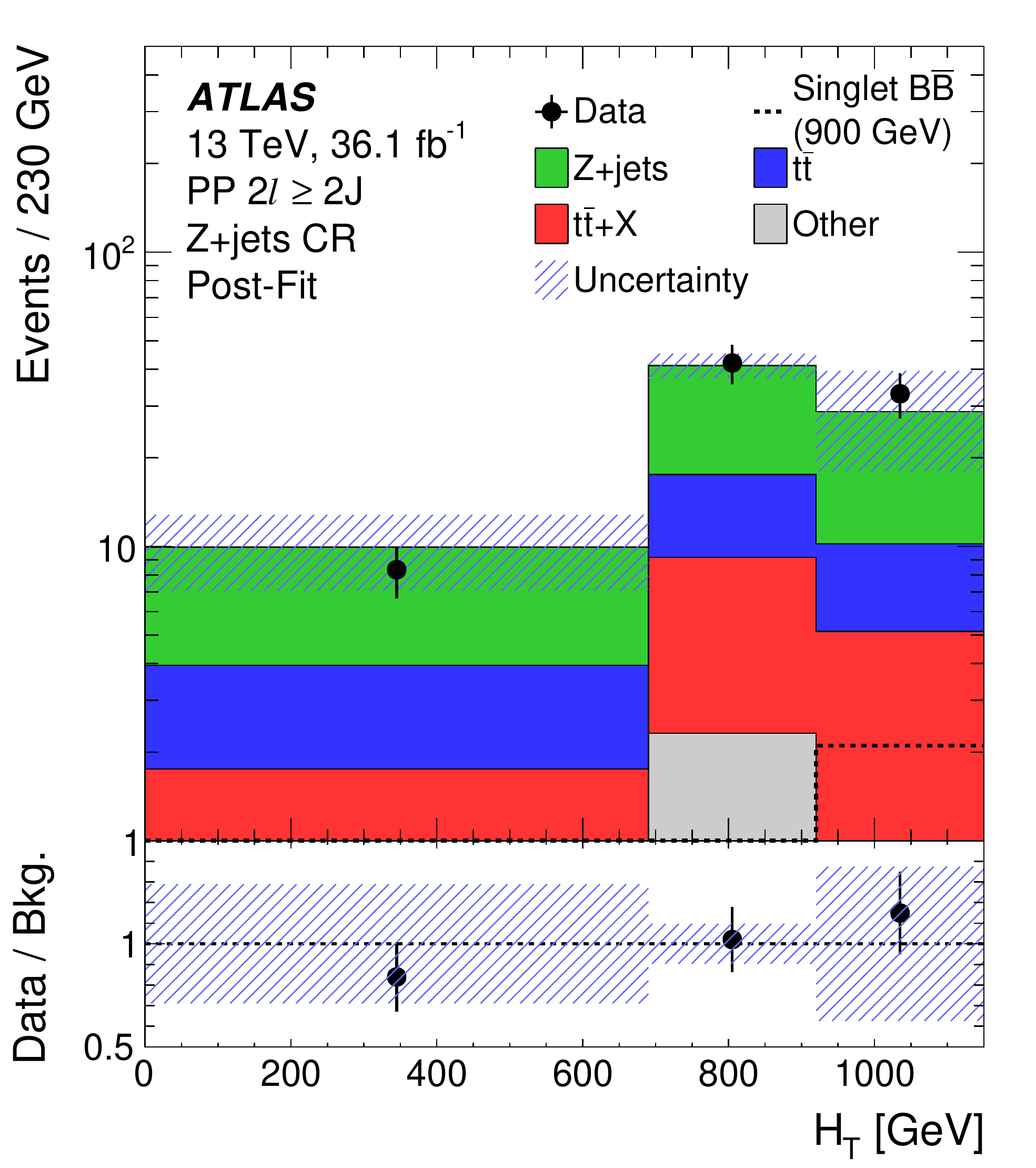}}\\
\subfloat[]{\includegraphics[width=.49\textwidth]{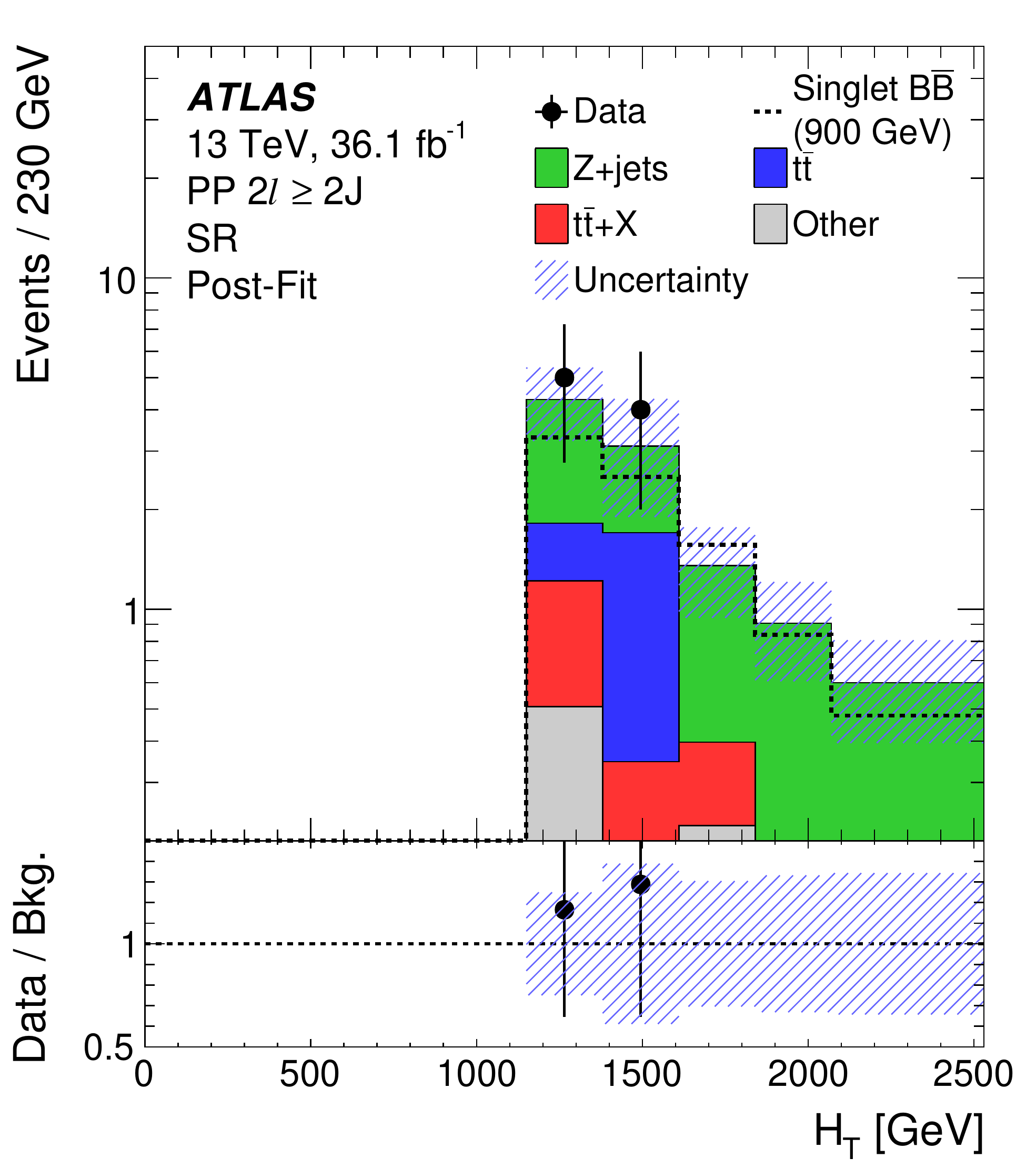}}
\caption{Comparison of the distribution of the scalar sum of \sjet\ transverse momenta, \htj, between data and the background prediction in (a) the \ttbar\ control region, (b) the $Z$+jets control region, and (c) the signal region of the pair-production \capdilboost\ channel. The background prediction is shown \emph{post-fit}, i.e.\ after the fit to the data $m_{Zb}$ distributions under the background-only hypothesis. The last bin contains the overflow. An upward or downward pointing triangle in the ratio plot indicates that the value of the ratio is beyond scale. An example distribution for a \BBbar\ signal in the singlet model with $\mVLQ = 900~\GeV$ is overlaid. The data are compatible with the background-only hypothesis.}
\label{fig:dilboost_htj}
\end{figure}

In \Fig{\ref{fig:dilboost_discr}}, the $m_{Zb}$ distribution is shown in the CRs and SR for data and the background prediction after the fit. The VLQ pair-production signal would be expected to result in an excess of data over the background prediction at large values of $m_{Zb}$, as shown in \Fig{\ref{fig:dilboost_discr_SR}}. The modeling of the main backgrounds was validated by comparing the distributions of kinematic variables and object multiplicities between data and background prediction in the respective CR. As an example, the \htj\ distribution is shown in \Fig{\ref{fig:dilboost_htj}} in the two CRs and in the SR. The background prediction is shown after the fit to the $m_{Zb}$ distribution. Good agreement between data and the background prediction is apparent in  kinematic variables in the CRs, validating the background prediction.

\subsection{Results: \tripair}
The observed number of events in the SR and the CRs and the expected number of events for the different background contributions are shown in \Tab{\ref{tab:trilepair_pre}} for the \tripair\ channel. Also shown is the expected number of events for \BBbar\ and \TTbar\ production in the singlet model for $\mVLQ = 900~\GeV$. The signal efficiency in the SR for these benchmarks is 0.31\% for \BBbar and 0.44\% for \TTbar production, and includes the branching ratios of the VLQ as well as of its decay products, including the decay $Z\rightarrow\ell^+\ell^-$.

\begin{table}[p]
\centering
        \caption{Observed number of events in data and \emph{pre-fit} expected number of signal and background events in the control regions and the signal region for the \tripair\ channel, i.e. before the fit to data. For the signal, the expected number of events for the \BBbar\ and \TTbar\ benchmark processes with $\mVLQ = 900~\GeV$ is shown for the singlet model. Statistical uncertainties from the limited size of MC samples and systematic uncertainties are added in quadrature. The uncertainty in the ratio of the observed and expected numbers of events contains the systematic uncertainties and the statistical uncertainty of the prediction from Poisson fluctuations.}
        \begin{tabular}{l|R{1cm}@{\hspace{1.5pt}}c@{\hspace{1.5pt}}p{1cm}|R{1cm}@{\hspace{1.5pt}}c@{\hspace{1.5pt}}p{1cm}|R{1cm}@{\hspace{1.5pt}}c@{\hspace{1.5pt}}p{1cm}}
                         \toprule

                 & \multicolumn{3}{c|}{Diboson CR} &  \multicolumn{3}{c|}{$t\bar{t}+X$ CR} &        \multicolumn{3}{c}{SR}       \\\midrule
Singlet \BBbar (900~GeV)&   \num{1.57} & $\pm$ & \num{0.31}  &   \num{1.26} & $\pm$ & \num{0.15}  &   \num{10.1} & $\pm$ & \num{0.6}   \\
Singlet \TTbar (900~GeV)&   \num{1.60} & $\pm$ & \num{0.30}  &   \num{1.64} & $\pm$ & \num{0.14}  &   \num{14.2} & $\pm$ & \num{0.7}   \\
\midrule
    $Z$+jets     &     \num{50} & $\pm$ & \num{80}    &     \num{11} & $\pm$ & \num{5}     &    \num{1.8} & $\pm$ & \num{2.8}   \\
   $t\bar{t}$    &     \num{7} & $\pm$ & \num{29}     &     \num{14} & $\pm$ & \num{13}    &    \num{0.7} & $\pm$ & \num{1.5}   \\
   Single top    &    \num{7.2} & $\pm$ & \num{2.0}   &     \num{26} & $\pm$ & \num{7}     &    \num{4.2} & $\pm$ & \num{1.1}   \\
  $t\bar{t}+X$   &     \num{23} & $\pm$ & \num{4}     &    \num{111} & $\pm$ & \num{15}    &     \num{47} & $\pm$ & \num{6}     \\
     Diboson     &   \num{1130} & $\pm$ & \num{280}   &    \num{120} & $\pm$ & \num{60}    &     \num{30} & $\pm$ & \num{14}    \\
    Triboson     &    \num{5.5} & $\pm$ & \num{0.5}   &   \num{0.43} & $\pm$ & \num{0.08}  &   \num{0.19} & $\pm$ & \num{0.04}  \\
\midrule
   Total Bkg.    &   \num{1220} & $\pm$ & \num{290}   &    \num{290} & $\pm$ & \num{60}    &     \num{84} & $\pm$ & \num{15}    \\
\midrule
      Data       &  \multicolumn{3}{c|}{\num{1150}}   &   \multicolumn{3}{c|}{\num{320}}   &    \multicolumn{3}{c}{\num{93}}    \\
\midrule
    Data/Bkg.    &        0.94 & $\pm$ & 0.23         &        1.12 & $\pm$ & 0.24         &        1.11 & $\pm$ & 0.24         \\
\bottomrule
        \end{tabular}
            \label{tab:trilepair_pre}

\vspace{1cm}
        \caption{Observed number of events in data and \emph{post-fit} expected number of background events in the control regions and the signal region for the \tripair\ channel, i.e.\ after the fit to the data \htjl\ distributions under the background-only hypothesis. The uncertainty in the expected number of events is the full uncertainty from the fit, from which the uncertainty in the ratio of the observed and expected numbers of events is calculated.}
        \begin{tabular}{l|R{1cm}@{\hspace{1.5pt}}c@{\hspace{1.5pt}}p{1cm}|R{1cm}@{\hspace{1.5pt}}c@{\hspace{1.5pt}}p{1cm}|R{1cm}@{\hspace{1.5pt}}c@{\hspace{1.5pt}}p{1cm}}
                         \toprule

                 &\multicolumn{3}{c|}{Diboson CR} &  \multicolumn{3}{c|}{$t\bar{t}+X$ CR} &        \multicolumn{3}{c}{SR}       \\\midrule
    $Z$+jets     &    \num{60} & $\pm$ & \num{60}    &     \num{12} & $\pm$ & \num{5}     &    \num{2.1} & $\pm$ & \num{2.1}   \\
   $t\bar{t}$    &     \num{5} & $\pm$ & \num{11}    &     \num{18} & $\pm$ & \num{8}     &    \num{0.4} & $\pm$ & \num{1.2}   \\
   Single top    &   \num{6.9} & $\pm$ & \num{2.0}   &     \num{29} & $\pm$ & \num{6}     &    \num{4.3} & $\pm$ & \num{1.1}   \\
  $t\bar{t}+X$   &     \num{23} & $\pm$ & \num{4}    &    \num{117} & $\pm$ & \num{14}    &     \num{49} & $\pm$ & \num{6}     \\
     Diboson     &   \num{1060} & $\pm$ & \num{70}   &    \num{137} & $\pm$ & \num{29}    &     \num{34} & $\pm$ & \num{7}     \\
    Triboson     &   \num{5.4} & $\pm$ & \num{0.4}   &   \num{0.43} & $\pm$ & \num{0.07}  &   \num{0.19} & $\pm$ & \num{0.04}  \\
\midrule
   Total Bkg.    &   \num{1160} & $\pm$ & \num{40}   &    \num{313} & $\pm$ & \num{21}    &     \num{90} & $\pm$ & \num{6}     \\
\midrule
      Data       &  \multicolumn{3}{c|}{\num{1150}}  &   \multicolumn{3}{c|}{\num{320}}   &    \multicolumn{3}{c}{\num{93}}    \\
\midrule
    Data/Bkg.    &        1.00 & $\pm$ & 0.04        &        1.02 & $\pm$ & 0.07         &        1.03 & $\pm$ & 0.07         \\
\bottomrule
        \end{tabular}
            \label{tab:trilepair_post}

\end{table}

A fit of the background prediction to the \htjl\ distributions in data was performed and the post-fit yields are shown in \Tab{\ref{tab:trilepair_post}}. The uncertainty in the background prediction was significantly reduced in all regions compared to the pre-fit value (\Tab{\ref{tab:trilepair_pre}}). The overall diboson ($\ttbar+X$) normalization is adjusted by a factor of \normfdibosonvvcrtril\ (\normfttvttvcrtril) in the diboson ($\ttbar+X$) CR. The ratios of the post-fit and pre-fit background yields are consistent with unity in all regions.

\begin{figure}[p]
\centering
\subfloat[]{\includegraphics[width=.49\textwidth]{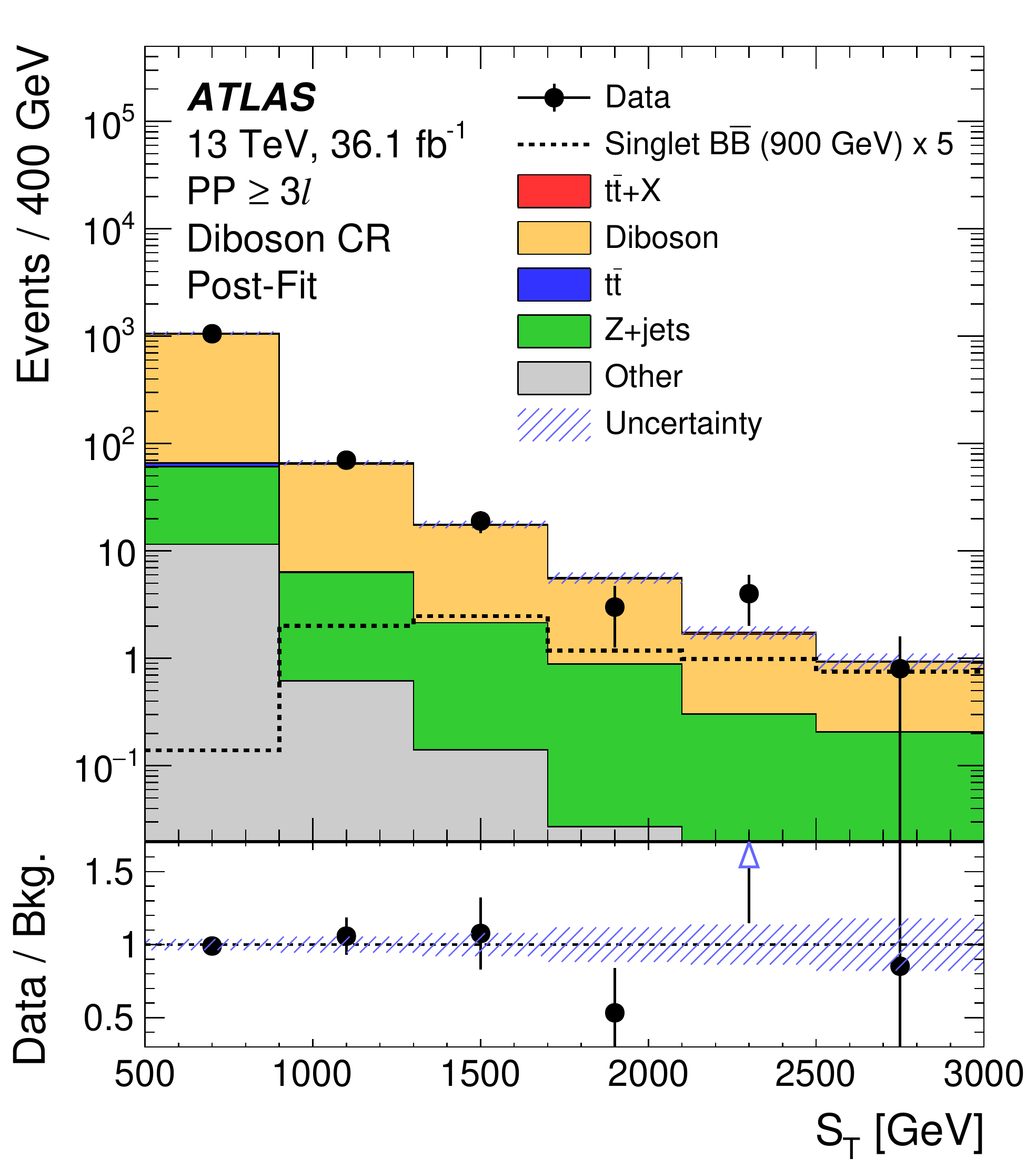}}
\subfloat[]{\includegraphics[width=.49\textwidth]{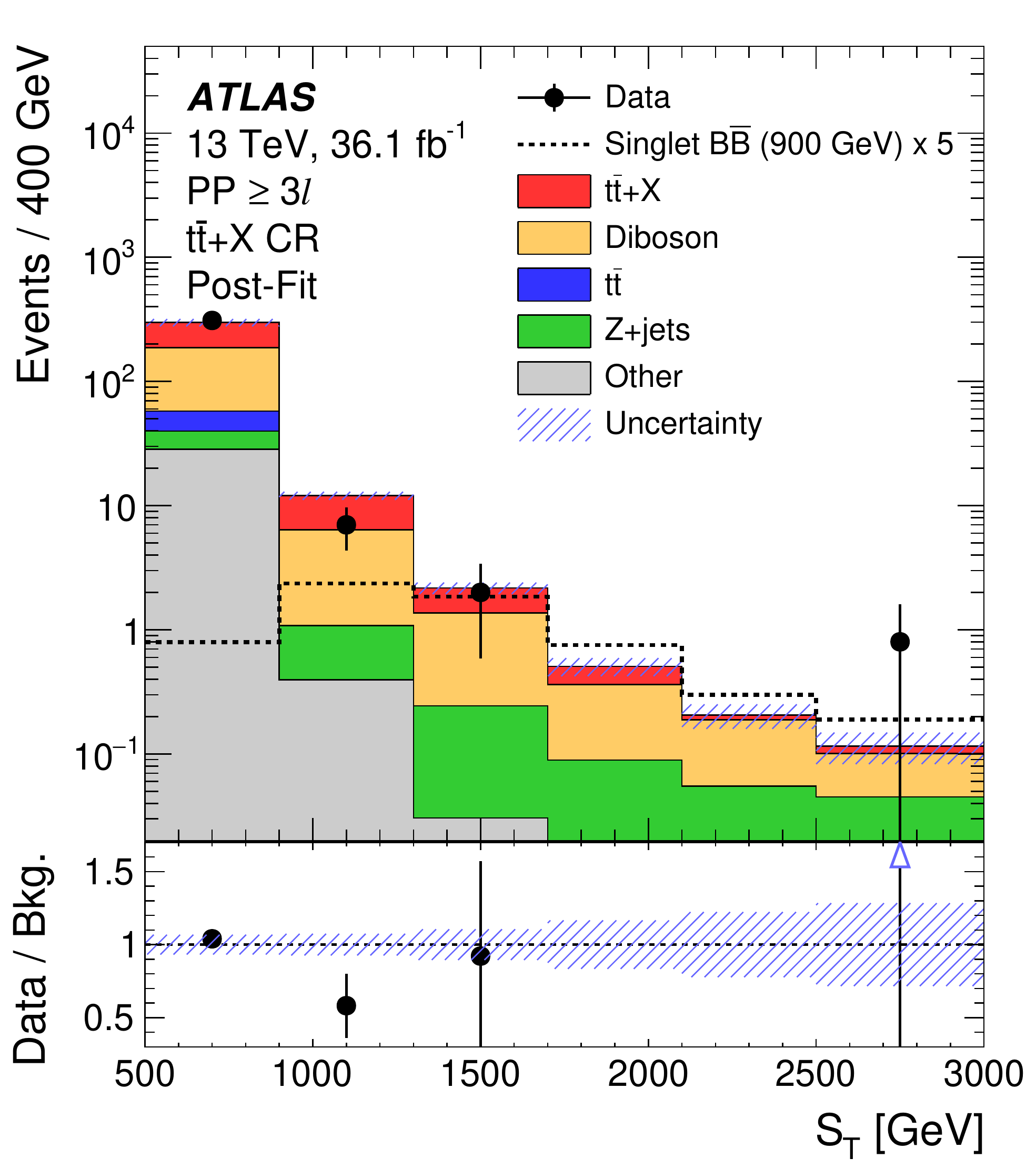}}\\
\subfloat[]{\includegraphics[width=.49\textwidth]{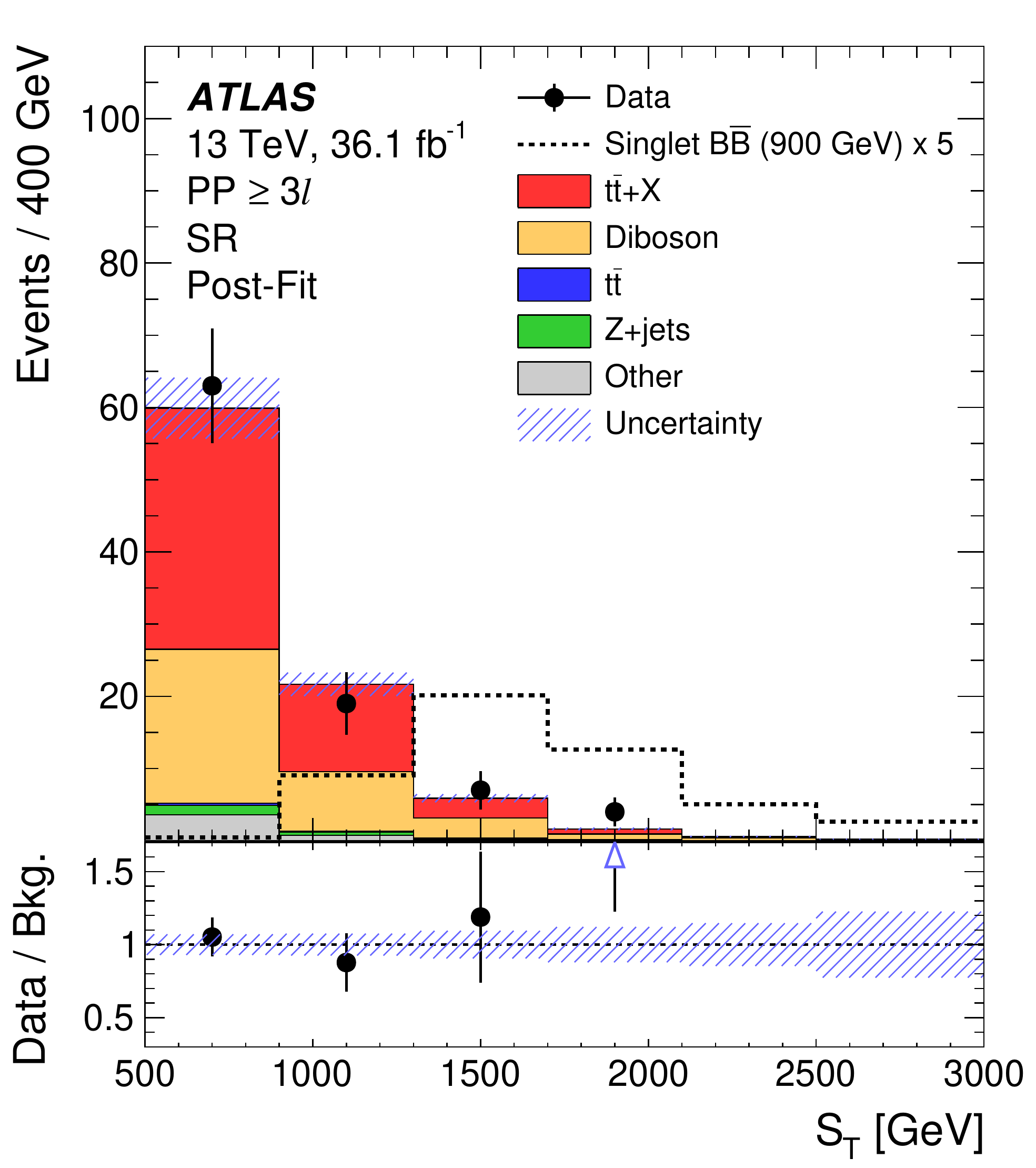}\label{fig:trilepair_discr_SR}}
\caption{Comparison of the distribution of the scalar sum of \sjet\ and lepton transverse momenta, \htjl, between data and the background prediction in (a) the diboson control region, (b) the $\ttbar+X$ control region, and (c) the signal region of the pair-production \captripair\ channel. The background prediction is shown \emph{post-fit}, i.e.\ after the fit to the data \htjl\ distributions under the background-only hypothesis. The last bin contains the overflow. An upward pointing triangle in the ratio plot indicates that the value of the ratio is beyond scale. An example distribution for a \BBbar\ signal in the singlet model with $\mVLQ = 900~\GeV$ is overlaid. For better visibility, it is multiplied by a factor of five. The data are compatible with the background-only hypothesis.}
\label{fig:trilepair_discr}
\end{figure}

\begin{figure}[p]
\centering
\subfloat[]{\includegraphics[width=.49\textwidth]{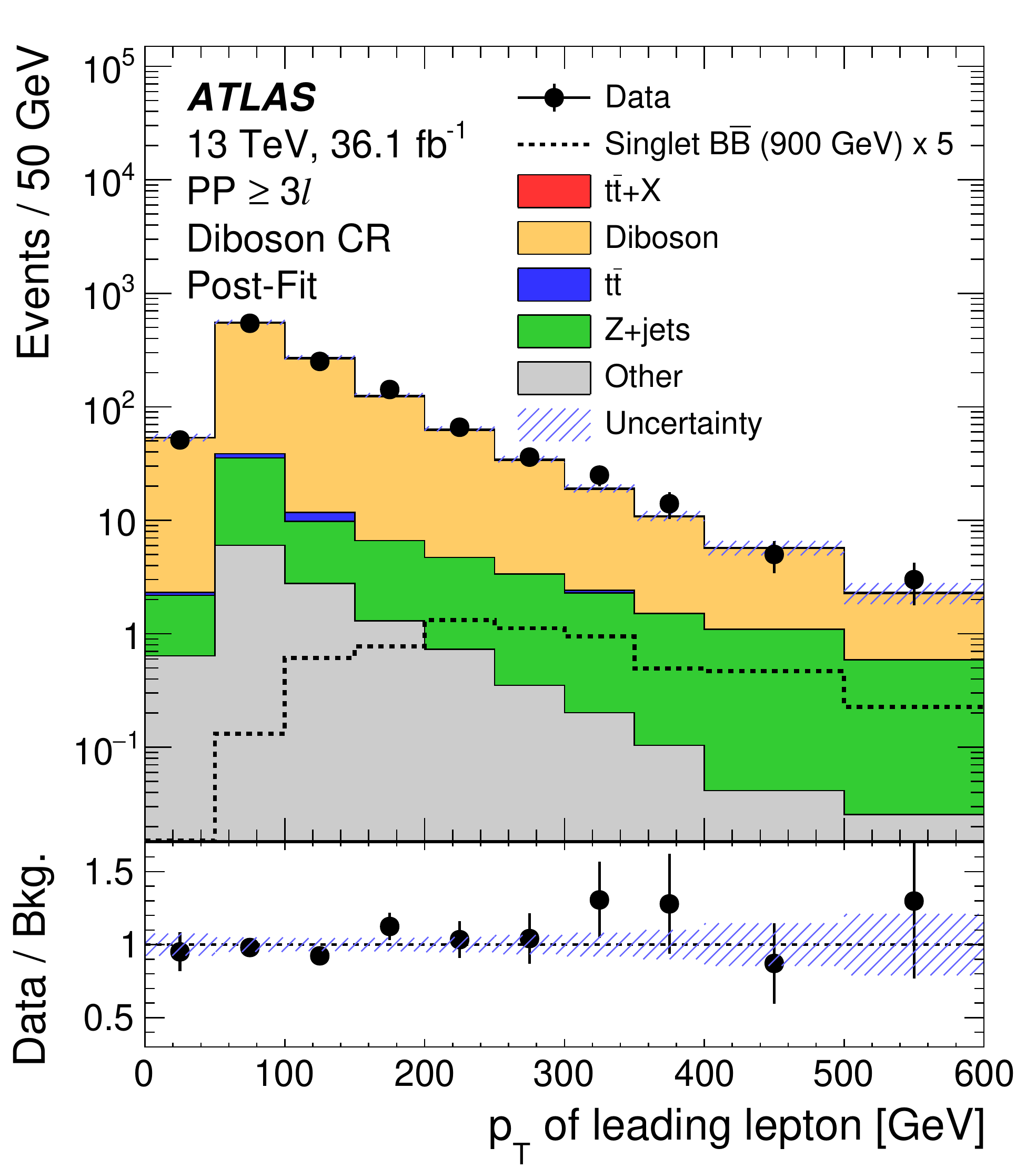}}
\subfloat[]{\includegraphics[width=.49\textwidth]{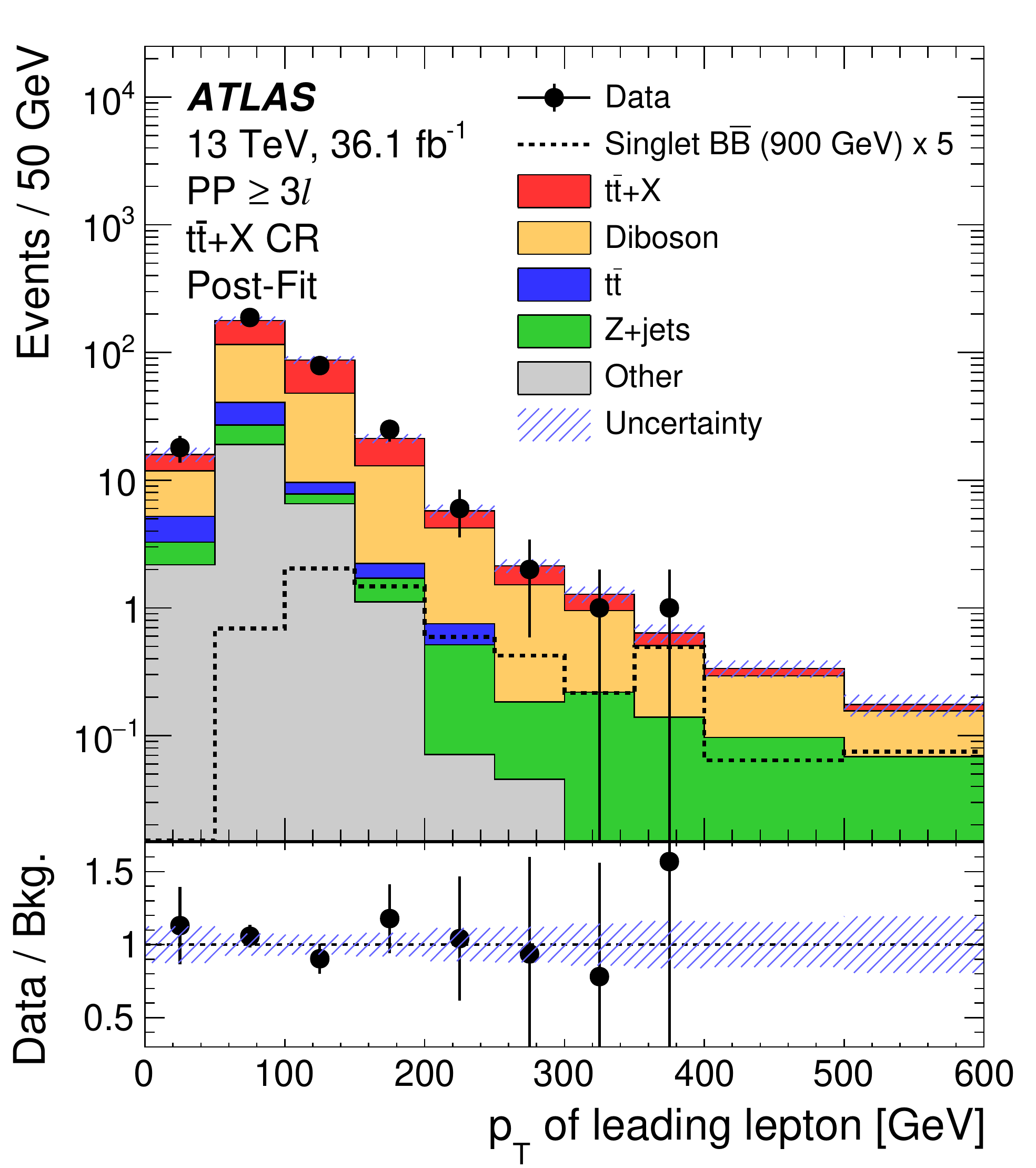}}\\
\subfloat[]{\includegraphics[width=.49\textwidth]{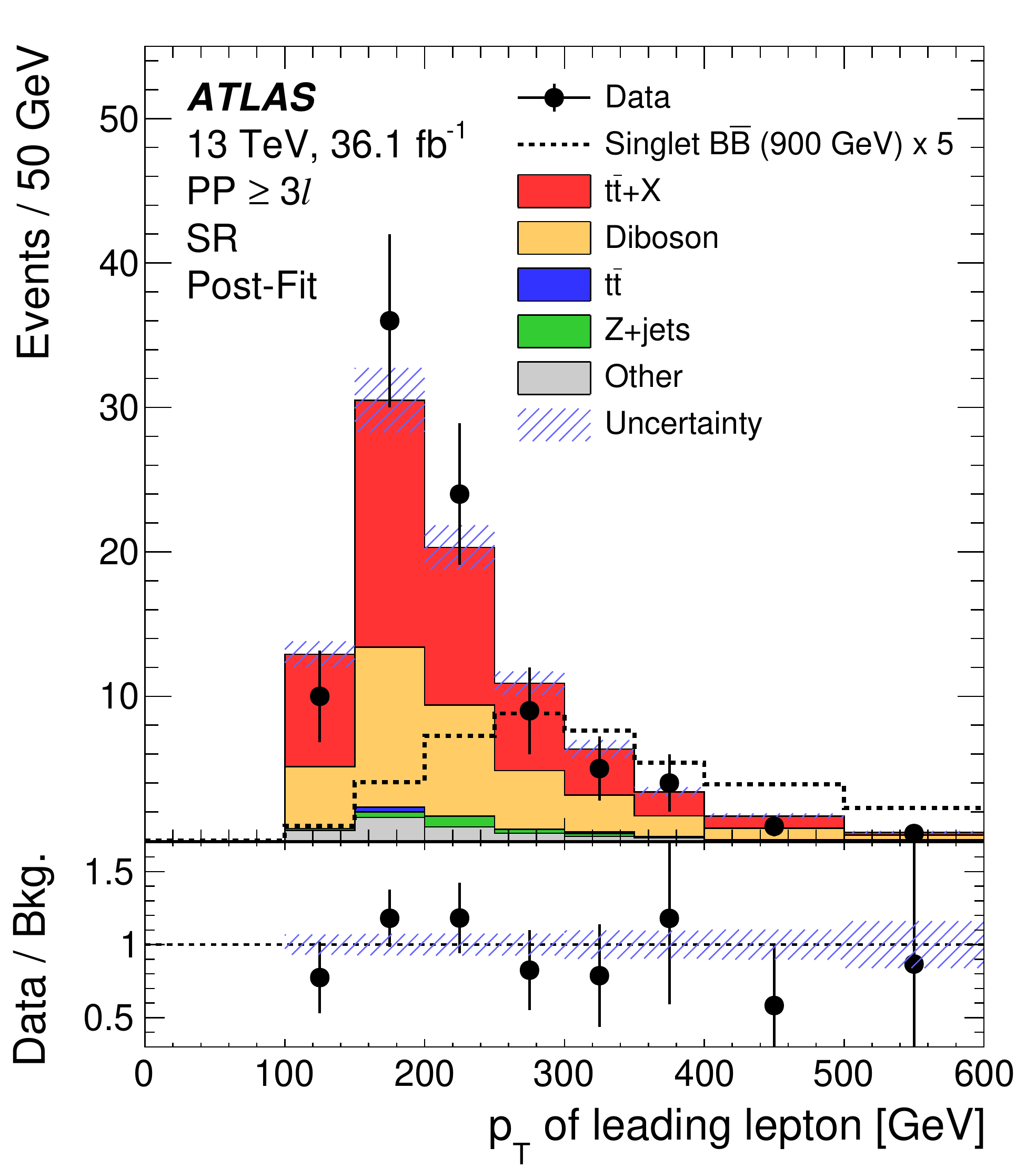}}
\caption{Comparison of the distribution of the transverse momentum of the highest-\pt\ lepton (leading lepton), $\max\pt^\ell$, between data and the background prediction in (a) the diboson control region, (b) the $\ttbar+X$ control region, and (c) the signal region of the pair-production \captripair\ channel. The background prediction is shown \emph{post-fit}, i.e.\ after the fit to the data \htjl\ distributions under the background-only hypothesis. The last bin contains the overflow. An example distribution for a \BBbar\ signal in the singlet model with $\mVLQ = 900~\GeV$ is overlaid. For better visibility, it is multiplied by a factor of five. The data are compatible with the background-only hypothesis.}
\label{fig:trilepair_leppt}
\end{figure}

In \Fig{\ref{fig:trilepair_discr}}, the \htjl\ distribution is shown in the CRs and SR for data and the background prediction after the fit. The VLQ pair-production signal would be expected to result in an excess of data over the background prediction at large values of \htjl, as shown in \Fig{\ref{fig:trilepair_discr_SR}}. The modeling of the main backgrounds was validated by comparing the distributions of kinematic variables and object multiplicities between data and background prediction in the respective CR. As an example, the distribution of the \pt\ of the highest-\pt\ lepton, $\max\pt^\ell$, in the event is shown in \Fig{\ref{fig:trilepair_leppt}} in the two CRs and in the SR. The background prediction is shown after the fit to the \htjl\ distribution. Good agreement between data and the background prediction is observed in both kinematic variables in the CRs, validating the background prediction.

\subsection{Results: \dilsing}
The observed number of events in the SR and the CRs and the expected number of events for the different background contributions are shown in \Tab{\ref{tab:dilepsing_pre}} for the \dilsing\ channel. Also shown is the expected number of events for \mbox{single-$T$-quark} production for $\mVLQ = 900~\GeV$ and $\kappa_T = 0.5$. The signal efficiency in the SR is $0.58\%$ for $T\rightarrow Zt$ decays produced via a $bWT$ coupling, and includes the branching ratios of the VLQ as well as of its decay products, including the decay $Z\rightarrow\ell^+\ell^-$.

\begin{table}[p]
\centering
        \caption{Observed number of events in data and \emph{pre-fit} expected number of signal and background events in the control regions and the signal region for the \dilsing\ channel, i.e.\ before the fit to data. For the signal, the expected number of events for the single-$T$-quark benchmark process with $\mVLQ = 900~\GeV$ and $\kappa_T = 0.5$ is shown. Statistical uncertainties from the limited size of MC samples and systematic uncertainties are added in quadrature. The uncertainty in the ratio of the observed and expected numbers of events contains the systematic uncertainties and the statistical uncertainty of the prediction from Poisson fluctuations.}
        \begin{tabular}{l|R{1cm}@{\hspace{1.5pt}}c@{\hspace{1.5pt}}p{1cm}|R{1cm}@{\hspace{1.5pt}}c@{\hspace{1.5pt}}p{1cm}|R{1cm}@{\hspace{1.5pt}}c@{\hspace{1.5pt}}p{1cm}}
                         \toprule

                                       &\multicolumn{3}{c|}{0-\btagged-jet CR} & \multicolumn{3}{c|}{$\geq 1$-\btagged-jet CR} &         \multicolumn{3}{c}{SR}        \\\midrule
Single-$T$ (900~GeV, $\kappa_T = 0.5$) &     \num{2.6} & $\pm$ & \num{0.4}    &    \num{13.7} & $\pm$ & \num{1.0}    &    \num{27.4} & $\pm$ & \num{3.4}    \\
\midrule
               $Z$+jets                &    \num{2300} & $\pm$ & \num{800}    &     \num{520} & $\pm$ & \num{130}    &     \num{130} & $\pm$ & \num{50}     \\
              $t\bar{t}$               &     \num{0.8} & $\pm$ & \num{0.7}    &     \num{3.4} & $\pm$ & \num{1.7}    &     \num{0.9} & $\pm$ & \num{0.9}    \\
              Single top               &    \num{0.64} & $\pm$ & \num{0.18}   &    \num{1.78} & $\pm$ & \num{0.22}   &     \num{2.5} & $\pm$ & \num{0.4}    \\
             $t\bar{t}+X$              &    \num{1.22} & $\pm$ & \num{0.23}   &     \num{8.5} & $\pm$ & \num{1.2}    &     \num{7.3} & $\pm$ & \num{1.3}    \\
                Diboson                &     \num{100} & $\pm$ & \num{140}    &      \num{30} & $\pm$ & \num{50}     &      \num{9} & $\pm$ & \num{12}      \\
               Triboson                &   \num{0.039} & $\pm$ & \num{0.014}  &   \multicolumn{3}{c|}{$<0.001$}  &   \num{0.005} & $\pm$ & \num{0.013}  \\
\midrule
              Total Bkg.               &    \num{2400} & $\pm$ & \num{800}    &     \num{570} & $\pm$ & \num{120}    &     \num{150} & $\pm$ & \num{50}     \\
\midrule
                 Data                  &   \multicolumn{3}{c|}{\num{2350}}    &    \multicolumn{3}{c|}{\num{495}}    &    \multicolumn{3}{c}{\num{124}}     \\
\midrule
               Data/Bkg.               &         0.96 & $\pm$ & 0.31          &         0.87 & $\pm$ & 0.19          &         0.81 & $\pm$ & 0.27          \\
\bottomrule
        \end{tabular}
            \label{tab:dilepsing_pre}

\vspace{1cm}
        \caption{Observed number of events in data and \emph{post-fit} expected number of background events in the control regions and the signal region for the \dilsing\ channel, i.e.\ after the fit to the data $m_{Zt}$ distributions under the background-only hypothesis. The uncertainty in the expected number of events is the full uncertainty from the fit, from which the uncertainty in the ratio of the observed and expected numbers of events is calculated.}
        \begin{tabular}{l|R{1cm}@{\hspace{1.5pt}}c@{\hspace{1.5pt}}p{1cm}|R{1cm}@{\hspace{1.5pt}}c@{\hspace{1.5pt}}p{1cm}|R{1cm}@{\hspace{1.5pt}}c@{\hspace{1.5pt}}p{1cm}}
                         \toprule

              &\multicolumn{3}{c|}{0-\btagged-jet CR} & \multicolumn{3}{c|}{$\geq 1$-\btagged-jet CR} &         \multicolumn{3}{c}{SR}        \\\midrule
   $Z$+jets   &    \num{2300} & $\pm$ & \num{100}    &     \num{480} & $\pm$ & \num{40}     &     \num{113} & $\pm$ & \num{13}     \\
  $t\bar{t}$  &     \num{0.8} & $\pm$ & \num{0.7}    &     \num{3.8} & $\pm$ & \num{1.6}    &     \num{1.0} & $\pm$ & \num{0.9}    \\
  Single top  &    \num{0.63} & $\pm$ & \num{0.18}   &    \num{1.77} & $\pm$ & \num{0.22}   &    \num{2.33} & $\pm$ & \num{0.28}   \\
 $t\bar{t}+X$ &    \num{1.27} & $\pm$ & \num{0.23}   &     \num{8.3} & $\pm$ & \num{1.1}    &     \num{6.8} & $\pm$ & \num{1.0}    \\
   Diboson    &      \num{40} & $\pm$ & \num{100}     &      \num{12} & $\pm$ & \num{34}     &       \num{4} & $\pm$ & \num{8}      \\
   Triboson   &   \num{0.038} & $\pm$ & \num{0.014}  &   \multicolumn{3}{c|}{$<0.001$}  &   \num{0.005} & $\pm$ & \num{0.013}  \\
\midrule
  Total Bkg.  &    \num{2400} & $\pm$ & \num{100}    &     \num{509} & $\pm$ & \num{34}     &     \num{127} & $\pm$ & \num{15}     \\
\midrule
     Data     &   \multicolumn{3}{c|}{\num{2350}}    &    \multicolumn{3}{c|}{\num{495}}    &    \multicolumn{3}{c}{\num{124}}     \\
\midrule
  Data/Bkg.   &         1.00 & $\pm$ & 0.04          &         0.97 & $\pm$ & 0.06          &         0.98 & $\pm$ & 0.11          \\
\bottomrule
        \end{tabular}
            \label{tab:dilepsing_post}

\end{table}

A fit of the background prediction to the $m_{Zt}$ distributions in data was performed and the post-fit yields are shown in \Tab{\ref{tab:dilepsing_post}}. The uncertainty in the background prediction was significantly reduced in all regions compared to the pre-fit value (\Tab{\ref{tab:dilepsing_pre}}). The overall $Z$+jets normalization was adjusted by factors of \normfzjetsnotnobcrtwols\ and \normfzjetsnotbcrtwols\ in the 0-\btagged-jet CR and the $\geq 1$-\btagged-jet CR, respectively. The ratios of the post-fit and pre-fit background yields are consistent with unity in all regions. The ratio for $Z$+jets production in the SR is \normfzjetssrtwols, which is consistent with unity within $2\sigma$ at most.

\begin{figure}[p]
\centering
\subfloat[]{\includegraphics[width=.49\textwidth]{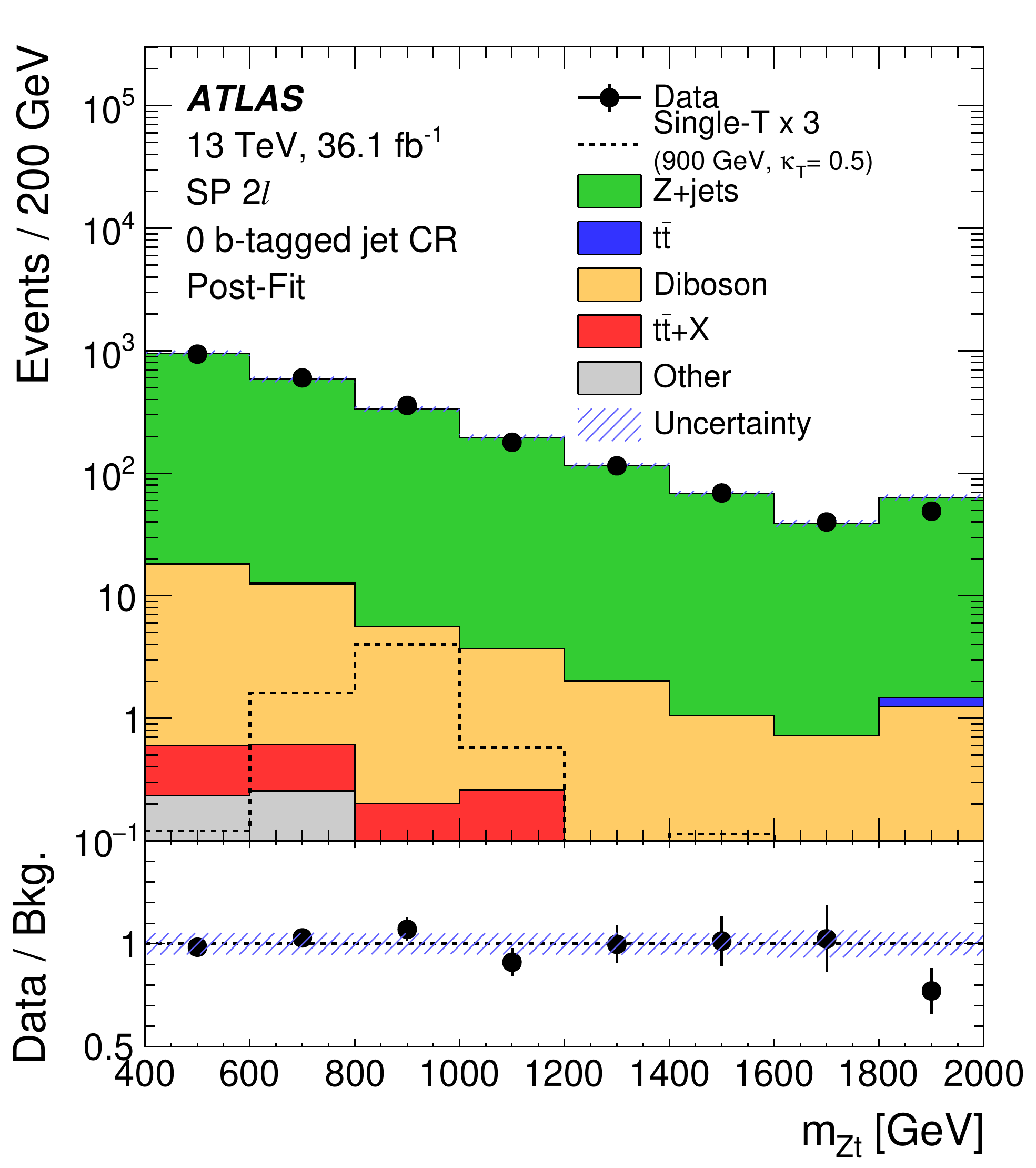}}
\subfloat[]{\includegraphics[width=.49\textwidth]{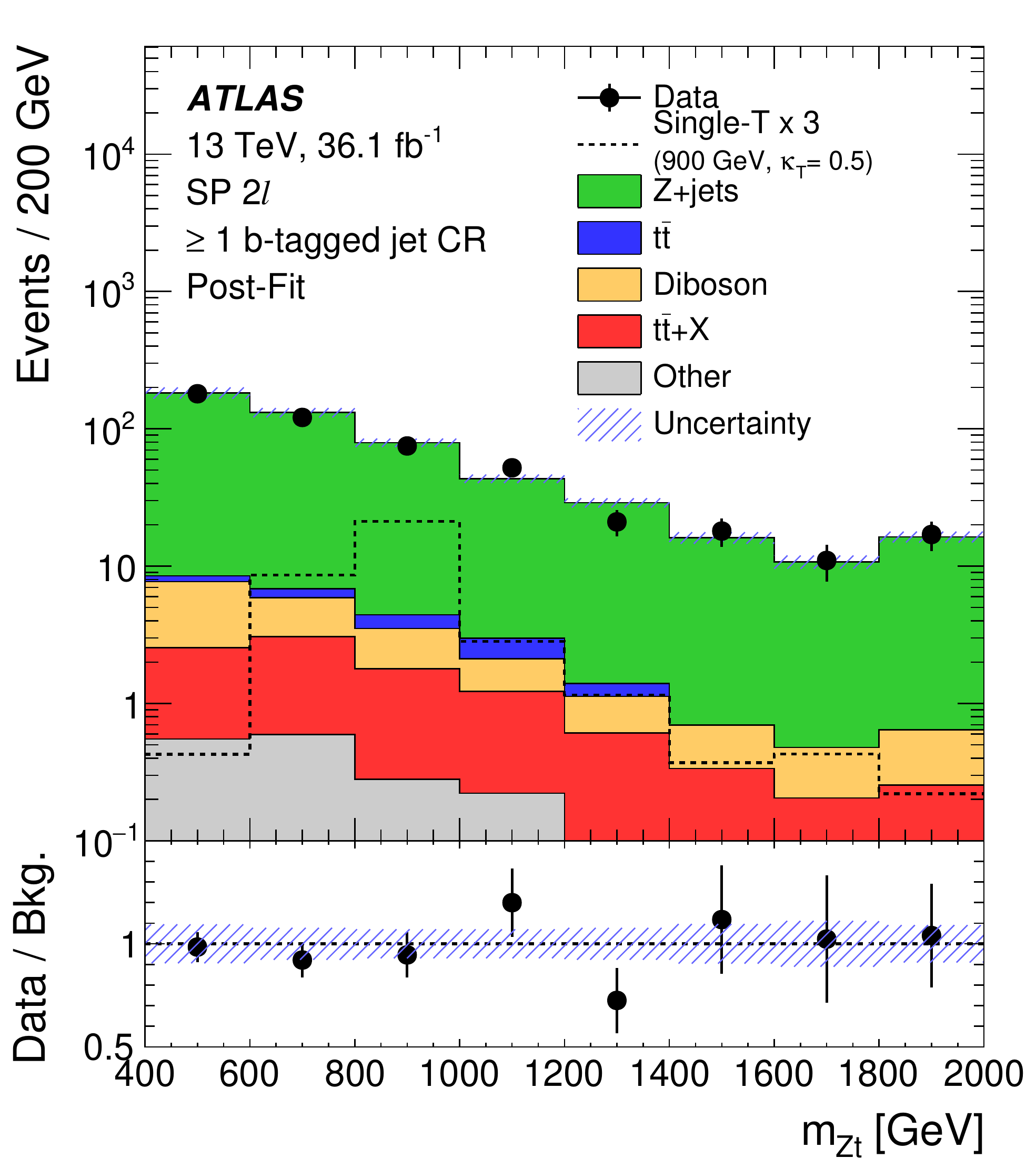}}\\
\subfloat[]{\includegraphics[width=.49\textwidth]{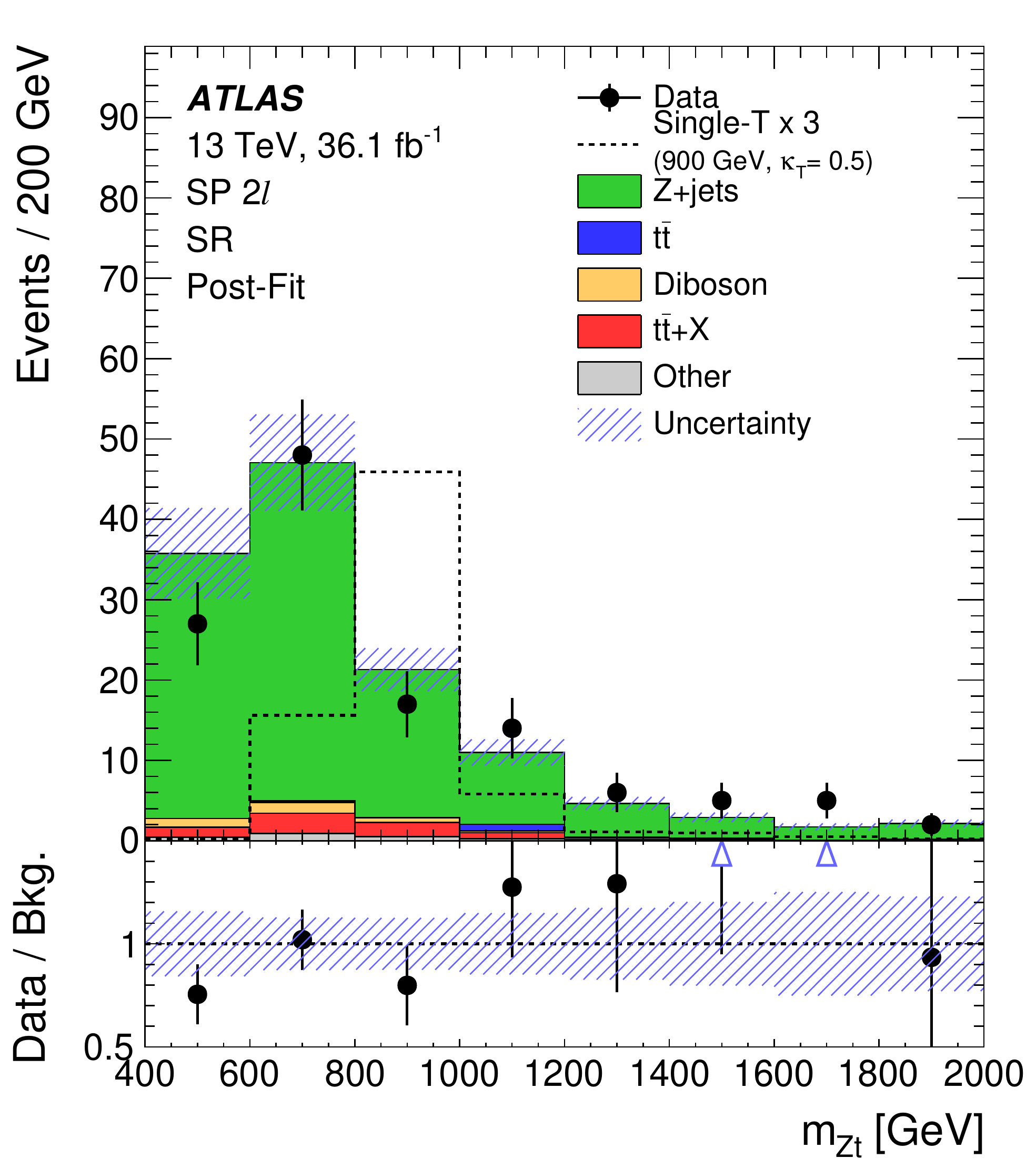}\label{fig:dilepsing_discr_SR}}
\caption{Comparison of the distribution of the invariant mass of the $Z$ boson candidate and the highest-\pt\ top-tagged \ljet, $m_{Zt}$, between data and the background prediction in (a) the 0-\btagged-jet control region, (b) the $\geq 1$-\btagged-jet control region, and (c) the signal region of the single-production \capdilsing\ channel. The background prediction is shown \emph{post-fit}, i.e.\ after the fit to the data $m_{Zt}$ distributions under the background-only hypothesis. The last bin contains the overflow. An upward pointing triangle in the ratio plot indicates that the value of the ratio is beyond scale. An example distribution for a single-$T$-quark signal with $\mVLQ = 900~\GeV$ and $\kappa_T = 0.5$ is overlaid. For better visibility, it is multiplied by a factor of three. The data are compatible with the background-only hypothesis.}
\label{fig:dilepsing_discr}
\end{figure}

\begin{figure}[p]
\centering
\subfloat[]{\includegraphics[width=.49\textwidth]{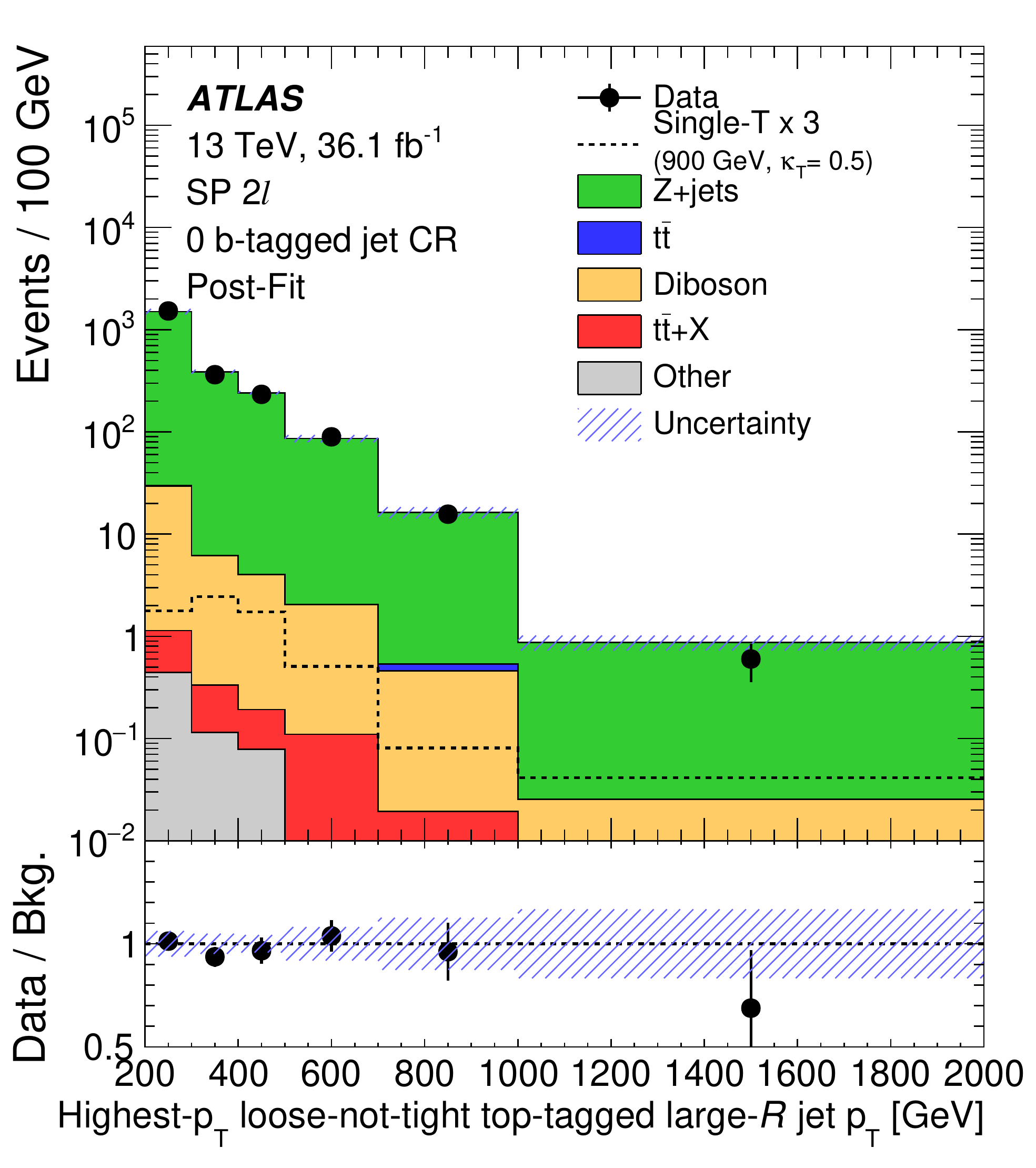}}
\subfloat[]{\includegraphics[width=.49\textwidth]{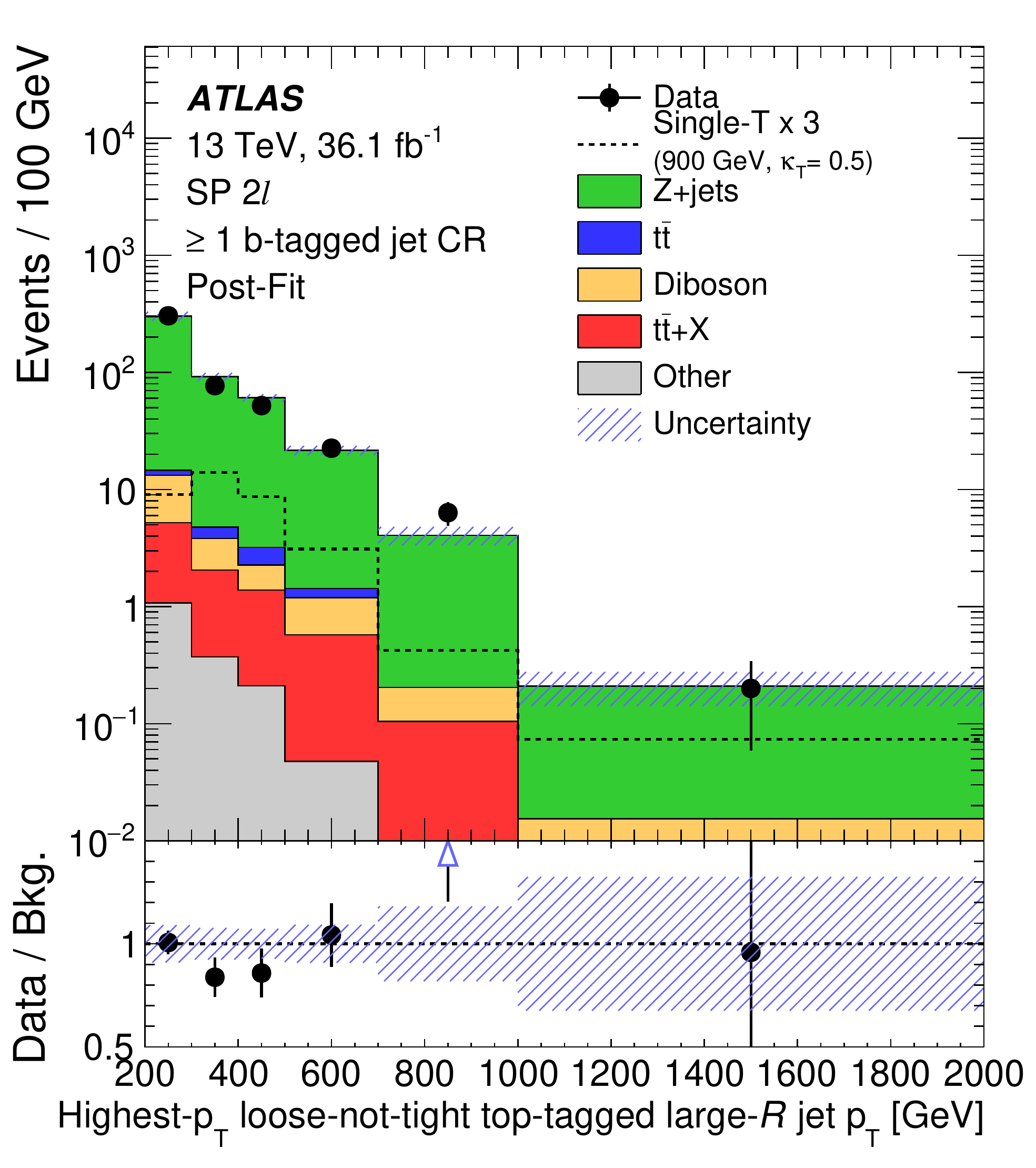}}\\
\subfloat[]{\includegraphics[width=.49\textwidth]{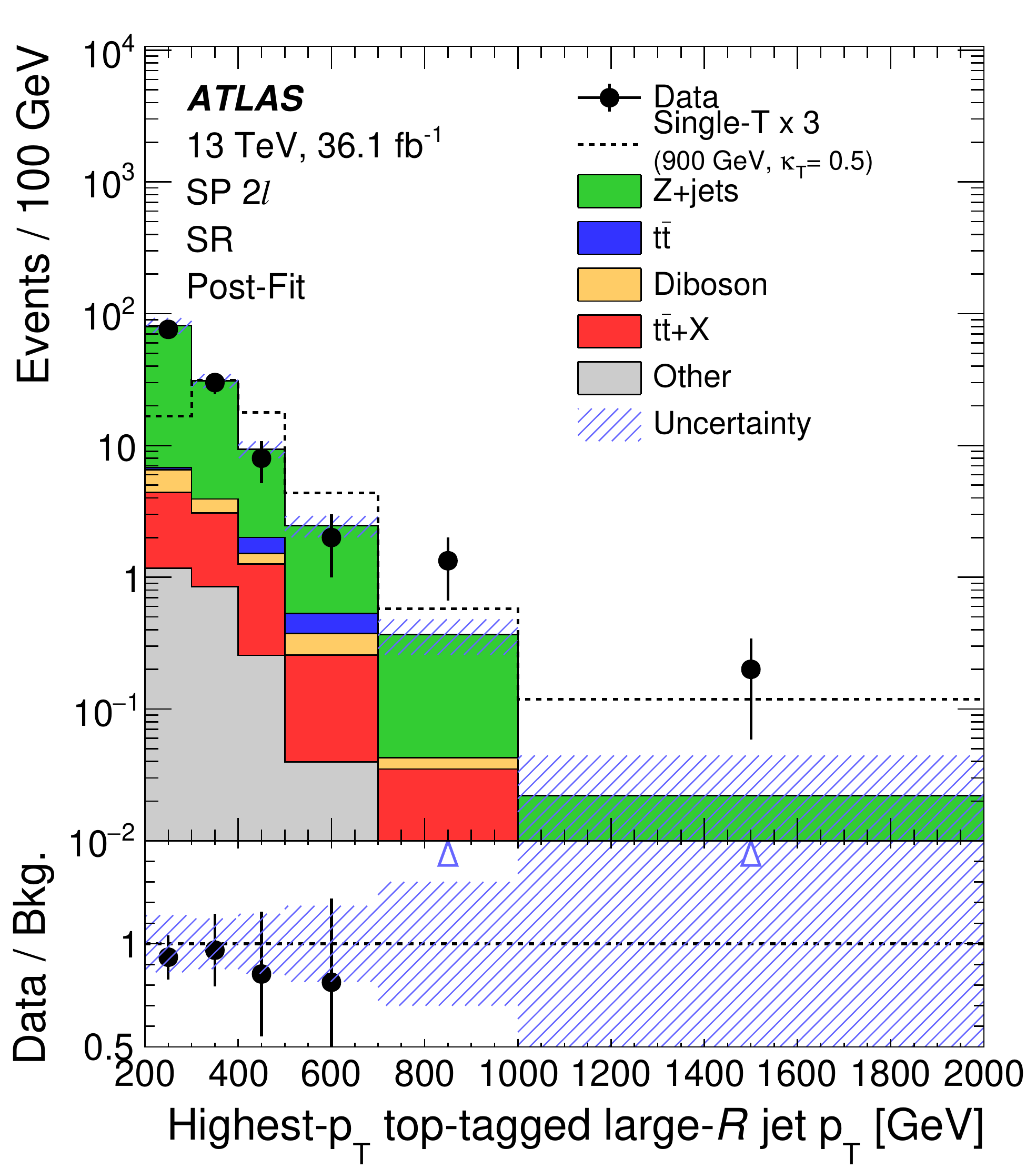}}
\caption{Comparison of the distribution of the transverse momentum of the highest-\pt\ top-tagged \ljet\ between data and the background prediction in (a) the 0-\btagged-jet control region, (b) the $\geq 1$-\btagged-jet control region, and (c) the signal region of the single-production \capdilsing\ channel. The background prediction is shown \emph{post-fit}, i.e.\ after the fit to the data $m_{Zt}$ distributions under the background-only hypothesis. The last bin contains the overflow. An upward pointing triangle in the ratio plot indicates that the value of the ratio is beyond scale. An example distribution for a single-$T$-quark signal with $\mVLQ = 900~\GeV$ and $\kappa_T = 0.5$ is overlaid. For better visibility, it is multiplied by a factor of three. The data are compatible with the background-only hypothesis.}
\label{fig:dilepsing_ljetpt}
\end{figure}

In \Fig{\ref{fig:dilepsing_discr}}, the $m_{Zt}$ distribution is shown in the CRs and SR for data and the background prediction after the fit. The VLQ single-production signal would be expected to result in an excess of data over the background prediction in the $m_{Zt}$ distribution, as shown in \Fig{\ref{fig:dilepsing_discr_SR}}. The modeling of the main background was validated by comparing the distributions of kinematic variables and object multiplicities between data and background prediction in the two CRs. As an example, the \pt\ distribution of the highest-\pt\ top-tagged \ljet\ in the event is shown in \Fig{\ref{fig:dilepsing_ljetpt}} in the two CRs and the SR. The background prediction is shown after the fit to the $m_{Zt}$ distribution. Contributions from VLQ single production would be expected at high values of the \ljet\ \pt. Good agreement between data and the background prediction is observed in both kinematic variables in the CRs, validating the background prediction.

\subsection{Results: \trising}
The observed number of events in the SR and the CRs and the expected number of events for the different background contributions are shown in \Tab{\ref{tab:trilepsing_pre}} for the \trising\ channel. Also shown is the expected number of events for single-$T$-quark production for $\mVLQ = 900~\GeV$ and $\kappa_T = 0.5$. Due to a low number of MC events for the single-$T$-quark signal in the \trising\ channel, in this channel the signal efficiency was interpolated as a function of $\mVLQ$ with a third-order polynomial describing the efficiencies estimated from MC simulations within the uncertainties. The resulting signal efficiency in the SR is $0.16\%$ for $T\rightarrow Zt$ decays produced via a $bWT$ coupling, and includes the branching ratios of the VLQ as well as of its decay products, including the decay $Z\rightarrow\ell^+\ell^-$.

\begin{table}[p]
\centering
        \caption{Observed number of events in data and \emph{pre-fit} expected number of signal and background events in the control regions and the signal region for the \trising\ channel, i.e.\ before the fit to data. For the signal, the expected number of events for the single-$T$-quark benchmark process with $\mVLQ = 900~\GeV$ and $\kappa_T = 0.5$ is shown. Statistical uncertainties from the limited size of MC samples and systematic uncertainties are added in quadrature. The uncertainty in the ratio of the observed and expected numbers of events contains the systematic uncertainties and the statistical uncertainty of the prediction from Poisson fluctuations.}
        \begin{tabular}{l|R{1cm}@{\hspace{1.5pt}}c@{\hspace{1.5pt}}p{1cm}|R{1cm}@{\hspace{1.5pt}}c@{\hspace{1.5pt}}p{1cm}|R{1cm}@{\hspace{1.5pt}}c@{\hspace{1.5pt}}p{1cm}}
                         \toprule

                         &\multicolumn{3}{c|}{Diboson CR} &  \multicolumn{3}{c|}{$t\bar{t}+X$ CR} &         \multicolumn{3}{c}{SR}        \\\midrule
Single-$T$ (900~GeV, $\kappa_T = 0.5$)&   \num{3.3} & $\pm$ & \num{0.5}   &   \num{1.78} & $\pm$ & \num{0.22}  &     \num{7.9} & $\pm$ & \num{0.6}    \\
\midrule
        $Z$+jets         &    \num{52} & $\pm$ & \num{29}    &      \num{9} & $\pm$ & \num{6}     &    \num{0.16} & $\pm$ & \num{0.10}   \\
       $t\bar{t}$        &   \num{7.1} & $\pm$ & \num{1.6}   &   \num{12.0} & $\pm$ & \num{2.7}   &   \multicolumn{3}{c}{$<0.001$}  \\
       Single top        &   \num{6.9} & $\pm$ & \num{0.9}   &   \num{18.9} & $\pm$ & \num{0.9}   &    \num{0.64} & $\pm$ & \num{0.11}   \\
      $t\bar{t}+X$       &     \num{22} & $\pm$ & \num{4}    &     \num{98} & $\pm$ & \num{15}    &     \num{5.6} & $\pm$ & \num{0.9}    \\
         Diboson         &   \num{1120} & $\pm$ & \num{260}  &    \num{110} & $\pm$ & \num{50}    &     \num{3.1} & $\pm$ & \num{1.4}    \\
        Triboson         &   \num{5.9} & $\pm$ & \num{0.4}   &   \num{0.46} & $\pm$ & \num{0.06}  &   \num{0.026} & $\pm$ & \num{0.007}  \\
\midrule
       Total Bkg.        &   \num{1210} & $\pm$ & \num{260}  &    \num{250} & $\pm$ & \num{50}    &     \num{9.5} & $\pm$ & \num{2.0}    \\
\midrule
          Data           &  \multicolumn{3}{c|}{\num{1145}}  &   \multicolumn{3}{c|}{\num{279}}   &     \multicolumn{3}{c}{\num{14}}     \\
\midrule
        Data/Bkg.        &        0.94 & $\pm$ & 0.20        &        1.13 & $\pm$ & 0.24         &          1.5 & $\pm$ & 0.6           \\
\bottomrule
        \end{tabular}
            \label{tab:trilepsing_pre}

\vspace{1cm}
        \caption{Observed number of events in data and \emph{post-fit} expected number of background events in the control regions and the signal region for the \trising\ channel, i.e.\ after the fit to the data \htjl\ distributions under the background-only hypothesis. The uncertainty in the expected number of events is the full uncertainty from the fit, from which the uncertainty in the ratio of the observed and expected numbers of events is calculated.}
        \begin{tabular}{l|R{1cm}@{\hspace{1.5pt}}c@{\hspace{1.5pt}}p{1cm}|R{1cm}@{\hspace{1.5pt}}c@{\hspace{1.5pt}}p{1cm}|R{1cm}@{\hspace{1.5pt}}c@{\hspace{1.5pt}}p{1cm}}
                         \toprule

                &\multicolumn{3}{c|}{Diboson CR} &  \multicolumn{3}{c|}{$t\bar{t}+X$ CR} &         \multicolumn{3}{c}{SR}        \\\midrule
    $Z$+jets    &    \num{55} & $\pm$ & \num{27}    &     \num{11} & $\pm$ & \num{6}     &    \num{0.17} & $\pm$ & \num{0.11}   \\
   $t\bar{t}$   &   \num{7.3} & $\pm$ & \num{3.4}   &     \num{15} & $\pm$ & \num{6}     &   \multicolumn{3}{c}{$<0.001$}  \\
   Single top   &   \num{7.0} & $\pm$ & \num{3.3}   &     \num{20} & $\pm$ & \num{10}    &    \num{0.68} & $\pm$ & \num{0.34}   \\
  $t\bar{t}+X$  &     \num{22} & $\pm$ & \num{4}    &    \num{110} & $\pm$ & \num{14}    &     \num{6.2} & $\pm$ & \num{0.8}    \\
    Diboson     &   \num{1060} & $\pm$ & \num{50}   &    \num{116} & $\pm$ & \num{25}    &     \num{3.2} & $\pm$ & \num{0.7}    \\
    Triboson    &   \num{6.0} & $\pm$ & \num{2.5}   &   \num{0.50} & $\pm$ & \num{0.17}  &   \num{0.031} & $\pm$ & \num{0.014}  \\
\midrule
   Total Bkg.   &   \num{1160} & $\pm$ & \num{40}   &    \num{280} & $\pm$ & \num{20}    &    \num{10.2} & $\pm$ & \num{1.1}    \\
\midrule
      Data      &  \multicolumn{3}{c|}{\num{1145}}  &   \multicolumn{3}{c|}{\num{279}}   &     \multicolumn{3}{c}{\num{14}}     \\
\midrule
   Data/Bkg.    &        0.99 & $\pm$ & 0.04        &        1.01 & $\pm$ & 0.07         &         1.37 & $\pm$ & 0.14          \\
\bottomrule
        \end{tabular}
            \label{tab:trilepsing_post}

\end{table}

A fit of the background prediction to the \htjl\ distributions in data was performed. The post-fit yields are shown in \Tab{\ref{tab:trilepsing_post}}. The uncertainty in the background prediction was significantly reduced in all regions compared to the pre-fit value (\Tab{\ref{tab:trilepsing_pre}}). The overall diboson ($\ttbar+X$) normalization was adjusted by a factor of \normfdibosonvvcrtrils\ (\normfttvttvcrtrils) in the diboson ($\ttbar+X$) CR. The ratios of the post-fit and pre-fit background yields are consistent with unity in all regions.

\begin{figure}[p]
\centering
\subfloat[]{\includegraphics[width=.49\textwidth]{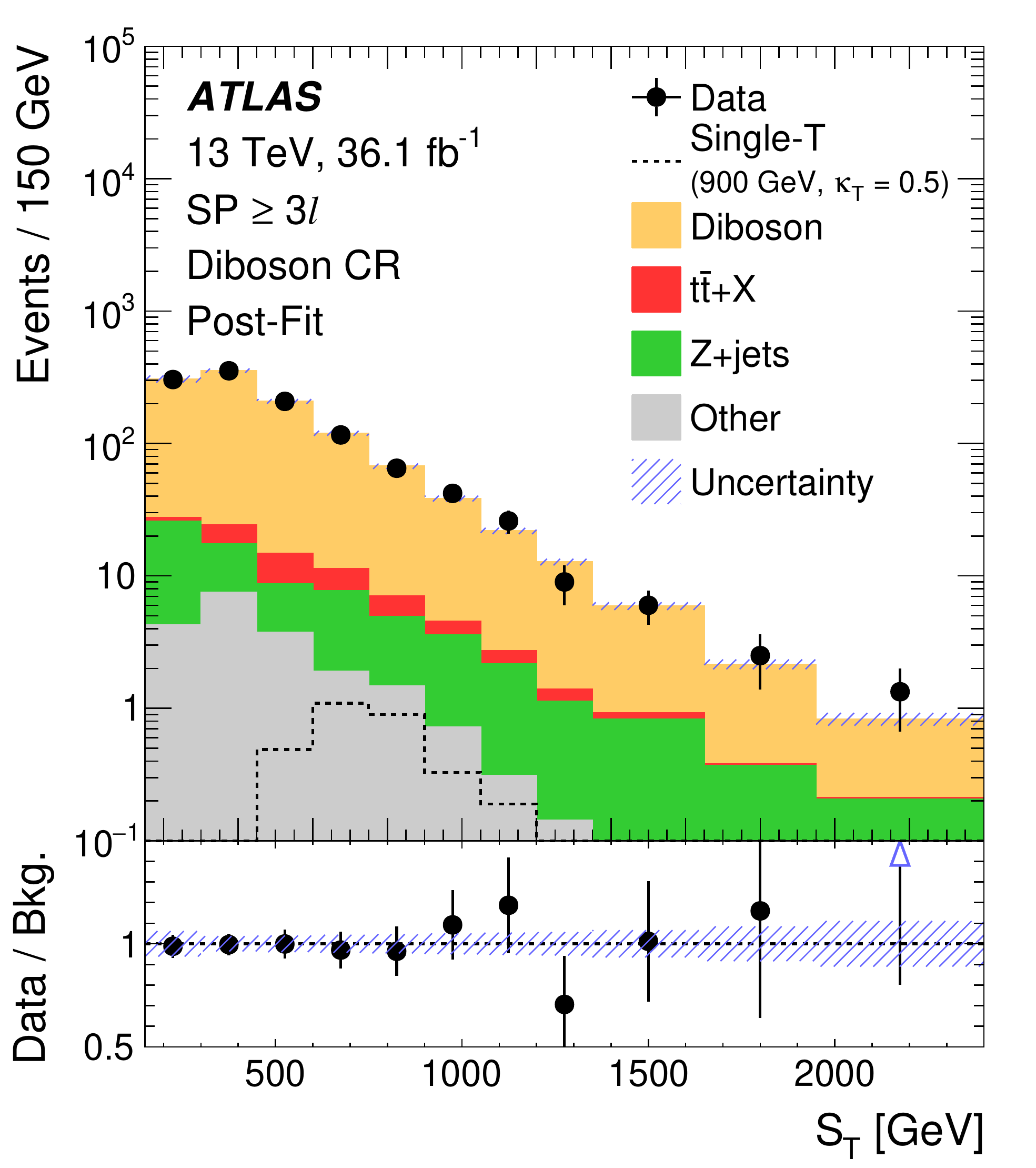}}
\subfloat[]{\includegraphics[width=.49\textwidth]{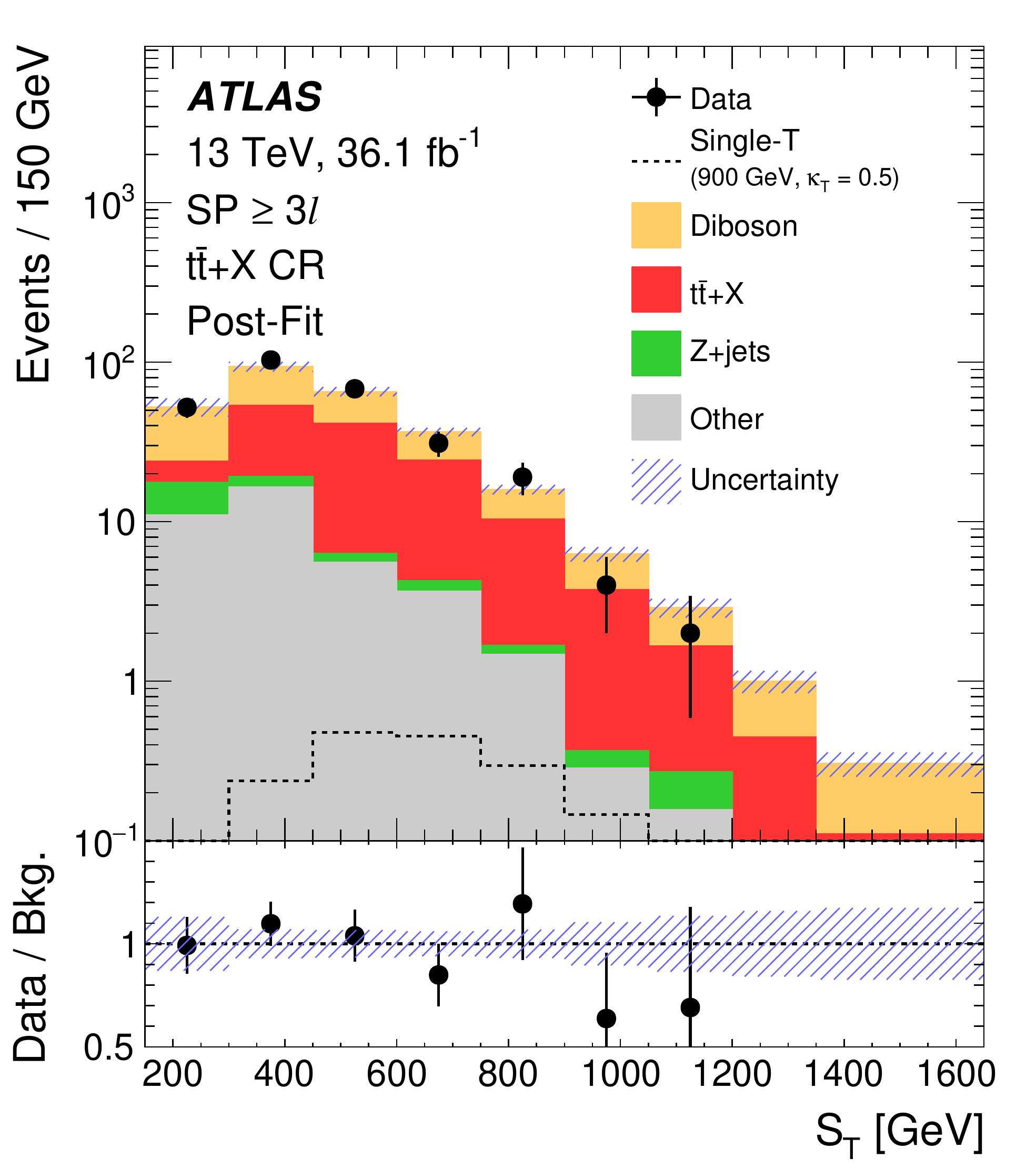}}\\
\subfloat[]{\includegraphics[width=.49\textwidth]{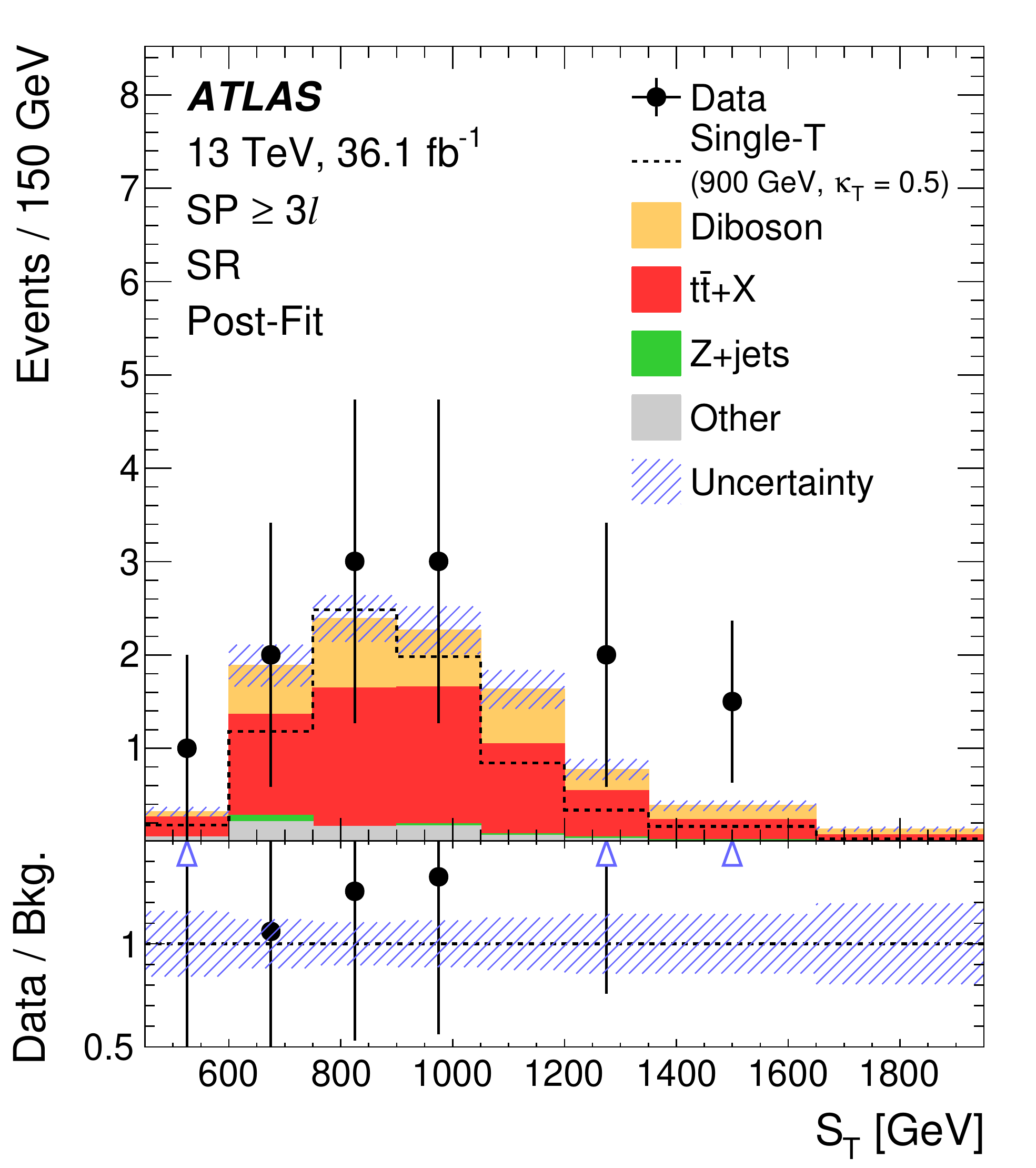}\label{fig:trilepsing_discr_SR}}
\caption{Comparison of the distribution of the scalar sum of \sjet\ and lepton transverse momenta, \htjl, between data and the background prediction in (a) the diboson control region, (b) the $\ttbar+X$ control region, and (c) the signal region of the single-production \captrising\ channel. The background prediction is shown \emph{post-fit}, i.e.\ after the fit to the data \htjl\ distributions under the background-only hypothesis. The last bin contains the overflow. An upward pointing triangle in the ratio plot indicates that the value of the ratio is beyond scale.  An example distribution for a single-$T$-quark signal with $\mVLQ = 900~\GeV$ and $\kappa_T = 0.5$ is overlaid. The data are compatible with the background-only hypothesis.}
\label{fig:trilepsing_discr}
\end{figure}

\begin{figure}[p]
\centering
\subfloat[]{\includegraphics[width=.49\textwidth]{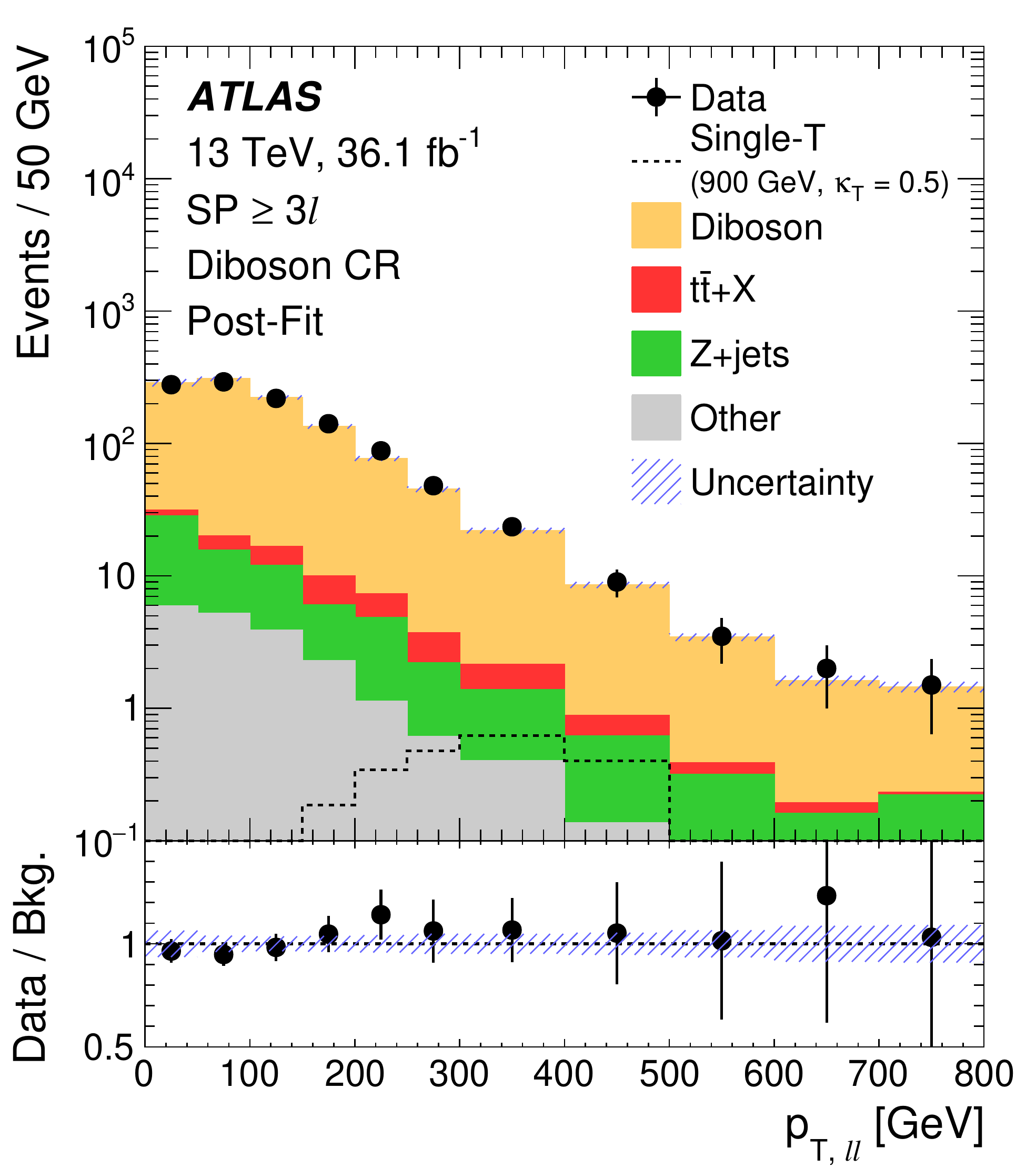}}
\subfloat[]{\includegraphics[width=.49\textwidth]{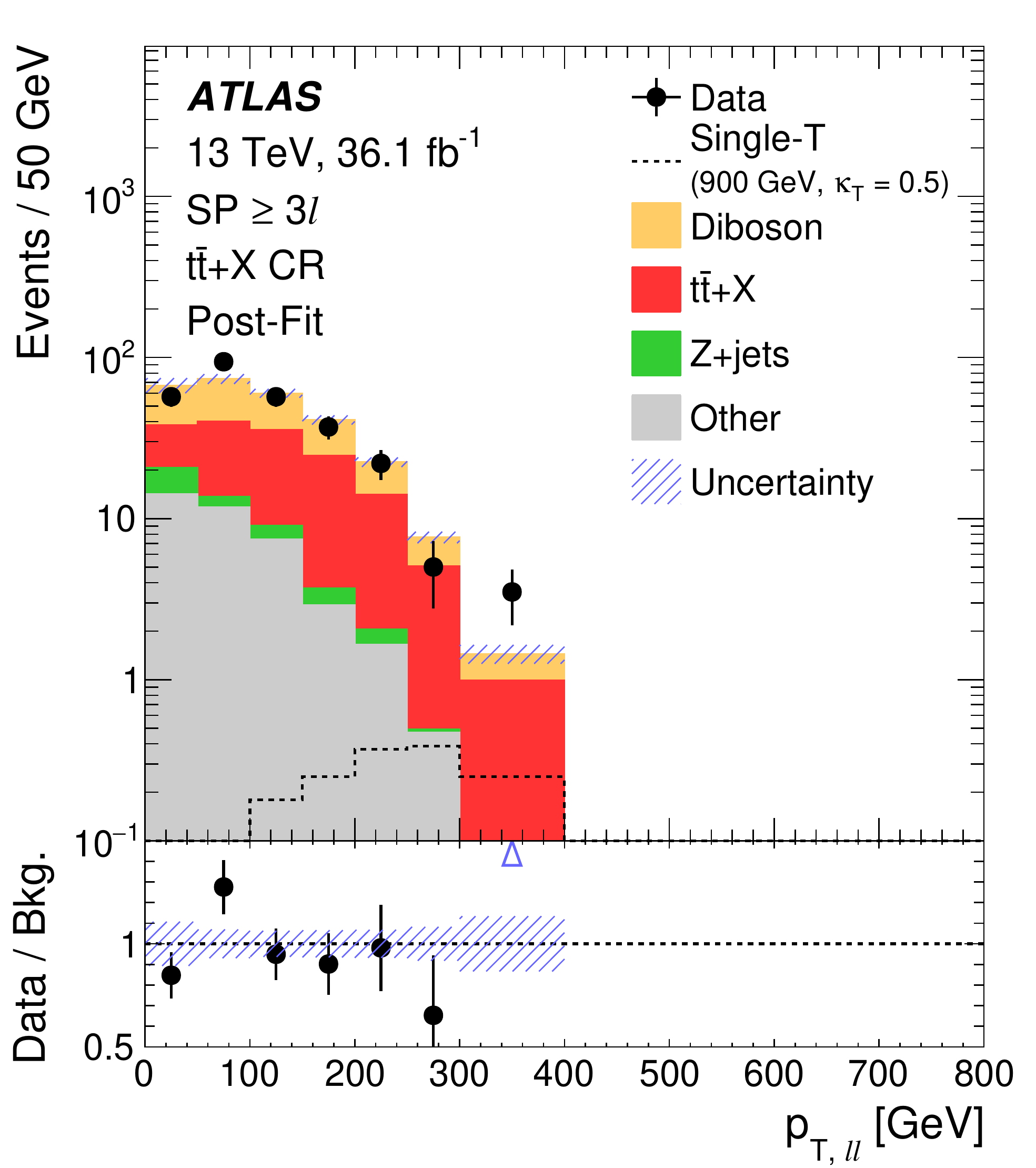}}\\
\subfloat[]{\includegraphics[width=.49\textwidth]{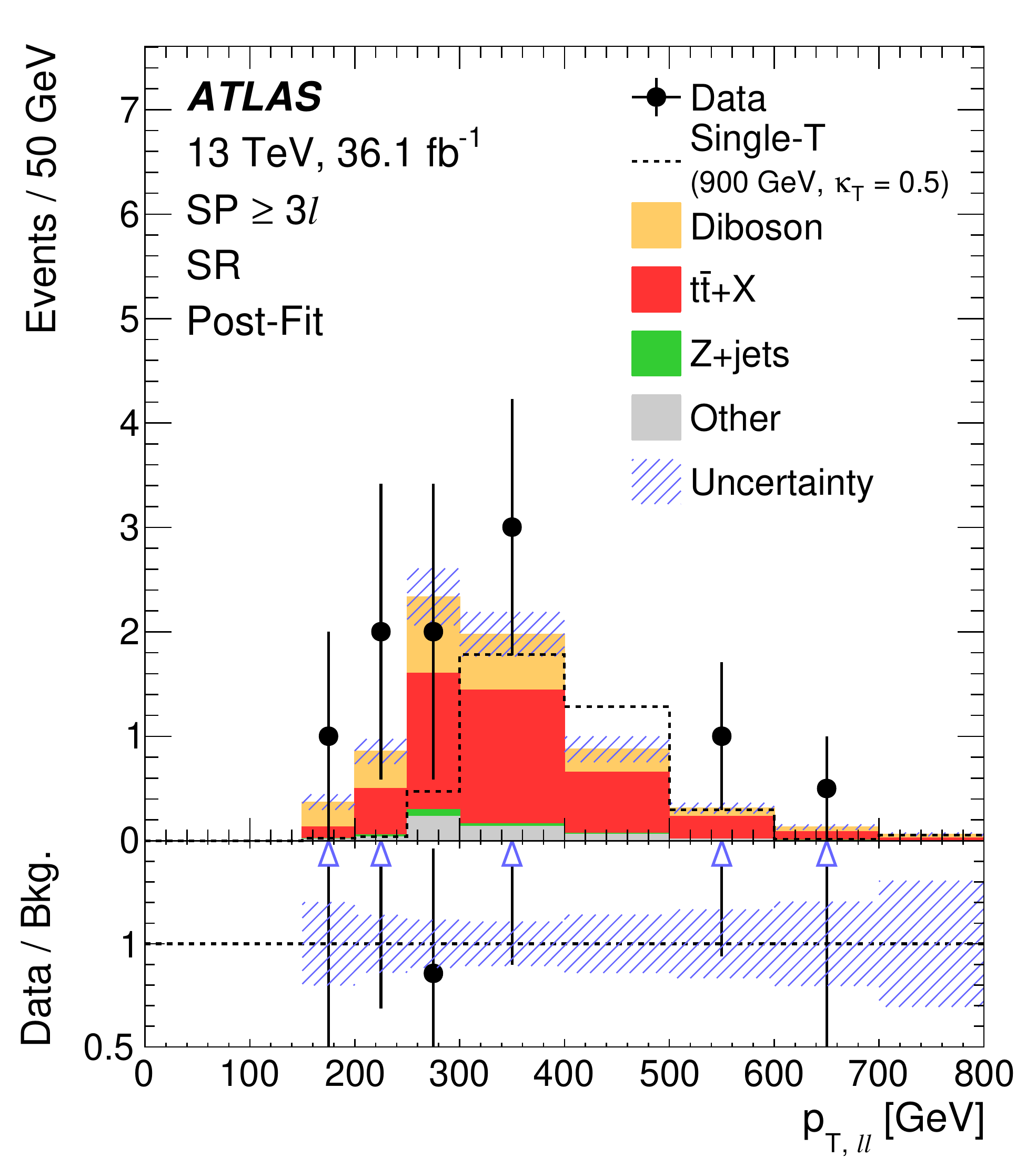}}
\caption{Comparison of the transverse momentum of the $Z$ boson candidate, \ptll, between data and the background prediction in (a) the diboson control region, (b) the $\ttbar+X$ control region, and (c) the signal region of the single-production \captrising\ channel. The background prediction is shown \emph{post-fit}, i.e.\ after the fit to the data \htjl\ distributions under the background-only hypothesis. The last bin contains the overflow. An upward pointing triangle in the ratio plot indicates that the value of the ratio is beyond scale. An example distribution for a single-$T$-quark signal with $\mVLQ = 900~\GeV$ and $\kappa_T = 0.5$ is overlaid. The data are compatible with the background-only hypothesis.}
\label{fig:trilepsing_Zpt}
\end{figure}

In \Fig{\ref{fig:trilepsing_discr}}, the \htjl\ distribution is shown in the CRs and SR for data and the background prediction after the fit. For large values of $\mVLQ$, VLQ single-production signal would be expected to result in an excess of data over the background prediction at large values of \htjl. An example for a lower value of $\mVLQ$ is shown in \Fig{\ref{fig:trilepsing_discr_SR}}. The modeling of the main backgrounds was validated by comparing the distributions of kinematic variables and object multiplicities between data and background prediction in the respective CR. As an example, the \ptll\ distribution is shown in \Fig{\ref{fig:trilepsing_Zpt}} in the two CRs and in the SR. The background prediction is shown after the fit to the \htjl\ distribution. Contributions from VLQ single production would be expected at high values of \ptll. Good agreement between data and the background prediction is observed in both kinematic variables in the CRs, validating the background prediction.

\FloatBarrier

\section{Interpretation}
No excess of data over the background-only hypothesis is observed and 95\% CL limits are set on the cross section ($\sigma$) as a function of the model parameters in the case of the pair-production channels, and on the cross section times branching ratio to $Zt$ [$\sigma \times \mathrm{BR}(T\rightarrow Zt)$] in the single-production channels. For this purpose, the profile-likelihood ratio $q_\mu = -2 \ln\left( L(\mu,\hat{\hat{\theta}})/L(\hat{\mu},\hat{\theta})\right)$ was used as the test statistic, with $\hat{\mu}$ and $\hat{\theta}$ the values of the signal strength and the set of NPs that maximize the likelihood under the constraint $0<\hat{\mu}<\mu$, and $\hat{\hat{\theta}}$ the set of NPs that maximizes the likelihood for a given value of $\mu$. The test statistic $q_\mu$ was evaluated with \textsc{RooFit}~\cite{2003physics...6116V,Roofitweb} and upper limits were derived from the probability distribution of $q_\mu$, evaluated with the asymptotic approximation~\cite{Cowan:2010js}. The limits were calculated with the CL$_\mathrm{s}$ method~\cite{Junk:1999kv,0954-3899-28-10-313}, excluding values of $\sigma$ at 95\% CL for the pair-production channels and $\sigma \times \mathrm{BR}(T\rightarrow Zt)$ for the single-production channels that resulted in a CL$_\mathrm{s}$ value $<0.05$.

The three pair-production channels were combined in order to enhance the sensitivity to \TTbar\ and \BBbar\ production. A combined binned likelihood fit was performed including all CRs and SRs of the three channels. The systematic uncertainties related to the luminosity, the leptons, \sjets, $b$-tagging, \ljets, \met, and the average number of interactions per bunch crossing are correlated among all channels, as are the uncertainties in the cross sections for the background processes. Residual systematic uncertainties related to the MC modeling of the different background processes and misidentified leptons, were treated separately for each channel. This was done in order to avoid NPs constrained in the CRs of one channel inadvertently constraining  kinematics in very different regions of other channels. It was verified that correlating these NPs in the combination of the channels has little impact on its sensitivity.

\begin{figure}[p]
\centering
\subfloat[]{\includegraphics[width=.49\textwidth]{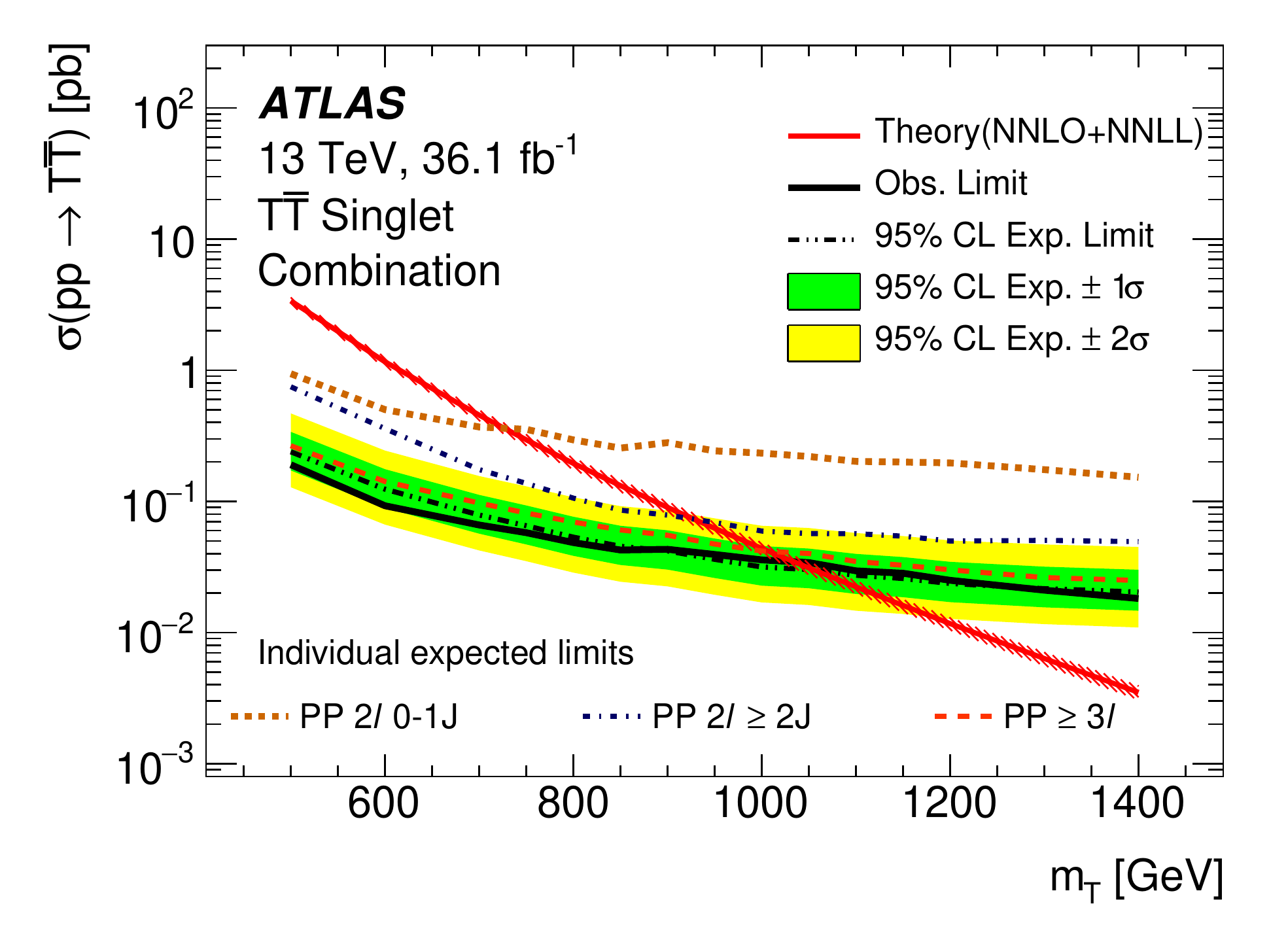}}
\subfloat[]{\includegraphics[width=.49\textwidth]{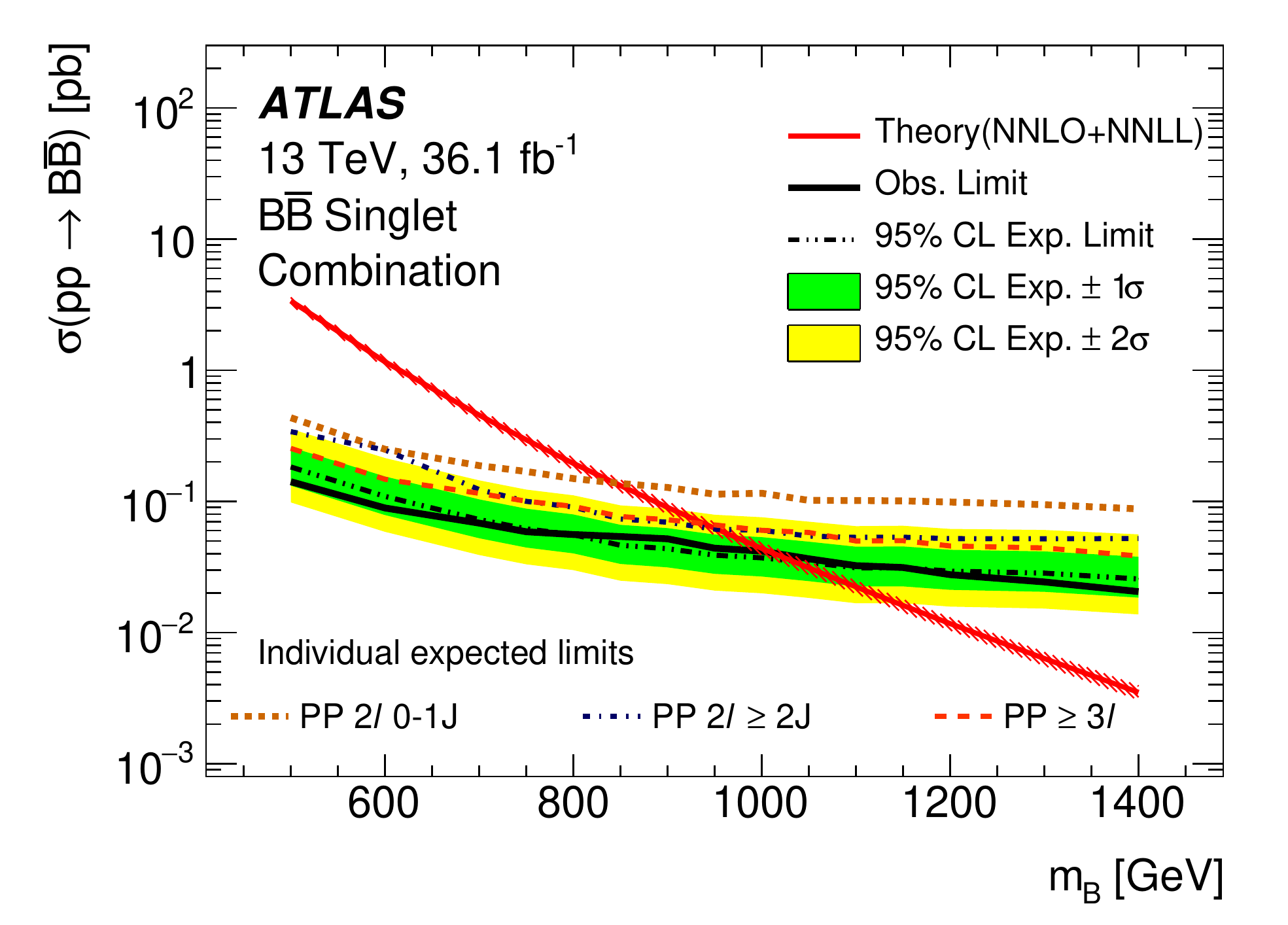}}\\
\subfloat[]{\includegraphics[width=.49\textwidth]{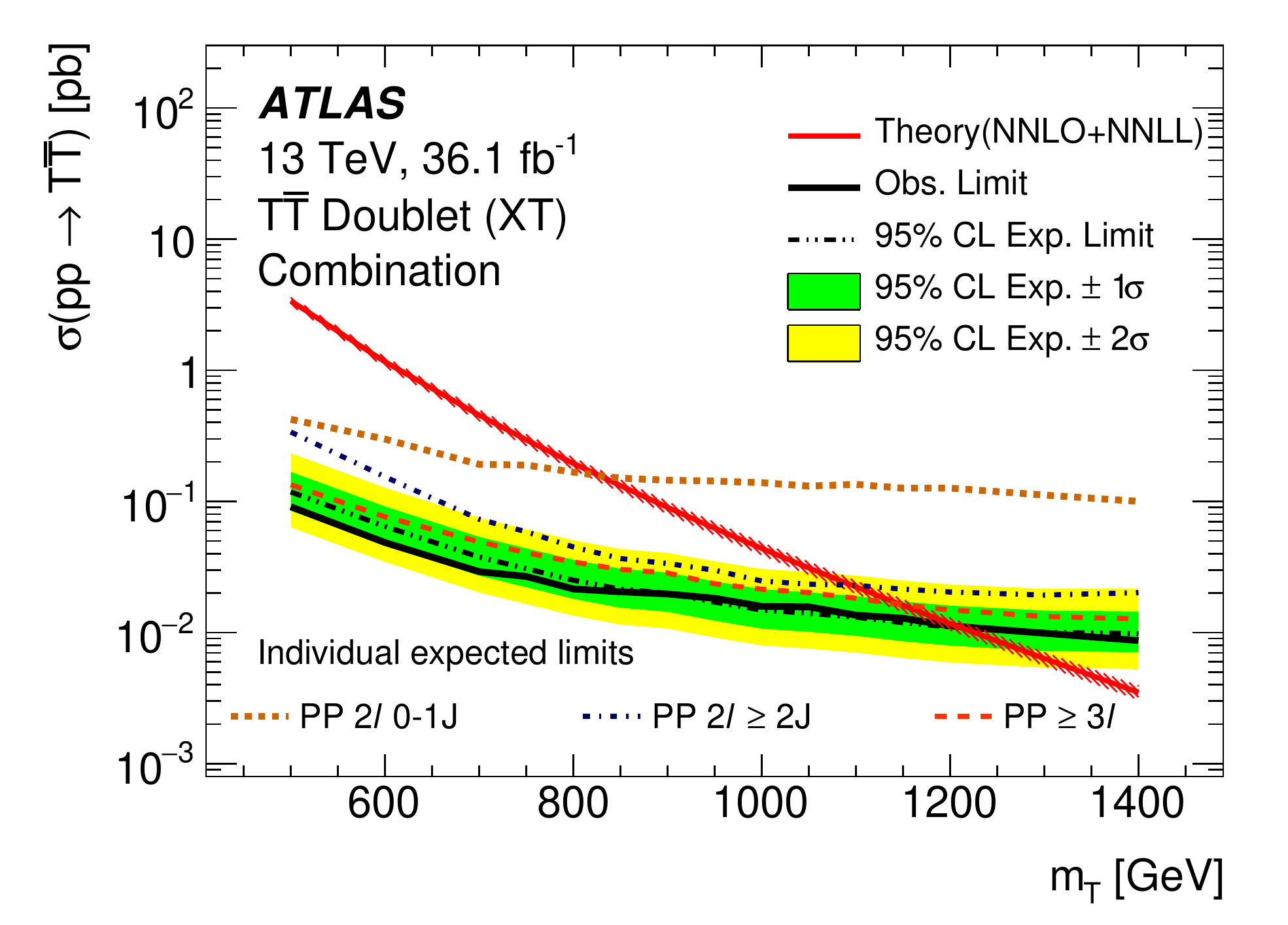}}
\subfloat[]{\includegraphics[width=.49\textwidth]{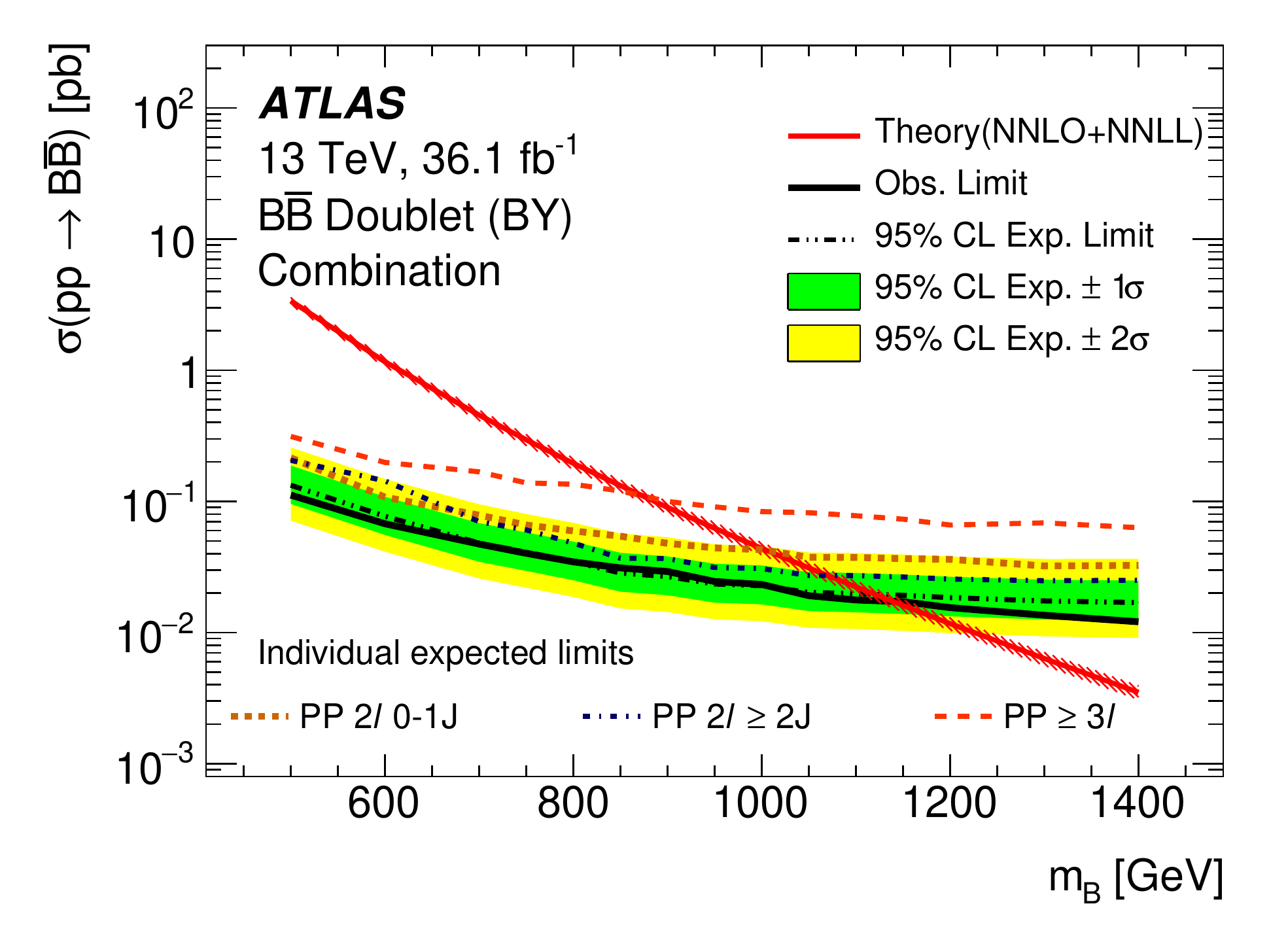}}\\
\subfloat[]{\includegraphics[width=.49\textwidth]{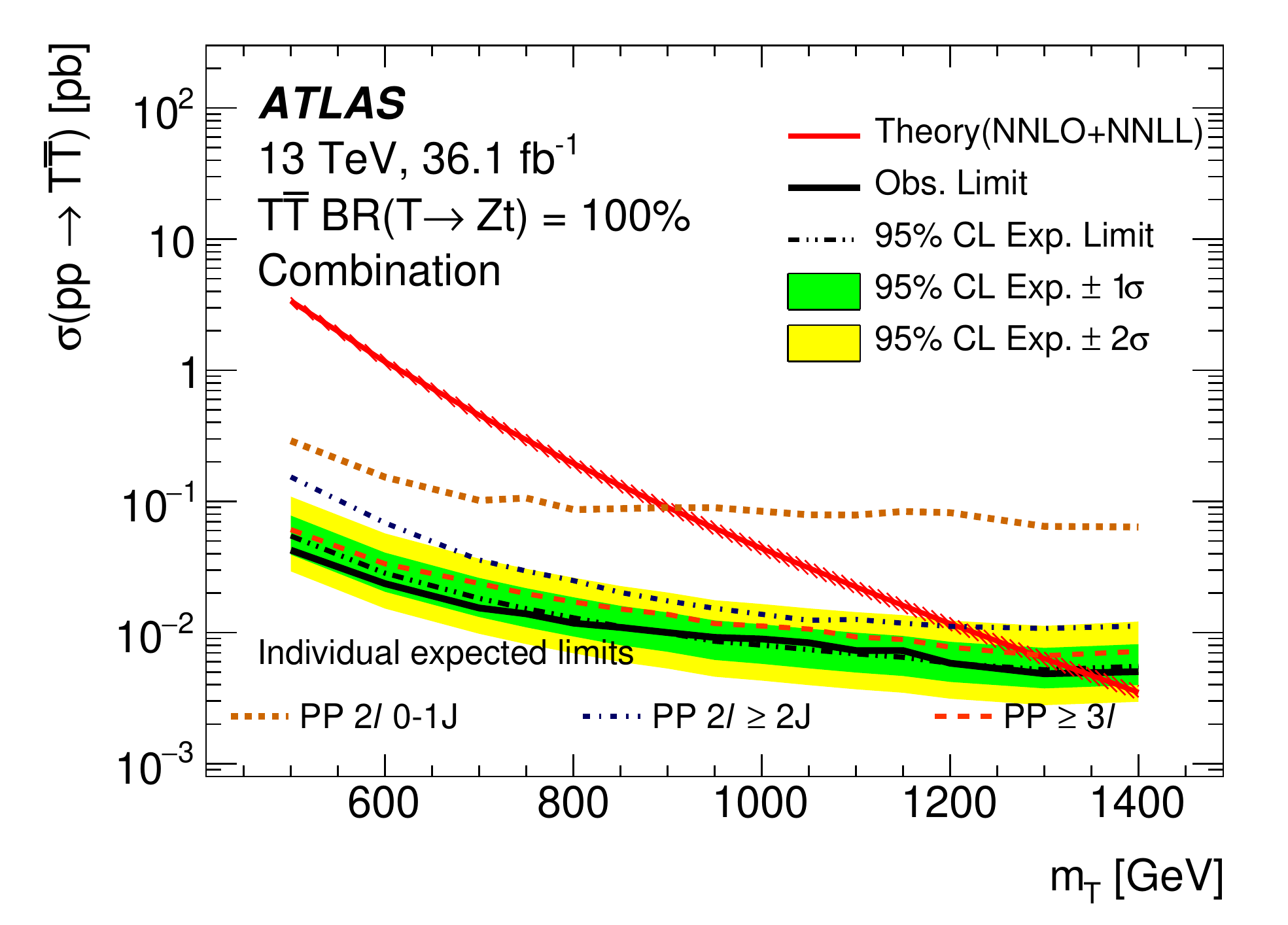}}
\subfloat[]{\includegraphics[width=.49\textwidth]{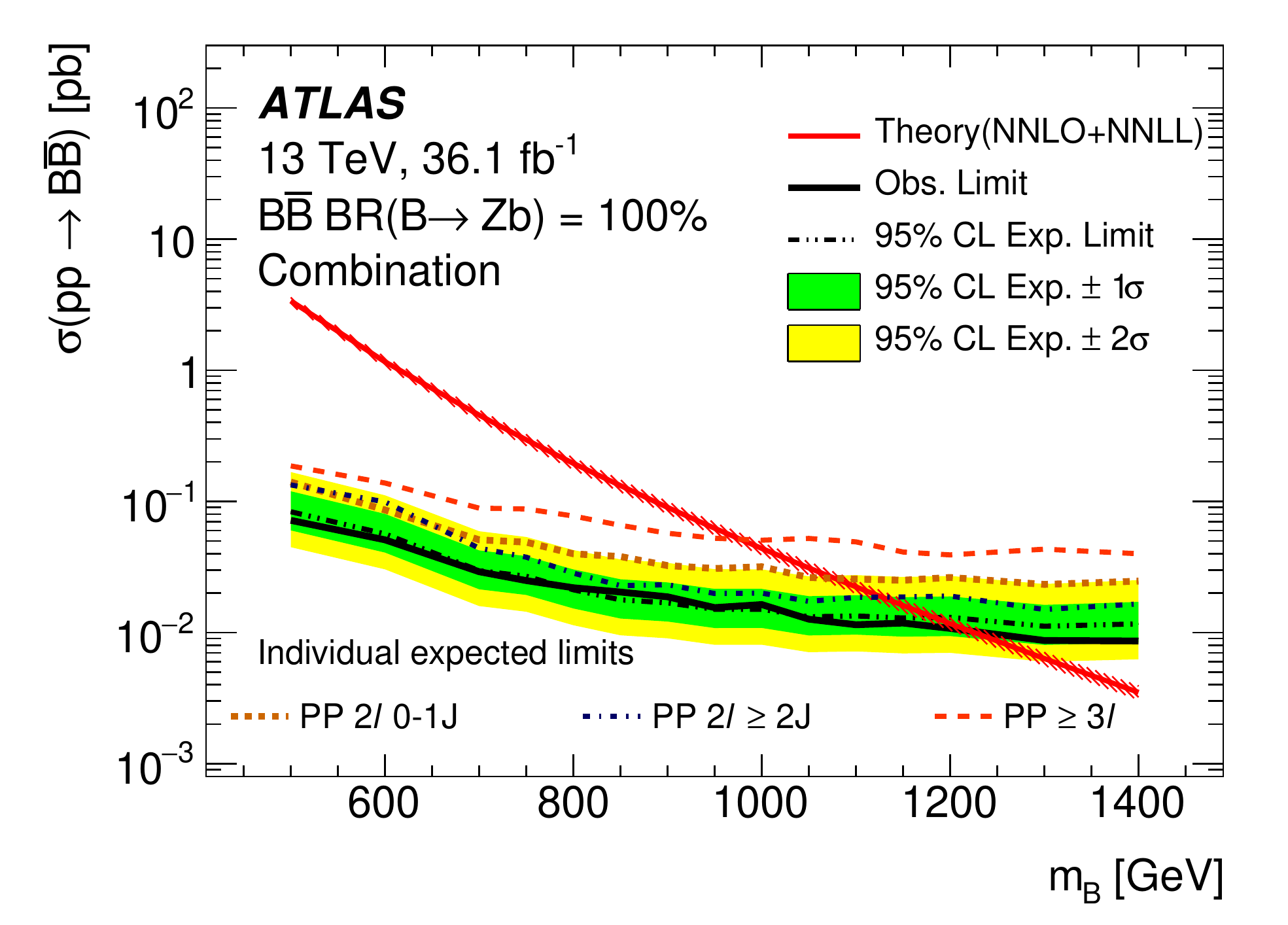}}
\caption{Upper limits at 95\% CL on the cross section of vectorlike quark pair production (PP) for (a) \TTbar\ in the singlet model, (b) \BBbar\ in the singlet model, (c) \TTbar\ in the doublet model, (d) \BBbar\ in the doublet model, (e) \TTbar\ with a BR of 100\% to $Zt$, and (f) \BBbar\ with a BR of 100\% to $Zb$. The expected limits are shown for the individual channels and for the combination of the channels, as are the observed limits for the combination. The expected cross section for pair production is also shown together with its uncertainty.}
\label{fig:limits_pair}
\end{figure}

\begin{table}[t]
\centering
\caption{Observed (expected) 95\% CL mass limits for the singlet and doublet benchmark models, as well as for the case of 100\% BR to $T\rightarrow Zt$ and $B\rightarrow Zb$ for the three pair-production channels and their combination.}
\begin{tabular}{c|c|c|c|c}
  \toprule
  Model & \dilres  & \dilboost  & \tripair & Combination \\
  \midrule
  \TTbar\ singlet         & 740  (720)~\GeV  & 950   (930)~\GeV & 950  (1010)~\GeV & 1030 (1060)~\GeV\\
  \TTbar\ doublet         & 850  (820)~\GeV  & 1100 (1100)~\GeV & 1090 (1150)~\GeV & 1210 (1210)~\GeV\\
  100\% $T\rightarrow Zt$ & 920  (900)~\GeV  & 1210 (1210)~\GeV & 1260 (1290)~\GeV & 1340 (1320)~\GeV\\\midrule
  \BBbar\ singlet         & 860  (840)~\GeV  & 930   (950)~\GeV & 890   (940)~\GeV & 1010 (1030)~\GeV\\
  \BBbar\ doublet         & 1040 (1000)~\GeV & 1060 (1070)~\GeV & 820   (880)~\GeV & 1140 (1120)~\GeV\\
  100\% $B\rightarrow Zb$ & 1110 (1080)~\GeV & 1120 (1130)~\GeV & 930   (980)~\GeV & 1220 (1180)~\GeV\\\bottomrule
\end{tabular}
\label{tab:limits_pair}
\end{table}

In \Fig{\ref{fig:limits_pair}}, the expected and observed upper limits on the pair-production cross section are shown as a function of $\mVLQ$ for $T$ and $B$ quarks with different assumptions for the BRs: for singlet BRs; doublet BRs; and the case of 100\% BR, to $Zt$ or $Zb$, respectively. Also shown are the expected upper limits on the cross section for the individual pair-production channels. The limits are compared with the predicted pair-production cross section, which results in lower limits on the mass of the $T$ and $B$ quarks in the different benchmark scenarios. These are summarized in \Tab{\ref{tab:limits_pair}} for the individual channels, as well as for their combination.

\begin{figure}[!ht]
\centering
\subfloat[]{\includegraphics[width=.49\textwidth]{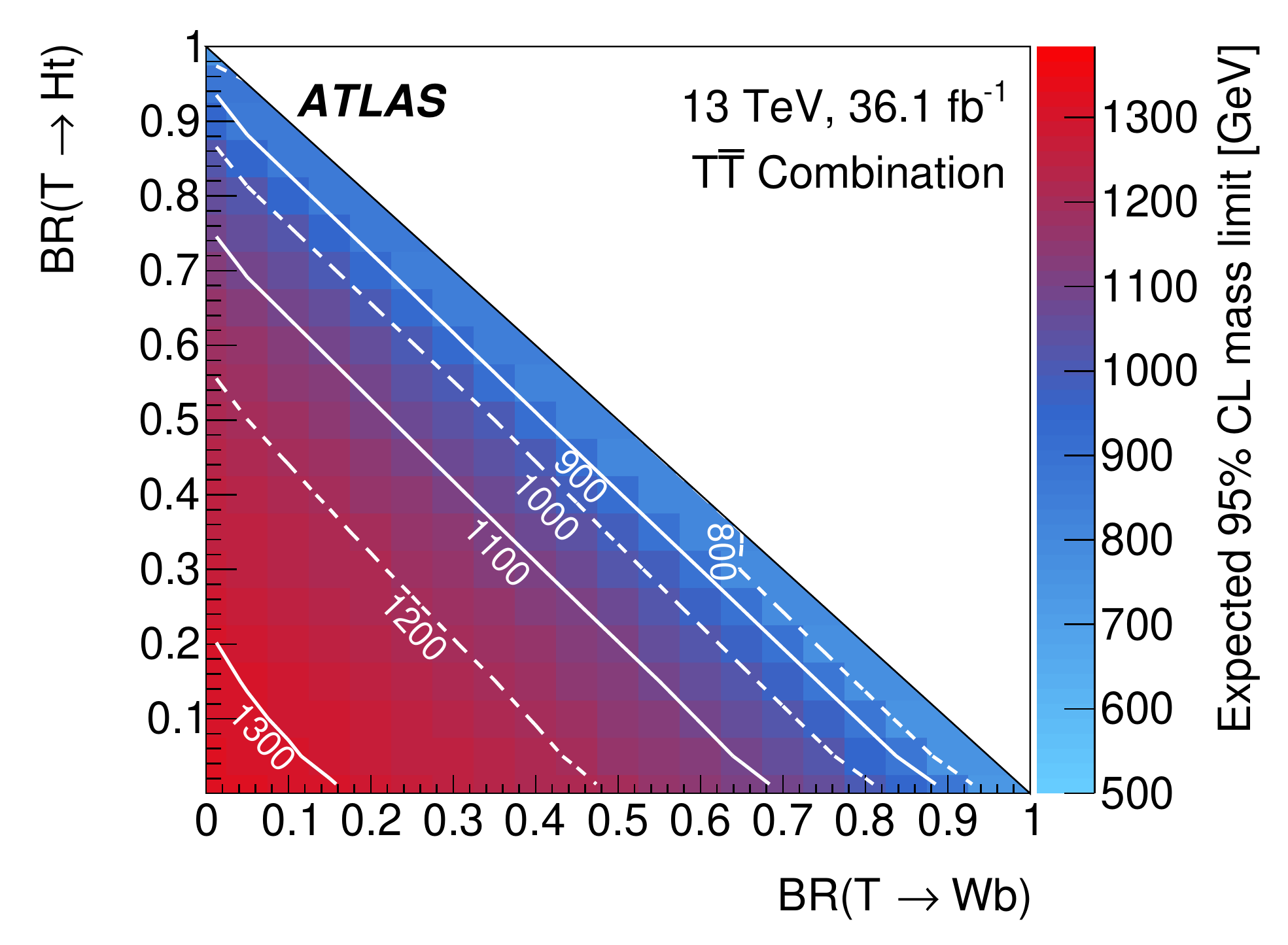}}
\subfloat[]{\includegraphics[width=.49\textwidth]{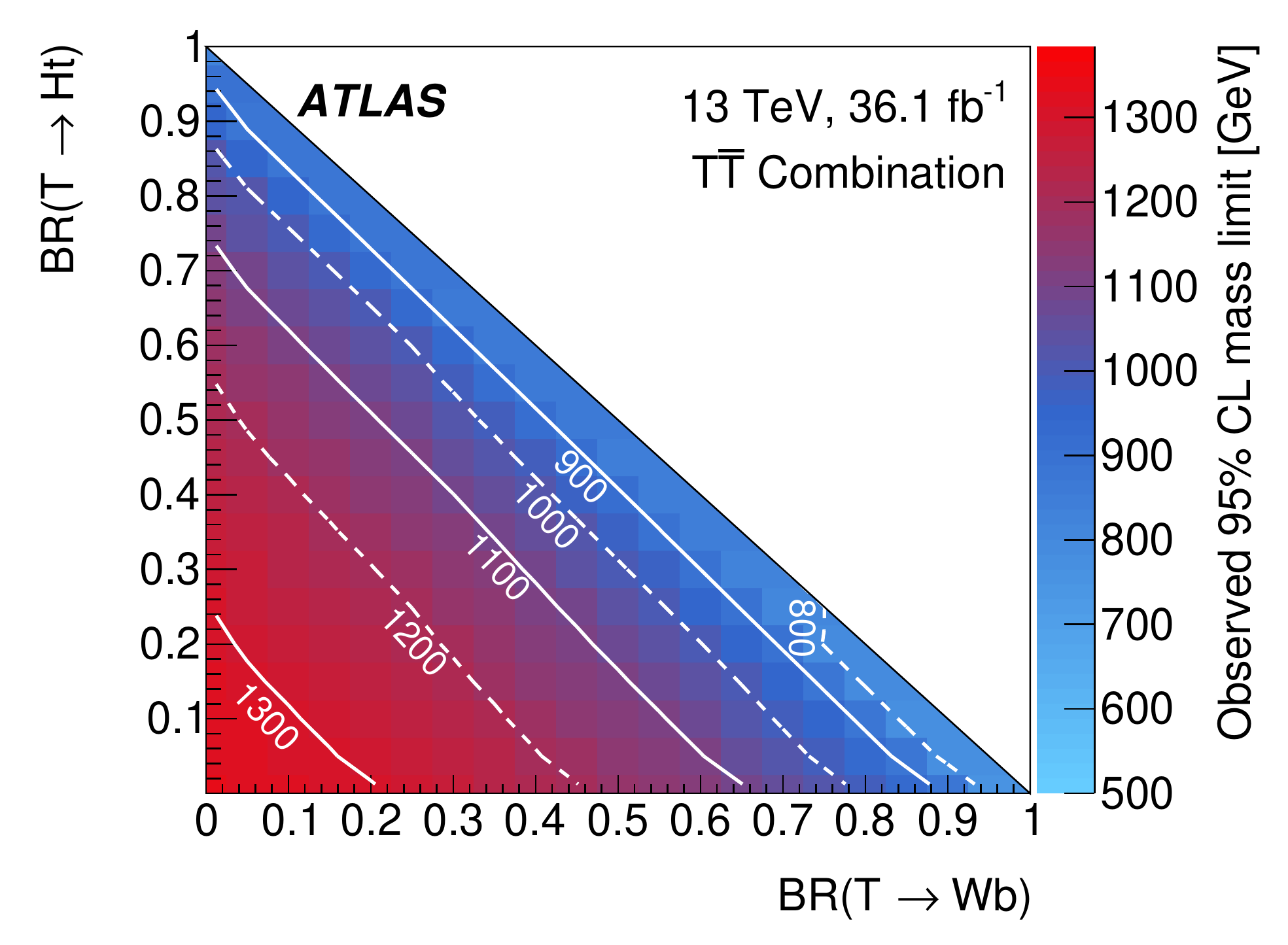}}\\
\subfloat[]{\includegraphics[width=.49\textwidth]{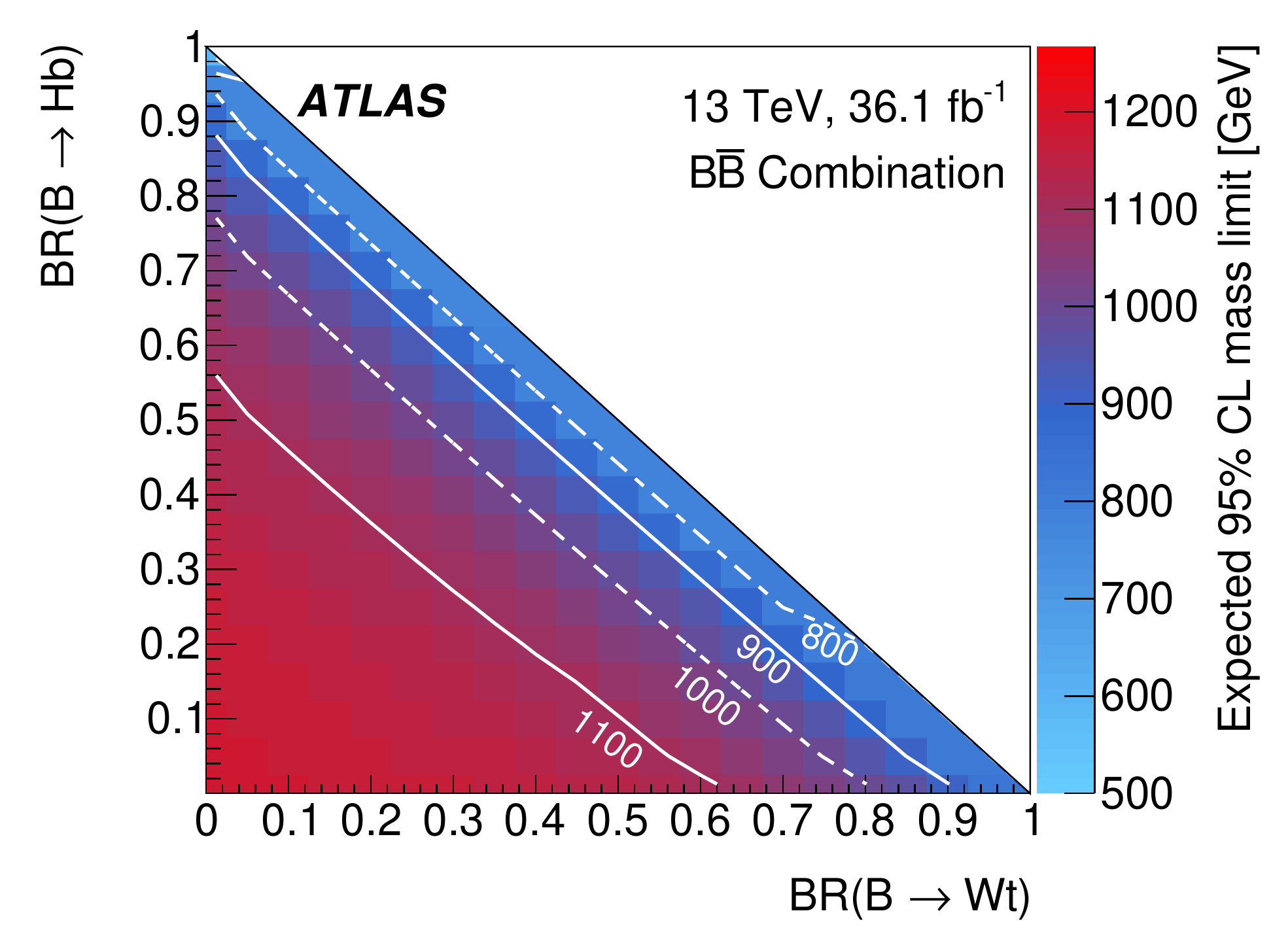}}
\subfloat[]{\includegraphics[width=.49\textwidth]{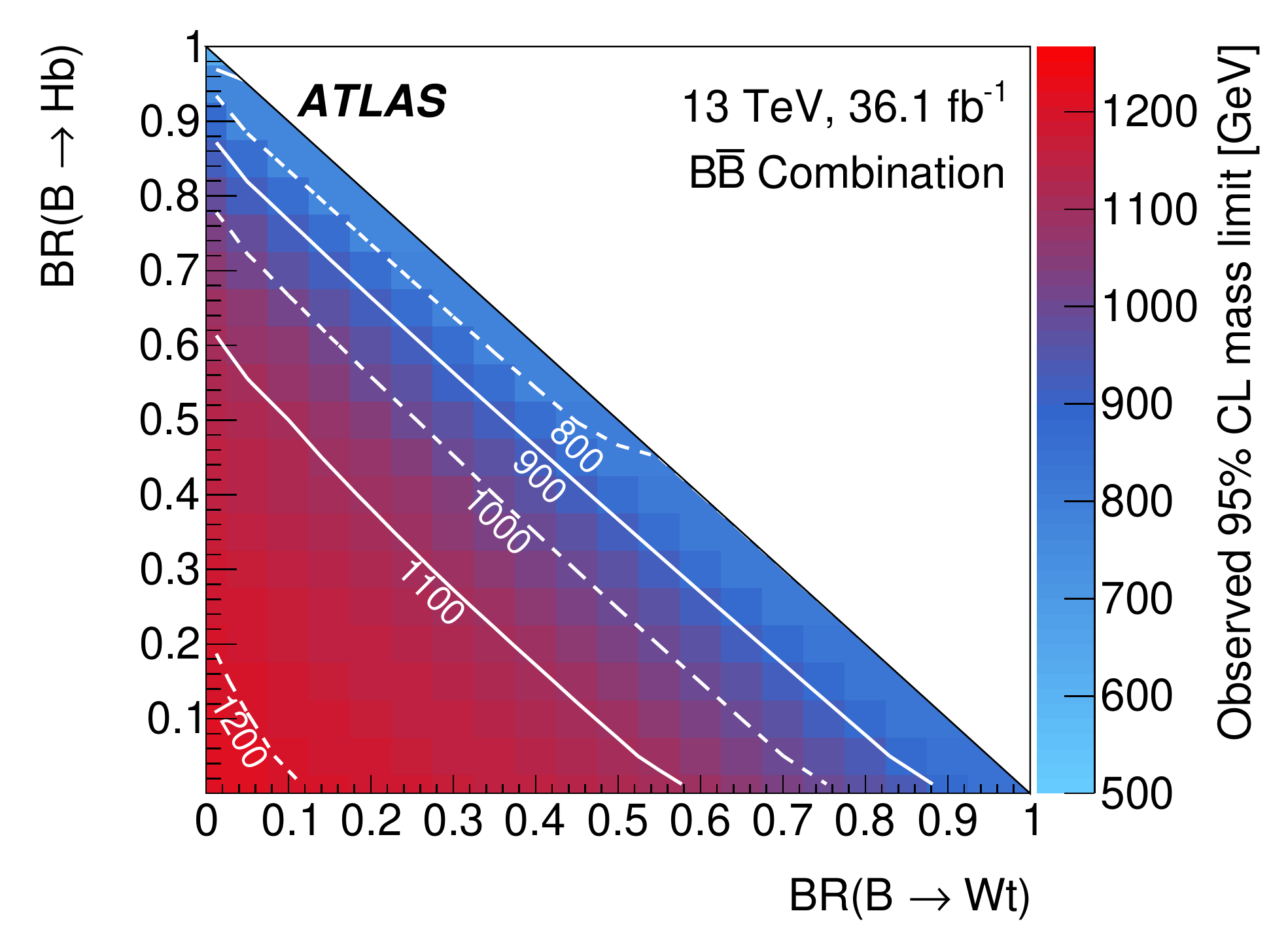}}
\caption{Expected (a,c) and observed (b,d) 95\% CL lower limits from the combination of the pair-production channels on the mass of vectorlike quarks for all combinations of BRs for (a,b) $T\rightarrow Zt$, $T\rightarrow Ht$, $T\rightarrow Wb$, and (c,d) $B\rightarrow Zb$, $B\rightarrow Hb$, $B\rightarrow Wt$, adding up to unity. The white lines are contours for fixed values of $\mVLQ$.}
\label{fig:limits_BR}
\end{figure}

All three pair-production channels contribute differently to the sensitivity of the analysis in the different benchmark scenarios. The \tripair\ channel is particularly important for the sensitivity to $T\bar{T}$ production, where it has the best sensitivity in all three cases shown in \Fig{\ref{fig:limits_pair}}. While the \dilboost\ channel also contributes significantly to the sensitivity of the analysis, the \dilres\ channel is less sensitive to $T\bar{T}$ production. At very low $T$-quark masses, the \dilboost\ channel loses sensitivity compared with the \tripair\ channel, because the decay products of the $T$ quarks are less boosted and result in fewer \ljets. For $B\bar{B}$ production, the sensitivity is driven by the \dilres\ and \dilboost\ channels in the case of the doublet and 100\% BR benchmarks, with less sensitivity from the \tripair\ channel. In general, the \dilres\ and \dilboost\ channels have a similar sensitivity to $B\bar{B}$ production, with the \dilboost\ becoming slightly more sensitive than the \dilres\ channel at higher $B$-quark masses. In the case of the singlet BRs for $B\bar{B}$ production, the \tripair\ channel contributes more to the sensitivity than for the doublet and 100\% BR cases. In the singlet case, the BR to $Zb$ is only $\approx 25$\% with $\approx 25$\% of the $B$ quarks decaying into $Hb$ and $\approx 50$\% decaying into $Wt$, while in the doublet and 100\% BR cases, the decay into $Wt$ is not allowed. Due to the significant probability of either the $W$ boson or the top quark decaying into a final state with an electron or muon, the \tripair\ channel is particularly sensitive to the final state $ZbWt$, which explains the high sensitivity of the \tripair\ channel to $B\bar{B}$ production in the singlet case.

In \Fig{\ref{fig:limits_BR}}, the expected and observed limits on the $T$-quark ($B$-quark) mass from the combination of the pair-production channels are shown as a function of the BRs to $Ht$ ($Hb$) and $Wb$ ($Wt$), where the BR into $Zt$ ($Zb$) is calculated by requiring that the BRs to these three decay modes add up to unity. It can be seen that the analysis is particularly sensitive to high BRs to $Zt$ or $Zb$ (lower left corners), but it also has good sensitivity for many other combinations of the three BRs.

Systematic uncertainties play a smaller role in the sensitivity of the different channels than the statistical uncertainties. In order to quantify the impact of systematic uncertainties, the limits obtained with the nominal analysis, which includes systematic uncertainties, are compared to an analysis where only statistical uncertainties are included. Compared to the nominal analysis, the expected upper limits on the pair-production cross section improve by approximately 30\%, 15\% and 10\% when systematic uncertainties are neglected in the \dilres\ channel, in the \dilboost\ channel, and in the \tripair\ channel, respectively, for $\mVLQ = 750~\GeV$. For higher VLQ masses, the effect of systematic uncertainties decreases further and reaches values of 10--15\%, 6--7\%, and 5\% in the three channels. In the \dilres\ and \dilboost\ channels, the main contributions come from the modeling uncertainties of the \zjets\ and \ttbar\ backgrounds and the \ljet\ resolution uncertainties. The NPs associated with these uncertainties each change the total background expectation in the SR before the fit by up to 9\% (for $\ttbar$ modeling in the SR with exactly one \ljet) to 24\% (for \ljet\ \pt resolution in the SR with no \ljet), depending on the uncertainty, the dilepton channel (\dilres\ or \dilboost), and the SR. In the \tripair\ channel, the modeling of the diboson background, in particular the uncertainty in the background from dibosons produced in association with $b$-quarks, is responsible for the main contributions to the systematic uncertainties. The impact of this uncertainty is 13\% on the total background expectation in the \tripair\ channel SR before the fit.

\begin{figure}[!ht]
\centering
\includegraphics[width=.6\textwidth]{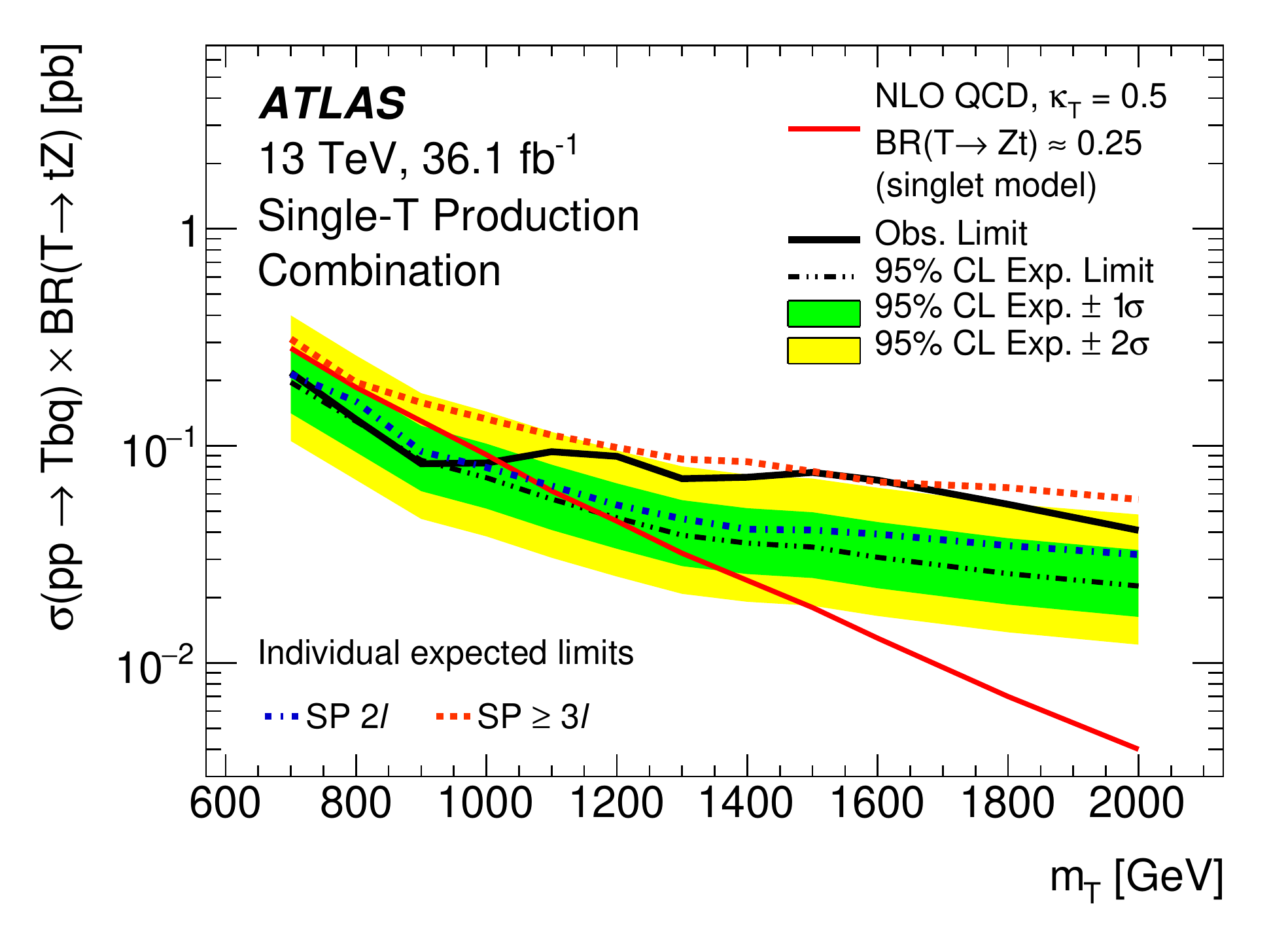}
\caption{Upper limits at 95\% CL on the cross section times BR to $Zt$ of single production (SP) of a $T$-quark. The expected limits are shown for the individual channels and for the combination of the channels, as are the observed limits for the combination. The expected cross section times BR to $Zt$ for single-$T$-quark production is also shown for a coupling $\kappa_T = 0.5$, which corresponds to a coupling of $c_W = \sqrt{c^2_{W,L} + c^2_{W,R}} = 0.45$ from Ref.~\cite{Matsedonskyi:2014mna}. The BR assumed here corresponds to the singlet benchmark model, i.e. $\approx 25$\%.}
\label{fig:limits_single1D}
\end{figure}

As in the case of the pair-production channels, the \dilsing\ and \trising\ channels were combined using the same correlation scheme for the NPs as for the pair-production channels described above. Possible interference effects with SM background processes were estimated to be small and were not taken into account in the interpretation. Only production via the coupling of the $T$ quark to the $W$ boson was considered. The signal efficiency in the SR of the \dilsing\ channel is similar for production via the coupling of the $T$ quark to the $Z$ boson and it is about a factor of two higher for the \trising\ channel. The expected signal cross section for production via the $Z$ boson is, however, roughly an order of magnitude smaller than for production via the $W$ boson for the same coupling value~\cite{Matsedonskyi:2014mna}, so that production via the $Z$ boson was neglected in this analysis.

In \Fig{\ref{fig:limits_single1D}}, the expected and observed upper limits on the single-$T$-quark production cross section times BR to $Zt$ are shown as a function of $m_T$. Also shown are the expected upper limits on the cross section times BR for the individual single-production channels. The observed limit deviates from the expected limit by about $2\sigma$ for high values of $m_T$, which is consistent with the upward fluctuations observed in the discriminating variables in both single-production channels (\Fig{\ref{fig:dilepsing_discr_SR}} and \Fig{\ref{fig:trilepsing_discr_SR}}). The limits are compared with the predicted single-production cross section times BR for the benchmark coupling of $\kappa_T = 0.5$ (at which the MC samples were produced) which corresponds to a coupling of $c_W = \sqrt{c^2_{W,L} + c^2_{W,R}} = 0.45$~\cite{Matsedonskyi:2014mna}. The \dilsing\ channel, which explicitly exploits the presence of high-\pt\ top quarks with top-tagging, is more sensitive than the \trising\ channel for all values of $\mVLQ$ studied, but the \trising\ channel contributes significantly to the combination of the two channels.

Similarly to the pair-production analysis, systematic uncertainties play a smaller role in the sensitivity of the different single-production channels than the statistical uncertainties. In order to quantify the impact of systematic uncertainties, the limits obtained with the nominal analysis, which includes systematic uncertainties, are compared to an analysis where only statistical uncertainties are included. Compared to the nominal analysis, the expected upper limits on the single-production cross section times BR improve by approximately 30\% in the \dilsing\ channel and approximately 10\% in the \trising\ channel for $\mVLQ = 900~\GeV$ when systematic uncertainties are neglected. At higher masses, the impact on the expected upper limits decreases to 5--10\% in both channels. The main contributions in the \dilsing\ channel originate from the uncertainties in the forward-jet modeling, the modeling of the \zjets\ background and in the \ljet\ mass resolution. The NPs associated with these uncertainties each change the total background expectation in the dilepton SR before the fit by up to 5--25\%. In the \trising\ channel, the main contributions to the systematic uncertainty arise from the $\ttbar+V$ theoretical cross section, misidentified leptons, and the modeling of the background from dibosons produced in association with $b$-quarks. The impact of these uncertainties is 1--12\% on the total background expectation before the fit in the \trising\ channel SR.

\begin{figure}[!ht]
\centering
\subfloat[]{\includegraphics[width=.49\textwidth]{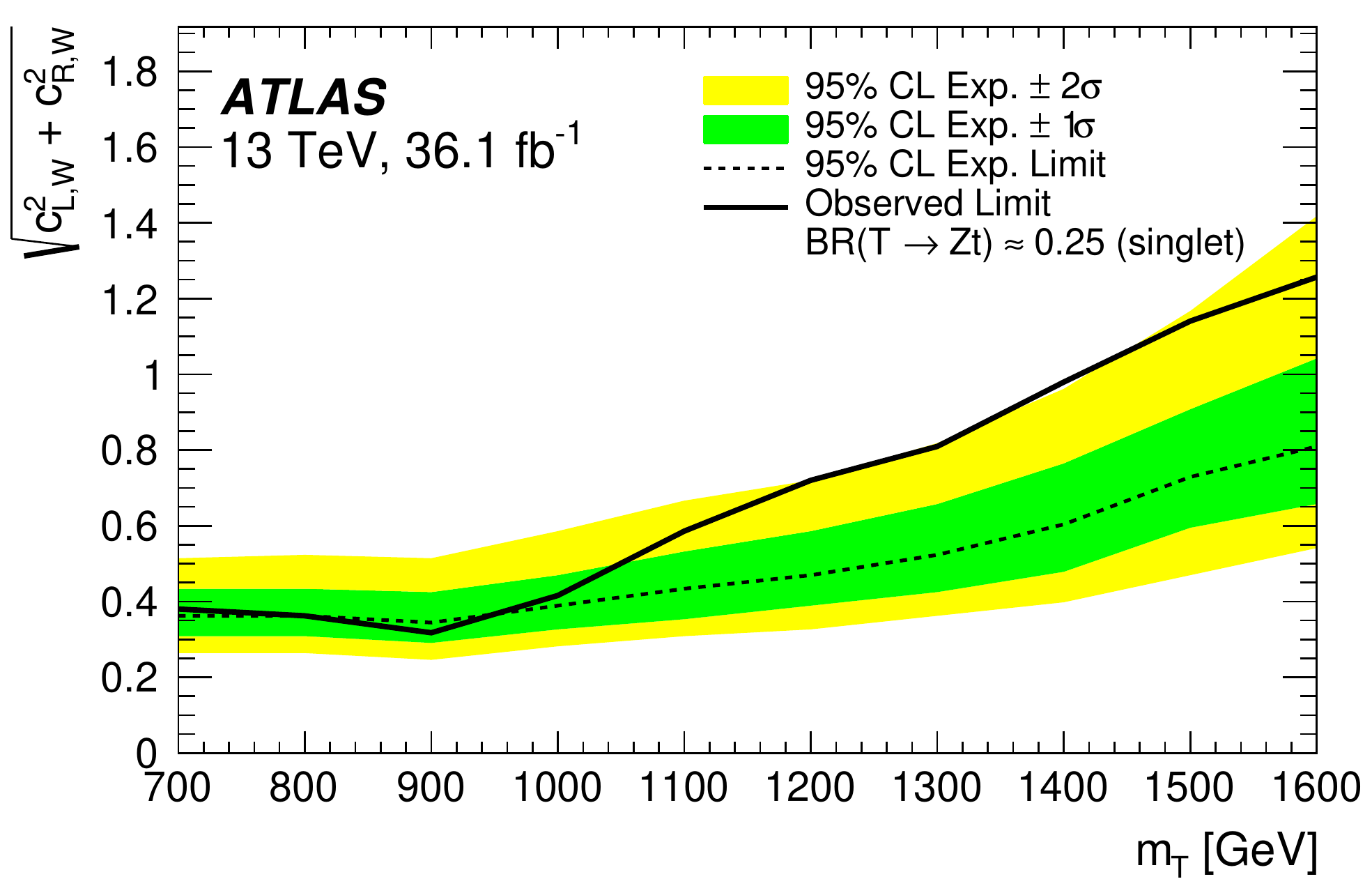}\label{fig:2D_a}}
\subfloat[]{\includegraphics[width=.49\textwidth]{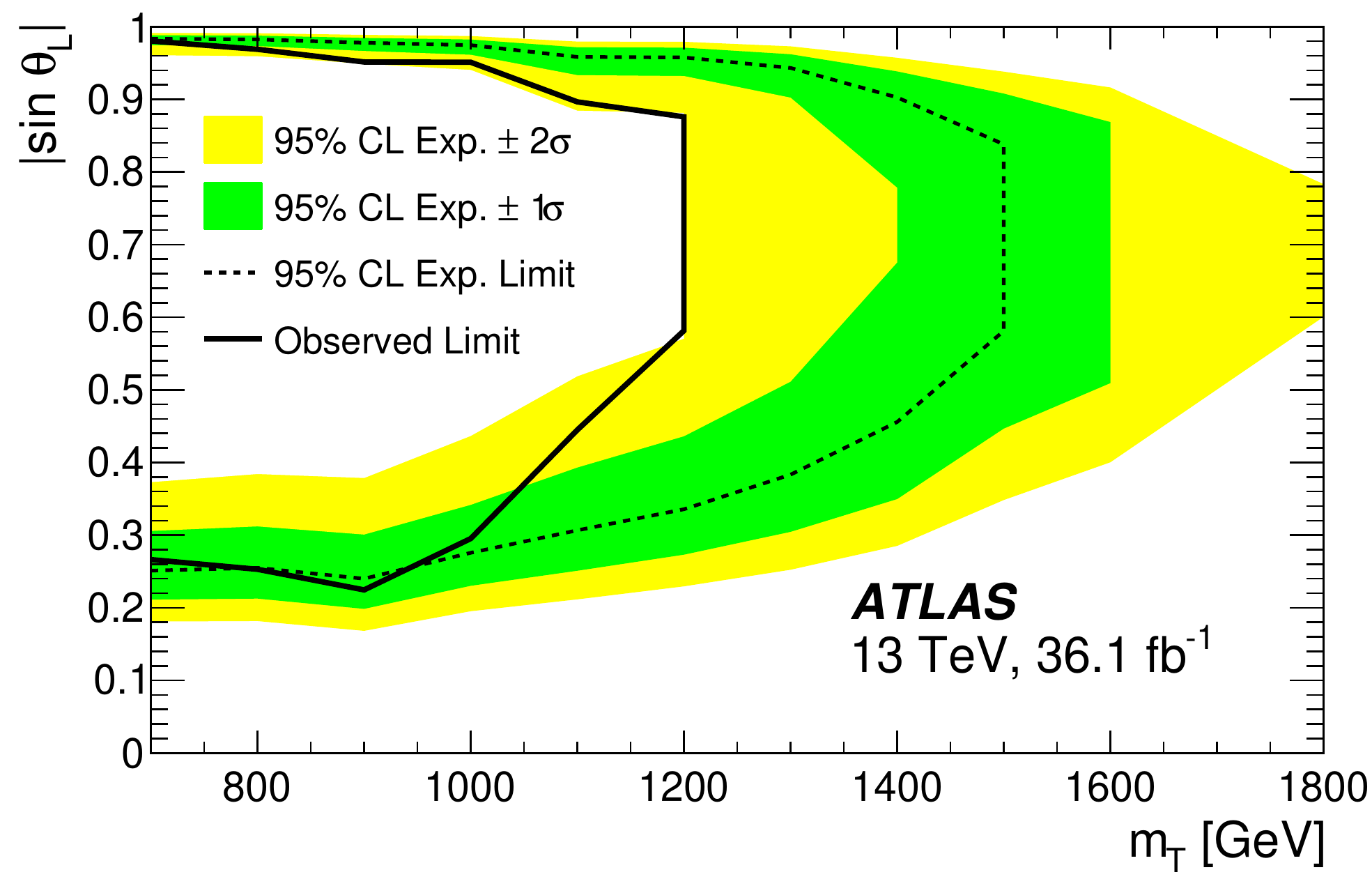}\label{fig:2D_b}}
\caption{Expected and observed 95\% CL limits from the combination of the single-production channels (a) on the coupling of the $T$ quark to SM particles, $c_W = \sqrt{c^2_{W,L} + c^2_{W,R}}$, from Ref.~\cite{Matsedonskyi:2014mna} assuming the singlet-model BR of $\approx 25$\%, and (b) on the mixing angle in the singlet model between the $T$ quark and the top quark, $|\sin\theta_L|$, from Ref.~\cite{Aguilar-Saavedra:2013qpa}, as a function of the mass of the $T$ quark, $m_T$. Values of $c_W$ larger than the observed limit are excluded, and values of $|\sin\theta_L|$ enclosed by the observed limit are excluded, i.e. for $m_T$ larger than $\approx 1200~\GeV$, no value of $|\sin\theta_L|$ is excluded.}
\end{figure}

\begin{figure}[!ht]
\centering
\subfloat[]{\includegraphics[width=.49\textwidth]{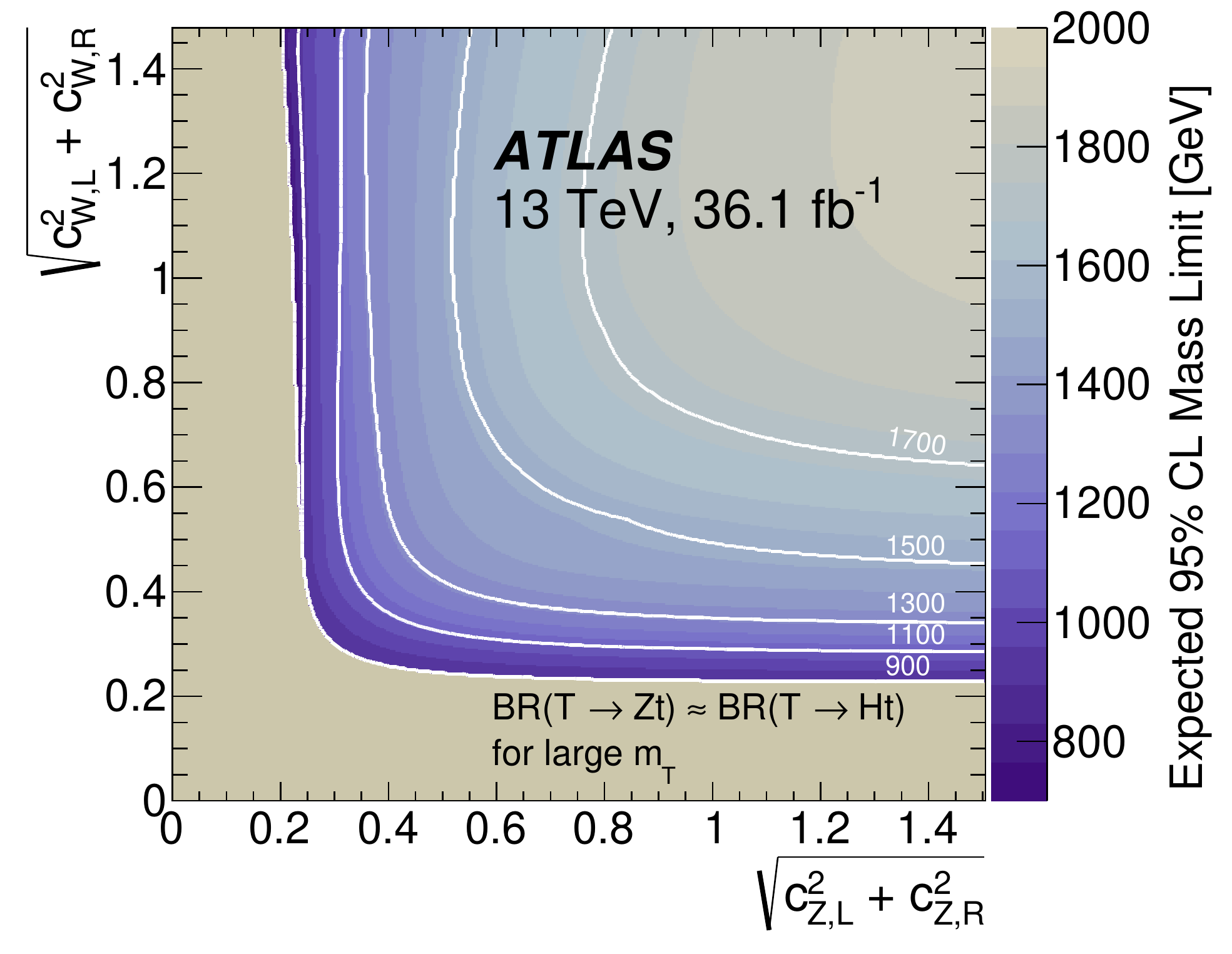}}
\subfloat[]{\includegraphics[width=.49\textwidth]{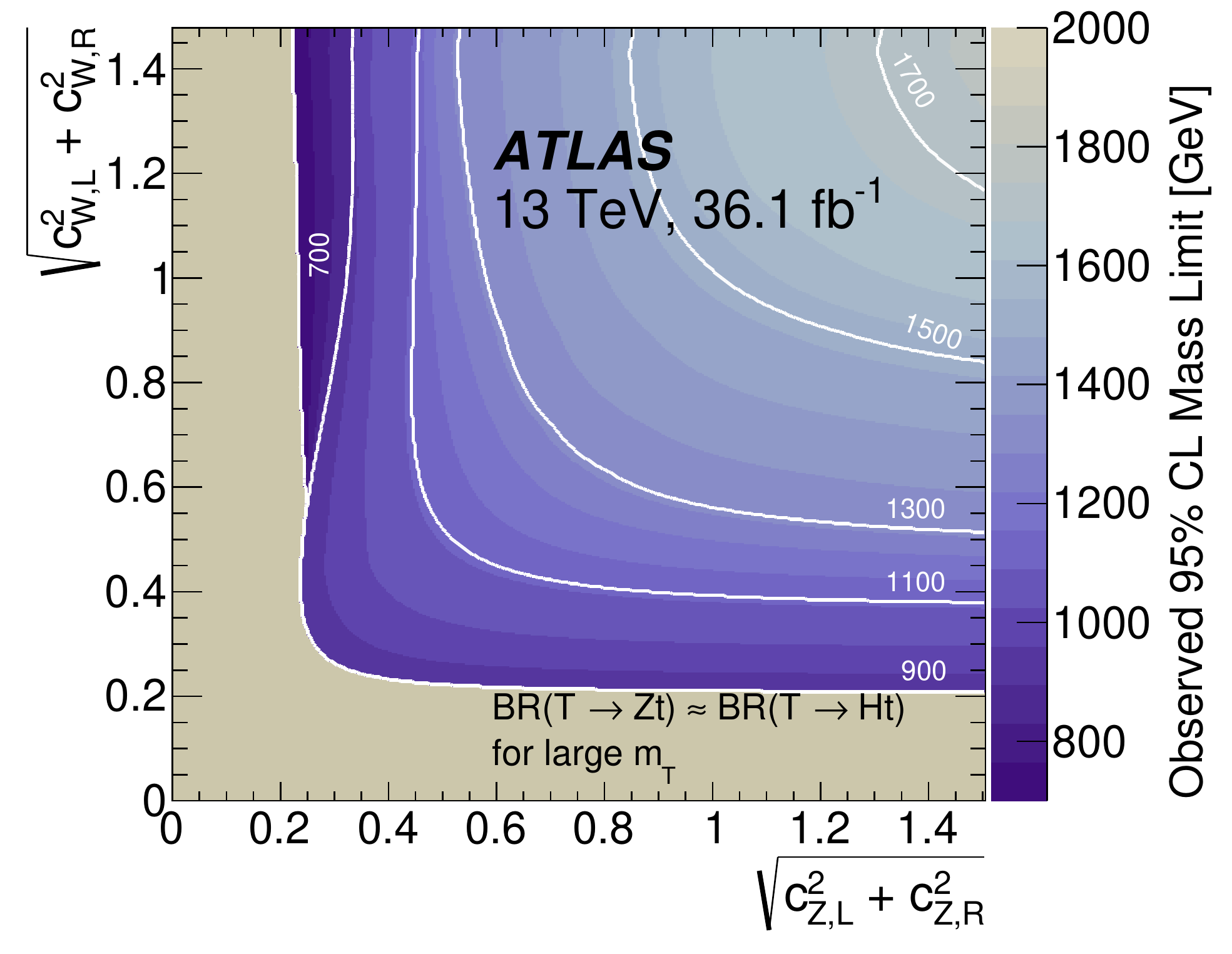}}
\caption{Expected (a) and observed (b) lower limit from the combination of the single-production channels on the mass of the $T$ quark as a function of the couplings of the $T$ quark to the $W$ boson, $\sqrt{c^2_{W,L} + c^2_{W,R}}$, and to the $Z$ boson, $\sqrt{c^2_{Z,L} + c^2_{Z,R}}$ with the assumption of equal BRs for $T\rightarrow Zt$ and $T\rightarrow Ht$ in the limit of large $T$-quark masses. The gray area corresponds to a region that is not excluded for any mass value tested because of the limited sensitivity of the analysis for very small $T$-quark masses. The white lines are contours for fixed values of $\mVLQ$.}
\label{fig:limits_single3D}
\end{figure}

The cross section for single-$T$-quark production does not only depend on the VLQ mass, but also on its coupling to SM quarks, in particular the coupling to $Wb$, which enters the lowest-order $t$-channel diagram for this process. A change in the coupling, however, also results in a change in the width of the $T$-quark mass distribution. The effect of the changing resonance width with the coupling is taken into account by reweighting the discriminating variable in the nominal samples with $\kappa_T = 0.5$ to different couplings, based on large MC samples that are generated without a detector simulation. The reweighting procedure was validated using samples that were generated with $\kappa_T =0.1$ and $1.0$ at $m_T = 900~\GeV$ including the detector simulation. Comparing distributions from these validation samples with distributions that were reweighted from the nominal samples with $\kappa_T = 0.5$ to values of $0.1$ and $1.0$, a small non-closure uncertainty of 3\% was assigned to the single-$T$-quark normalization and this has negligible impact on the sensitivity of the analysis.

Expected and observed limits on the coupling as a function of $m_T$ are shown in \Fig{\ref{fig:2D_a}}, assuming the singlet model with $c_W = \sqrt{2} \frac{m_W}{m_Z} c_Z$, which results in the singlet BR to $Zt$ of $\approx 25$\% over the mass range studied in this analysis. For low values of $m_T$, couplings larger than 0.3--0.4 are excluded. For larger masses, the lower limits on the coupling increases because the single-production cross section decreases for a given coupling value with increasing $m_T$. The coupling $c_W$ can also be expressed in terms of a mixing angle with the top quark in the singlet model, $|\sin\theta_L|$, as defined in Ref.~\cite{Aguilar-Saavedra:2013qpa} by $c_W = \sqrt{2} |\sin\theta_L|$ and $c_Z = \frac{m_Z}{m_W}|\sin\theta_L\cos\theta_L|$. Since the mixing angle enters not only in the production cross section but also in the calculation of the BRs, the expected and observed limits shown in \Fig{\ref{fig:2D_b}} show a lower and an upper branch. For a given mass, only values of $|\sin\theta_L|$ between these two branches are excluded. For values of $m_T$ larger than $\approx 1200~\GeV$, no value of the mixing angle can be excluded.

The limits presented in \Fig{\ref{fig:2D_a}} are only valid for BRs as predicted for the singlet model. In order to lift this assumption, the results of the search for single-$T$-quark production are interpreted in terms of couplings to the $W$, $Z$ and Higgs boson, $c_W$, $c_Z$ and $c_H$, with the assumption that the BRs to $Zt$ and $Ht$ are equal in the large-$m_T$ limit, as it is the case in many multiplets~\cite{Aguilar-Saavedra:2013qpa}. This assumption defines the value of $c_H$ for given values of $c_W$ and $c_Z$. In \Fig{\ref{fig:limits_single3D}}, expected and observed limits on the $T$-quark mass are shown as a function of $c_W$ and $c_Z$. Again, no distinction between left- and right-handed couplings is made because the analysis is not sensitive to differences in the chirality of the couplings. Therefore, $c_W$ and $c_Z$ are defined as the sum in quadrature of left- and right-handed couplings. It can be seen that for large values of $c_W$ and $c_Z$, $T$-quark masses smaller than $1600~\GeV$ can be excluded, while for very small values of the couplings no limits can be set, as indicated by the gray area.

\FloatBarrier

\section{Conclusions}
A search for vectorlike quarks is presented, which uses 36.1~\ifb\ of $pp$ collision data taken with the ATLAS detector at the Large Hadron Collider at $\sqrt{s} = 13~\TeV$. Five channels are used for sensitivity to the production of vectorlike quarks with at least one vectorlike quark decaying into a $Z$ boson and a third-generation Standard Model quark.

Three channels are optimized for sensitivity to the pair production of vectorlike quarks. Two dilepton channels make use of \ljets\ to discriminate the signal from the Standard Model background, and a trilepton channel uses the presence of an additional charged lepton to define a signal-enriched region. The modeling of the main background processes was validated in background-dominated regions and no excess over the background-only expectation was found in the search regions. The three channels were combined and upper limits on the cross section for the pair production of vectorlike quarks were set at 95\% CL as a function of the mass of the vectorlike quark. These limits were interpreted as lower limits on the masses of vectorlike quarks, yielding $m_T > 1030~\GeV$ ($m_T > 1210~\GeV$) and $m_B > 1010~\GeV$ ($m_B > 1140~\GeV$) in the singlet (doublet) model, significantly exceeding the existing limits from Run 1. In the case of 100\% branching ratio for $T\rightarrow Zt$ ($B\rightarrow Zb$), lower limits of $m_T > 1340~\GeV$ ($m_B > 1220~\GeV$) were set.

Two channels were optimized for sensitivity to the single production of vectorlike quarks. A dilepton channel makes use of \ljets\ and top-tagging to separate the signal from the background, and a trilepton channel uses the presence of an additional charged lepton to suppress the background. The modeling of the main background processes was validated in background-dominated regions and no excess over the background-only expectation was found in the search regions. The two channels were combined and 95\% CL upper limits were set on the coupling of vectorlike quarks to Standard Model quarks as a function of the mass of the vectorlike quark. The corresponding limits on the production cross section times branching fraction into $Zt$ are in the range 0.16--0.18~pb at $m_T = 700~\GeV$ and decrease to 0.03--0.05~pb at $m_T = 2000~\GeV$, depending on the value of the coupling in the range $\kappa_T = 0.1$--1.6.

The results presented in this paper significantly tighten the existing bounds on the pair production of vectorlike $T$ and $B$ quarks that decay with a large branching ratio into a $Z$ boson and a third-generation quark and they present competitive bounds on the single production of vectorlike $T$ quarks that decay into a $Z$ boson and a top quark. The results on the pair-production search were combined~\cite{Aaboud:2018pii} with other searches by the ATLAS Collaboration~\cite{Aaboud:2017qpr,Aaboud:2017zfn,Aaboud:2018xuw,WtX,Aaboud:2018xpj,Aaboud:2018wxv}, in order to improve the overall sensitivity to vectorlike $T$ and $B$ quarks.

\section*{Acknowledgments}


We thank CERN for the very successful operation of the LHC, as well as the
support staff from our institutions without whom ATLAS could not be
operated efficiently.

We acknowledge the support of ANPCyT, Argentina; YerPhI, Armenia; ARC, Australia; BMWFW and FWF, Austria; ANAS, Azerbaijan; SSTC, Belarus; CNPq and FAPESP, Brazil; NSERC, NRC and CFI, Canada; CERN; CONICYT, Chile; CAS, MOST and NSFC, China; COLCIENCIAS, Colombia; MSMT CR, MPO CR and VSC CR, Czech Republic; DNRF and DNSRC, Denmark; IN2P3-CNRS, CEA-DRF/IRFU, France; SRNSFG, Georgia; BMBF, HGF, and MPG, Germany; GSRT, Greece; RGC, Hong Kong SAR, China; ISF and Benoziyo Center, Israel; INFN, Italy; MEXT and JSPS, Japan; CNRST, Morocco; NWO, Netherlands; RCN, Norway; MNiSW and NCN, Poland; FCT, Portugal; MNE/IFA, Romania; MES of Russia and NRC KI, Russian Federation; JINR; MESTD, Serbia; MSSR, Slovakia; ARRS and MIZ\v{S}, Slovenia; DST/NRF, South Africa; MINECO, Spain; SRC and Wallenberg Foundation, Sweden; SERI, SNSF and Cantons of Bern and Geneva, Switzerland; MOST, Taiwan; TAEK, Turkey; STFC, United Kingdom; DOE and NSF, United States of America. In addition, individual groups and members have received support from BCKDF, CANARIE, CRC and Compute Canada, Canada; COST, ERC, ERDF, Horizon 2020, and Marie Sk{\l}odowska-Curie Actions, European Union; Investissements d' Avenir Labex and Idex, ANR, France; DFG and AvH Foundation, Germany; Herakleitos, Thales and Aristeia programmes co-financed by EU-ESF and the Greek NSRF, Greece; BSF-NSF and GIF, Israel; CERCA Programme Generalitat de Catalunya, Spain; The Royal Society and Leverhulme Trust, United Kingdom. 

The crucial computing support from all WLCG partners is acknowledged gratefully, in particular from CERN, the ATLAS Tier-1 facilities at TRIUMF (Canada), NDGF (Denmark, Norway, Sweden), CC-IN2P3 (France), KIT/GridKA (Germany), INFN-CNAF (Italy), NL-T1 (Netherlands), PIC (Spain), ASGC (Taiwan), RAL (UK) and BNL (USA), the Tier-2 facilities worldwide and large non-WLCG resource providers. Major contributors of computing resources are listed in Ref.~\cite{ATL-GEN-PUB-2016-002}.

\printbibliography

\clearpage
 
\begin{flushleft}
{\Large The ATLAS Collaboration}

\bigskip

M.~Aaboud$^\textrm{\scriptsize 34d}$,    
G.~Aad$^\textrm{\scriptsize 99}$,    
B.~Abbott$^\textrm{\scriptsize 125}$,    
O.~Abdinov$^\textrm{\scriptsize 13,*}$,    
B.~Abeloos$^\textrm{\scriptsize 129}$,    
D.K.~Abhayasinghe$^\textrm{\scriptsize 91}$,    
S.H.~Abidi$^\textrm{\scriptsize 164}$,    
O.S.~AbouZeid$^\textrm{\scriptsize 39}$,    
N.L.~Abraham$^\textrm{\scriptsize 153}$,    
H.~Abramowicz$^\textrm{\scriptsize 158}$
% ...
\bigskip
\\

% ...
$^{13}$Institute of Physics, Azerbaijan Academy of Sciences, Baku; Azerbaijan.\\
% ...
$^{34}$$^{(a)}$Facult\'e des Sciences Ain Chock, R\'eseau Universitaire de Physique des Hautes Energies - Universit\'e Hassan II, Casablanca;$^{(b)}$Centre National de l'Energie des Sciences Techniques Nucleaires (CNESTEN), Rabat;$^{(c)}$Facult\'e des Sciences Semlalia, Universit\'e Cadi Ayyad, LPHEA-Marrakech;$^{(d)}$Facult\'e des Sciences, Universit\'e Mohamed Premier and LPTPM, Oujda;$^{(e)}$Facult\'e des sciences, Universit\'e Mohammed V, Rabat; Morocco.\\
% ...
$^{39}$Niels Bohr Institute, University of Copenhagen, Copenhagen; Denmark.\\
% ...
$^{91}$Department of Physics, Royal Holloway University of London, Egham; United Kingdom.\\
% ...
$^{99}$CPPM, Aix-Marseille Universit\'e, CNRS/IN2P3, Marseille; France.\\
% ...
$^{125}$Homer L. Dodge Department of Physics and Astronomy, University of Oklahoma, Norman OK; United States of America.\\
% ...
$^{129}$LAL, Universit\'e Paris-Sud, CNRS/IN2P3, Universit\'e Paris-Saclay, Orsay; France.\\
% ...
$^{153}$Department of Physics and Astronomy, University of Sussex, Brighton; United Kingdom.\\
% ...
$^{158}$Raymond and Beverly Sackler School of Physics and Astronomy, Tel Aviv University, Tel Aviv; Israel.\\
% ...
$^{164}$Department of Physics, University of Toronto, Toronto ON; Canada.\\
% ...
$^{a}$ Also at Borough of Manhattan Community College, City University of New York, NY; United States of America.\\
$^{b}$ Also at California State University, East Bay; United States of America.\\
$^{c}$ Also at Centre for High Performance Computing, CSIR Campus, Rosebank, Cape Town; South Africa.\\
$^{d}$ Also at CERN, Geneva; Switzerland.\\
$^{e}$ Also at CPPM, Aix-Marseille Universit\'e, CNRS/IN2P3, Marseille; France.\\
% ...
$^{*}$ Deceased

\end{flushleft}


\end{document}